\documentclass[11pt,a4paper]{article}
\pdfoutput=1
\usepackage[pdftex]{graphics}
\usepackage{jheppub}
\usepackage{amsmath,amssymb,amsfonts}
\usepackage{enumitem}
\usepackage{multirow}
\usepackage{array,booktabs}
\usepackage{slashed}

\usepackage{accents}
\usepackage{mathrsfs}
\usepackage{mathtools}

\usepackage[usenames,dvipsnames]{xcolor}
\definecolor{labelcolor}{RGB}{194, 175, 116}
\definecolor{rmkcolor}{RGB}{215,30,5}
\definecolor{feyntext}{RGB}{20,125,233}
\definecolor{lred}{RGB}{255,130,130}
\definecolor{llred}{RGB}{255,160,160}
\definecolor{skyblue}{RGB}{34,139,230}
\definecolor{navy}{rgb}{0,0,0.7}
\definecolor{purple}{RGB}{171,1,207}
\definecolor{lgreen}{RGB}{231, 242, 43}
\definecolor{lgray}{RGB}{135, 135, 145}
\definecolor{recur}{RGB}{231, 242, 43}
\usepackage[export]{adjustbox}
\usepackage[final]{showlabels} % use "inline" or "final"

\newcommand{\fref}[1]{Fig.\,\ref{#1}}
\renewcommand{\eqref}[1]{Eq.\,(\ref{#1})}
\newcommand{\eqrefs}[2]{Eqs.\,(\ref{#1}) and (\ref{#2})}
\newcommand{\Sec}[1]{Sec.\,\ref{#1}}
\newcommand{\Secs}[2]{Secs.\,\ref{#1} and \ref{#2}}
\newcommand{\App}[1]{App.\,\ref{#1}}
\newcommand{\rcite}[1]{Ref.\,\cite{#1}}
\newcommand{\rrcite}[1]{Refs.\,\cite{#1}}
\def\mem{\hspace{0.1em}}
\def\hem{\hspace{0.05em}}
\def\nem{\hspace{-0.1em}}
\def\hnem{\hspace{-0.05em}}
\def\hhem{\hspace{0.025em}}
\def\hhnem{\hspace{-0.025em}}

\def\snem{\hspace{-0.075em}}

\def\qiq{{\quad\implies\quad}}

\def\a{\alpha}
\def\b{\beta}
\def\c{{\gamma}}
\def\d{{\delta}}
\def\e{\epsilon}
\def\ve{\varepsilon}
\def\m{\mu}
\def\n{\nu}
\def\r{\rho}
\def\s{\sigma}
\def\k{\kappa}

\def\t{\tau}
\def\z{\zeta}
\def\bpsi{{\smash{\bar{\psi}}\kern0.02em\vphantom{\psi}}}
 \def\mathe{{\scalebox{1.01}[1]{$\mathrm{e}$}}}

\def\I{{\mathcal{I}\hhem\hem}}

\def\mtimes{{\mem\times\mem}}
\def\mdot{{\mem\cdot\mem}}
\def\mwedge{{\mem\wedge\mem\hhem}}
\def\swedge{{\mem{\wedge}\,}}

\def\modot{{\mem\odot\mem}}
\def\mtensor{{\mem\otimes\mem}}
\def\mplus{{\mem+\mem}}

\def\mlra{{\mem\leftrightarrow\mem}}

\def\tensor\otimes

\def\MPl{{M_\text{Pl}}}

\newcommand{\wrap}[1]{{\smash{#1}\vphantom{\b}}}

\def\lsq{{
		\kern-0.037em
		\adjustbox{scale=0.90,valign=c}{$
			{
				\adjustbox{raise=-0.09em}{$\lfloor$}
				\llap{\reflectbox{\rotatebox[origin=c]{180}{$\lfloor$}}}
			}
			$}
		\kern-0.04em
}}
\def\rsq{{
		\kern-0.04em
		\adjustbox{scale=0.90,valign=c}{$
			{
				\rlap{\reflectbox{\rotatebox[origin=c]{180}{$\rfloor$}}} 
				\adjustbox{raise=-0.09em}{$\rfloor$}
			}
			$}
		\kern-0.037em
}}

\def\O{\mathcal{O}}
\def\Z{\mathcal{Z}}

\def\A{\mathcal{A}}
\def\I{\mathcal{I}}
\def\S{\mathcal{S}}

\def\F{\mathcal{F}}

\def\ten{\mathcal{N}}
\def\den{\mathcal{D}}

\newcommand{\BB}[1]{\Big(\,{#1}\,\Big)}
\newcommand{\bb}[1]{\bigg(\,{#1}\,\bigg)}

\newcommand{\bbsq}[1]{\bigg[\,{#1}\,\bigg]}

\newcommand{\bigbig}[1]{\big(\mem{#1}\mem\big)}

\let\oldexp\exp
\renewcommand{\exp}{\oldexp\nem}

\def\i{{\texttt{i}}}

\def\P{\mathcal{P}}

\newcommand{\Ket}[1]{{\hem\big|\hem{#1}\big\rangle}}
\newcommand{\Bra}[1]{{\big\langle{#1}\hem\big|\hem}}
\newcommand{\BraKet}[2]{{\big\langle{#1}\hem\big|\hem{#2}\big\rangle}}

\newcommand{\dbar}{
	d\kern-.20em\makebox[0pt][l]{\kern0.01em\adjustbox{raise=-0.01em}{\scalebox{1.4}[1.0]{$\bar{}$}}\kern-0.01em}\kern.20em
}
\newcommand{\deltabar}{
	\delta\kern-.20em\makebox[0pt][l]{\adjustbox{raise=-0.01em}{\scalebox{1.0}[1.0]{$\bar{}$}}}\kern.20em
}

\usepackage{bm}
\def\bmz{\bm{\zeta}}
\def\dz{\delta\zeta}

\def\bmx{\bm{x}}
\def\dx{{\delta x}}

\def\bmp{\bm{p}}
\def\ddp{{\delta p}}

\def\dphi{{\delta \phi}}
\def\dtphi{{\delta \tphi}}

\usepackage{dsfont}

\DeclareMathOperator{\vol}{vol}
\def\Gauge{{\text{Gauge}}}
\newcommand{\D}[1]{\mathcal{D}\hnem{#1}\mem}

\def\M{{\mathcal{M}}}

\newcommand{\expval}[1]{
	\big\langle\hem{
		#1
	}\hem\big\rangle
}

\let\oldcap\cap
\renewcommand{\cap}{{\,\oldcap\,}}

\setlist[itemize]{
	label=\adjustbox{scale=0.7}{$\bullet$}, itemsep=0pt,topsep=0px
}
\setlist[enumerate]{
	itemsep=0pt,topsep=3px
}

\usepackage{tikz}
\usetikzlibrary{calc} % to use relative coordinates
\usetikzlibrary{shapes.geometric} % to draw regular polygons
\usetikzlibrary{positioning} % to use right=of 
\usetikzlibrary{fit} % for fit size
\usepackage[a]{esvect} % arrow styling %f
\usetikzlibrary{decorations.pathmorphing}
\tikzset{every node/.style = {inner sep = 0pt, outer sep = 0, minimum size = 0}}
\tikzset{t/.style = {
		inner sep = 1.5pt, outer sep =1.5pt, minimum size = 1pt,
		font = \small, text = feyntext
}}
\tikzset{T/.style = {
		inner sep = 1.5pt, outer sep =1.5pt, minimum size = 1pt,
		font = \small
}}

\tikzset{linear/.style = {draw, line width = 1.2pt}}
\tikzset{wiggly/.style = {draw, line width = 1.2pt,
		decorate, decoration={snake, 
			amplitude=1.5pt, segment length=5.0pt, post length=0pt, pre length=0pt
		}
}}
\tikzset{sdot/.style = {circle, fill=black, inner sep=0pt, outer sep=0pt, minimum size=1.0pt}}
\tikzset{tdot/.style = {circle, fill=black, inner sep=0pt, outer sep=0pt, minimum size=1.5pt}}
\tikzset{cdot/.style = {circle, draw=black, fill=black, inner sep=0pt, outer sep=0pt, minimum size=2.0pt, line width=1.2pt}}
\tikzset{odot/.style = {circle, draw=black, fill=lred, inner sep=0pt, outer sep=0pt, minimum size=4.0pt, line width=1.2pt}}

\tikzset{gdot/.style = {circle, draw=black, fill=lgray, inner sep=0pt, outer sep=0pt, minimum size=4.0pt, line width=1.2pt}}

\tikzset{wdot/.style = {circle, draw=black, fill=white, inner sep=0pt, outer sep=0pt, minimum size=6.0pt, line width=1.2pt}}
\tikzset{bdot/.style = {
		circle, fill=black, inner sep=0pt, outer sep=0pt, minimum size=6.0pt, line width=1.2pt,
		opacity=0.8,
		postaction={
			circle, draw=black, inner sep=0pt, outer sep=0pt, minimum size=6.0pt, line width=1.2pt
			,pattern={mylines[size=1.5pt, line width=0.8pt, angle=45]},
			pattern color=white,
			opacity=1.0,
		}
}}

\tikzset{poly2/.style={
		append after command={\pgfextra{
				\filldraw[fill=lgray, draw=black, line width=1.2pt]
				($(\tikzlastnode.center) - (0,0.344146em)$) arc[
				start angle=-35, end angle=35, radius= 0.6em
				] arc[
				start angle=-35, end angle=35, radius=-0.6em
				] -- cycle;
		}}
}}
\tikzset{poly1/.style={
		append after command={\pgfextra{
				\filldraw[fill=lgray, draw=black, line width=1.2pt]
				($(\tikzlastnode.center)$)
				--
				($(\tikzlastnode.center) + ( 45:0.0722)$)
				arc[
				start angle=135, end angle=-135, radius=0.0722
				]
				-- 
				($(\tikzlastnode.center) + (-45:0.0722)$)
				-- cycle;
		}}
}}

\tikzset{poly3/.style = {
		draw,
		regular polygon, 
		regular polygon sides = 3,
		line width = 1.2pt,
		fill = lgray,
		minimum size = 5.8pt,
		rotate=30
}}
\tikzset{poly4/.style = {
		draw,
		regular polygon, 
		regular polygon sides = 4,
		line width = 1.2pt,
		fill = lgray,
		minimum size = 5.8pt
}}

\tikzset{yblob/.style = {
regular polygon,
regular polygon sides = 4,
draw=black, fill=llred, inner sep=0pt, outer sep=0pt, minimum size=15pt, line width=1.2pt,
font = \footnotesize
}}
\tikzset{squ/.style = {
		regular polygon,
		regular polygon sides = 4,
		draw=black, fill=white, inner sep=0pt, outer sep=0pt, minimum size=12pt, line width=1.2pt
}}
\tikzset{zsqu/.style = {
		regular polygon,
		regular polygon sides = 4,
		draw=black, fill=white, inner sep=0pt, outer sep=0pt, minimum size=14pt, line width=1.2pt,
		font = \footnotesize
}}
\tikzset{ysqu/.style = {
		regular polygon,
		regular polygon sides = 4,
		draw=black, fill=recur, inner sep=0pt, outer sep=0pt, minimum size=14pt, line width=1.2pt,
		font = \footnotesize
}}

\usetikzlibrary{patterns}
\usetikzlibrary{patterns.meta}
\tikzdeclarepattern{
	name=mylines,
	parameters={
		\pgfkeysvalueof{/pgf/pattern keys/size},
		\pgfkeysvalueof{/pgf/pattern keys/angle},
		\pgfkeysvalueof{/pgf/pattern keys/line width},
	},
	bounding box={
		(0,-0.5*\pgfkeysvalueof{/pgf/pattern keys/line width}) and
		(\pgfkeysvalueof{/pgf/pattern keys/size},
		0.5*\pgfkeysvalueof{/pgf/pattern keys/line width})},
	tile size={(\pgfkeysvalueof{/pgf/pattern keys/size},
		\pgfkeysvalueof{/pgf/pattern keys/size})},
	tile transformation={rotate=\pgfkeysvalueof{/pgf/pattern keys/angle}},
	defaults={
		size/.initial=5pt,
		angle/.initial=45,
		line width/.initial=.4pt,
	},
	code={
		\draw [line width=\pgfkeysvalueof{/pgf/pattern keys/line width}]
		(0,0) -- (\pgfkeysvalueof{/pgf/pattern keys/size},0);
	},
}
\tikzset{arb-blob/.style = {
		circle, fill=lgray, inner sep=0pt, outer sep=0pt, minimum size=10pt, line width=0pt,
		postaction={
			circle, draw=black, inner sep=0pt, outer sep=0pt, minimum size=8pt, line width=1.2pt
			,pattern={mylines[size=1.5pt, line width=0.8pt, angle=45]},
			pattern color=white
		}
}}
\tikzset{arb-blob-red/.style = {
		circle, fill=lred, inner sep=0pt, outer sep=0pt, minimum size=10pt, line width=0pt,
		postaction={
			circle, draw=black, inner sep=0pt, outer sep=0pt, minimum size=8pt, line width=1.2pt
			,pattern={mylines[size=1.5pt, line width=0.8pt, angle=45]},
			pattern color=white
		}
}}

\tikzset{gray-blob/.style = {
		circle, fill=lgray, inner sep=0pt, outer sep=0pt, minimum size=8pt, draw, line width=1.2pt
}}

\usetikzlibrary{arrows, decorations.markings}
\makeatletter
\def\pgf@lib@dec@parsenum#1{%
	\gdef\pgf@lib@dec@computed@width{0 pt}%
	\tsx@pgf@lib@dec@parsenum#1+endmarker+%
	\ifdim\pgf@lib@dec@computed@width<0pt\relax%
	\pgfmathparse{\pgfdecoratedpathlength\pgf@lib@dec@computed@width}
	\edef\pgf@lib@dec@computed@width{\pgfmathresult pt}%
	\fi%
}

\def\tsx@pgf@lib@dec@parsenum@endmarker{endmarker}

\def\tsx@pgf@lib@dec@parsenum#1+{
	\def\temp{#1}%
	\ifx\temp\tsx@pgf@lib@dec@parsenum@endmarker%
	\else%
	\tsx@pgf@lib@dec@parsenum@one{#1}%
	\expandafter\tsx@pgf@lib@dec@parsenum%
	\fi%
}

\def\tsx@pgf@lib@dec@parsenum@one#1{%
	\pgfmathparse{#1}%
	\ifpgfmathunitsdeclared%
	\pgfmathparse{\pgf@lib@dec@computed@width + \pgfmathresult pt}%
	\else%
	\pgfmathparse{\pgf@lib@dec@computed@width + \pgfmathresult*\pgfdecoratedpathlength*1pt}%
	\fi%
	\edef\pgf@lib@dec@computed@width{\pgfmathresult pt}%
}
\makeatother

\usetikzlibrary{arrows.meta}
\tikzset{
	lin/.style = {
		draw, line width=1.2pt
	}
}
\tikzset{
	prop/.style = {
		draw, line width=1.2pt, 
		decoration = { 
			markings, 
			mark = at position 0.5 + 3.2pt with {
				\arrow{>[length=5pt,width=5pt]}
			}
		},
		postaction = {decorate}
	}
}
\tikzset{
	dprop/.style = {
		draw, line width=1.2pt,
		dotted, 
		line cap=round,
		decoration = { 
			markings, 
			mark = at position 0.5 + 3.2pt with {
				\arrow{>[length=5pt,width=5pt]}
			}
		},
		postaction = {decorate}
	}
}
\tikzset{arrp/.style={-{Circle  [length=3.2pt,width=3.2pt]}}}
\tikzset{arrv/.style={-{Circle  [length=4pt,width=4pt, open]}}}

\tikzset{
	proper/.style = {
		draw, line width=1.2pt, 
		decoration = { 
			markings, 
			mark = at position 0.5 + 3.2pt with {
				\arrow{>[length=5pt,width=5pt]}
			}
		},
		postaction = {decorate}
	}
}

\tikzset{
	fprop/.style = {
		draw, line width=1.2pt, 
		decoration = { 
			markings, 
			mark = at position 0.5 + 5.4pt with {
				\arrow{Triangle[length=10pt,width=6pt,fill=lgray]}
			}
		},
		postaction = {decorate}
	}
}
\tikzset{
	qprop/.style = {
		draw, line width=1.2pt, 
		decoration = { 
			markings, 
			mark = at position 0.5 + 3.95pt with {
				\arrow{Triangle[length=6pt,width=3.6pt]}
			}
		},
		postaction = {decorate}
	}
}
\tikzset{wprop/.style = {
		draw, line width = 1.2pt,
		line cap = round,
		line join = round,
		decorate, decoration={
			zigzag,
			amplitude=0.8pt,
			segment length=3.0pt, 
			post length=0pt, pre length=0pt
		}
}}

\def\pin{%
	\adjustbox{valign=c}{\begin{tikzpicture}[]
				\node (o) at (0,0) {};
				\node (a) at (-0.08,0.12) {};
				\node (b) at (0.08,0.12) {};
				\draw[linear] (a)--(o)--(b);
		\end{tikzpicture}}%
}
\def\vpin{\phantom{\pin}}

\def\vph{\vphantom{{K_5)}^2}}

\tikzset{empty/.style = {inner sep = 0pt, outer sep = 0, minimum size = 0}}
\tikzset{w/.style = {inner sep = 1pt, outer sep = 2pt, minimum size = 12pt, anchor = west,	
	font = \small
}}

\newcommand{\kk}[1]{{k{\kern-0.17em}k_{#1}}}
\newcommand{\kf}[1]{{k{\kern-0.17em}f_{#1}}}
\newcommand{\ff}[1]{{f{\kern-0.27em}f_{#1}}}

\newcommand{\K}[1]{{K_{#1}}}

\def\R{\mathbb{R}}
\def\so{\mathfrak{so}}

\def\g{\mathfrak{g}}  
\def\tg{\tilde{\mathfrak{g}}}

\def\sdiff{\mathfrak{diff}}  
            
\def\tphi{\widetilde{\phi}}

\def\su{\mathfrak{su}}

\def\ta{{\smash{\tilde{a}}}{}}
\def\tb{{\smash{\tilde{b}}}{}}
\def\tc{{\smash{\tilde{c}}}{}}
\def\td{{\smash{\tilde{d}}}{}}

\def\tf{{\smash{\tilde{f}}}{}}

\title{
	Worldline Formalism in Phase Space
}

 \author{Joon-Hwi Kim}

\affiliation{Walter Burke Institute for Theoretical Physics, California Institute of Technology,\\ Pasadena, CA 91125}

\abstract{
	We implement the worldline formalism in phase space
	to compute scattering amplitudes.
	First, 
	the Feynman rules exhibit several useful universal features,
	reflecting elements of the symplectic geometry of the phase space target.
	Next,
	noncompact
	worldline topologies automate LSZ reductions
	in accordance with the boundary conditions available in phase space,
	provided correct identifications of the associated moduli spaces.
	Further, employing noncanonical coordinates
	cubicizes the Feynman rules and
	manifests gauge invariance.
	As a result,
	our phase space implementation could
	provide a framework optimized for computing scattering amplitudes
	while
	retaining the nice features of
	the original 
	formalism.
	For an explicit demonstration,
	we compute the multi-photon Compton amplitudes
	up to six points
	in the classical limit.
	Compton amplitudes in
	Yang-Mills theory and gravity
	are also computed
	in a uniform fashion
	by
	supposing the backgrounds of nonlinearly superposed plane waves.
}

\emailAdd{joonhwi@caltech.edu}

\begin{document}
	
	\setlength{\parindent}{1.38em}
	
	\begin{flushright}
		CALT-TH 2025-003
	\end{flushright}
	\maketitle
	
	\bibliographystyle{unsrt}
	\renewcommand*{\bibfont}{\footnotesize}
	
	\newpage

\section{Introduction}
\label{sec:Intro}

In quantum field theory,
one interprets Feynman diagrams 
as representations of
virtual particles propagating in spacetime.
Can this interpretation, however, be established in a direct fashion?
As is well-known,
the worldline formalism 
\cite{Feynman:1948ur,Feynman:1950ir,Schwinger:1951nm,vH,polchinski1985worldlineformalism,Bern:1990cu,Bern:1991aq,Strassler:1992zr,Schubert:2001he,schubert2012lectures,schwartz2014qft,witten2015every}
provides a direct realization of the idea that
Feynman diagrams visualize spacetime trajectories of particles.
In particular,
the worldline formalism establishes that
the partition function
for a first-quantized relativistic particle
traveling from one spacetime point to another
derives an off-shell propagator in field theory.
Moreover, the formalism 
provides a powerful technique for
computing
loop amplitudes and effective actions.
It is also referred to as the string-inspired formalism,
as it was approached as the infinite tension counterpart of string theory computations
\cite{Bern:1990cu,Bern:1991aq,Strassler:1992zr,Schubert:2001he}.
See \rcite{Schubert:2001he}
for a comprehensive review on the subject.

To extract tree-level scattering amplitudes
from the worldline formalism,
one imports the Lehmann–Symanzik–Zimmermann (LSZ) reduction formula
as an external input from field theory.
For instance,
the multi-photon Compton amplitudes
are obtained by
first computing the off-shell propagators
in the backgrounds of superposed electromagnetic plane waves
and then amputating two external legs
via LSZ reduction.

However,
this post-processing might not be 
entirely trivial
in practice.
Besides the technicalities of
Gaussian and Fourier integrals,
a central tension lies in the fact that
the primary output of the worldline formalism
is an off-shell propagator in position space,
whereas scattering amplitudes are on-shell objects examined in the momentum space.
Thus, if one's goal is to obtain tree-level scattering amplitudes,
the original worldline formalism
might have offered a slightly complex workflow.

To circumvent such difficulties,
a natural option one can think of
is to employ phase space actions,
which will offer 
momentum as a basic variable on the worldline.
For instance,
one can compute the transition amplitude of the particle
from a momentum eigenstate to another.
While this approach 
can let us drop the Fourier integrals,
it still does not automate the LSZ reductions,
meaning that we have computed an off-shell quantity.
In fact,
from physical grounds,
a definite-momentum state
shall rather be realized at the boundaries (asymptotic infinities),
implying noncompact worldline topologies.

Meanwhile,
in the modern literature,
the techniques of scattering amplitudes
have found fruitful applications
in describing astrophysical bodies
\cite{Kosower:2018adc,%
Cheung:2018wkq,Bern:2019nnu,Bern:2019crd,Bern:2021dqo,Driesse:2024xad,Levi:2022rrq,Levi:2015msa,Levi:2018nxp,Porto:2016pyg,Porto:2008tb,Porto:2006bt,Porto:2005ac,Goldberger:2004jt,Ben-Shahar:2023djm,Scheopner:2023rzp,gmoov,ambikerr0,ambikerr1,Kim:2024grz,%
Mogull:2020sak,Kopp:2022acm,Bohnenblust:2025gir,Comberiati:2022cpm,Wang:2022ntx,Haddad:2024ebn,Jakobsen:2021zvh,Shi:2021qsb,%
Kim:2024svw,Kim:2025hpn,Gonzo:2024zxo,Damgaard:2021ipf},
which is of a direct relevance
to the gravitational wave physics at LIGO
\cite{LIGOScientific:2016aoc,LIGOScientific:2017vwq,LIGOScientific:2021qlt,Buonanno:2014aza}.
Notably, worldline effective theories
for compact objects
have been one of the key directions
in this development
\cite{Levi:2022rrq,Levi:2015msa,Levi:2018nxp,Porto:2016pyg,Porto:2008tb,Porto:2006bt,Porto:2005ac,Goldberger:2004jt,Ben-Shahar:2023djm,Scheopner:2023rzp,gmoov,ambikerr0,ambikerr1,Kim:2024grz,%
Mogull:2020sak,Kopp:2022acm,Bohnenblust:2025gir,Comberiati:2022cpm,Wang:2022ntx,Haddad:2024ebn,Jakobsen:2021zvh,Shi:2021qsb}.
In the same context,
works
\cite{Mogull:2020sak,Kopp:2022acm,Bohnenblust:2025gir,Comberiati:2022cpm,Wang:2022ntx,Haddad:2024ebn,Jakobsen:2021zvh,Shi:2021qsb}
have established
a worldline perturbation theory
based on infinitely extended worldlines
that is
optimized for extracting classical observables.
An investigation along this direction
has introduced the idea of 
classical eikonal
for an in-in formalism
\cite{Kim:2024grz,Kim:2024svw,Kim:2025hpn,Gonzo:2024zxo,Damgaard:2021ipf},
enriching a manifestly classical understanding
of the framework.

It should be clarified, however,
that
the framework established by
works \cite{Mogull:2020sak,Kopp:2022acm,Bohnenblust:2025gir,Comberiati:2022cpm,Wang:2022ntx,Haddad:2024ebn,Jakobsen:2021zvh,Shi:2021qsb}
may convey a slightly different flavor
than the original, string-inspired formalism
\cite{Feynman:1948ur,Feynman:1950ir,Schwinger:1951nm,vH,polchinski1985worldlineformalism,Bern:1990cu,Bern:1991aq,Strassler:1992zr,Schubert:2001he,schubert2012lectures,schwartz2014qft,witten2015every}. 
It exploits infinitely extended worldlines,
whereas 
the original formalism
describes a sigma model
from a finite interval $[0,1]$ to spacetime.
It also best adapts to an in-in scattering problem
and does not strictly necessitate a path integral construction,
while the original worldline formalism
is inherently an in-out formalism
that computes a path integral.

Nevertheless, 
an explanation for
the relation 
between the two frameworks
has been provided by
the work \cite{Mogull:2020sak}.
In the original worldline formalism,
the complete sum-over-histories for the relativistic particle
is implemented by an integral over the moduli space of intervals,
parameterized by the Schwinger proper time \cite{Schwinger:1951nm}.
This is neccessitated by the strict definition and faithful evaluation 
of the path integral,
summing over all one-dimensional dynamical metrics (einbeins) on the interval topology
so that the Hamiltonian constraint is imposed.
The result shown by \rcite{Mogull:2020sak}
is that,
at the level of integrals,
LSZ reduction on one leg
amounts to sending one integral bound to infinity
while discarding the moduli parameter integral.
Similarly, 
for the infinitely extended domain,
\rcite{Mogull:2020sak}
observes that
dropping a one-dimensional delta function,
encoding the common support of mass-shell conditions,
reproduces the field-theoretic scattering amplitude.

Notably,
this suggests that
the LSZ reductions in the post-processing procedure of the original formalism
could be automated by
choosing noncompact worldline topologies.
However,
it seems that
a top-down derivation of
this observation
from a faithful implementation of the einbein path integral
in terms of the moduli spaces
has not been provided in
\rcite{Mogull:2020sak}.

In this article,
we hope to revisit and clarify these issues
while providing a working version of phase space worldline formalism
optimized for computing scattering amplitudes.

We start from a top-down angle.
In
\Sec{General.mech},
the phase space worldline formalism
is introduced 
as a sigma model to a symplectic manifold.
We then discuss the choice of worldline field basis
in \Sec{General.kinkan},
which draws a dichotomy between
canonical and noncanonical coordinates.
Employing canonical coordinates in the phase space
loses a unique charm of the 
original formalism 
in terms of a simplicity of Feynman rules
and obscures gauge invariance
in intermediate steps.
We advocate employing noncanonical coordinates
due to
the kinetic, i.e., gauge-covariant, momentum.
This necessitates the idea of
what we call the symplectic perturbation theory,
where interactions are solely encoded by modified Poisson brackets.

Next,
\Secs{General.variation}{General.feyn}
provide a bird's-eye view
on perturbation theory in phase space.
By deriving the Feynman rules in full generality,
we observe universal features
that are useful and insightful to remember and explicitly articulate,
as is summarized below.
\begin{itemize}
\vskip5pt
	\item 
		The propagator
		encodes the Poisson bracket.
	\item 
		The vertices fall in two classes:
		Hamiltonian and symplectic.
		The symplectic vertices come in two types: regular and pinched.
	\item
		Vertices sourcing one worldline fluctuation
		arise from leading corrections to the Hamiltonian equations of motion.
		Especially, the symplectic vertex of valence one
		is quickly derived by
		a formula analogous to the Lorentz force.
\end{itemize}
\vskip3pt
Although
the idea of utilizing first-order actions
is not entirely new
(consider models
endowed with fermions
\cite{Brink:1976sz,Brink:1976uf,Berezin:1976eg},
color degrees of freedom
\cite{Corradini:2016czo,Edwards:2017nvs,Bastianelli:2015iba,Ahmadiniaz:2015xoa},
or spin
\cite{Gibbons:1993ap,Haddad:2024ebn,Jakobsen:2021zvh,%
Levi:2022rrq,Levi:2015msa,Levi:2018nxp,Porto:2016pyg,Porto:2008tb,Porto:2006bt,Porto:2005ac,Goldberger:2004jt,Ben-Shahar:2023djm,Scheopner:2023rzp,gmoov,ambikerr0,ambikerr1,Kim:2024grz}),
it seems that the above insights
have been not clearly stated in the literature.

\Sec{WLF} 
then
zooms into the specifics of the worldline formalism.
Notably, it is established that
noncompact worldline topologies implement
LSZ reductions purely within the first-quantized framework.
With the proper identifications of the moduli spaces,
the strict definition of the partition function 
of a relativistic particle
yields
off-shell propagators,
partly on-shell propagators,
and
scattering amplitudes
for
the interval, half-line, and full line
topologies, respectivley.
These are naturally discovered from
various boundary conditions in phase space,
since definite-momentum states
are naturally a boundary notion.
Consequently,
a path integral derivation of
the observations made by \rcite{Mogull:2020sak}
is established.

To elaborate,
recall that
the original worldline formalism
integrates over the Schwinger proper time
for a bulk-to-bulk propagation.
For a bulk-to-boundary propagation,
there is no such moduli parameter
since all half-lines are diffeomorphic to each other:
they can be arbitrarily stretched or rescaled
just like how rooms are created in a Hilbert hotel.
For infinite worldlines,
there is even a residual redundancy
due to rigid translations.
A proper understanding of 
this point
confirms the relationship
argued in
\rcite{Mogull:2020sak}
in a precise fashion.

Finally,
by building upon these observations,
\Sec{QED}
demonstrates how 
scattering amplitudes
can be efficiently obtained by computing
bulk-to-boundary propagators in plane wave backgrounds.
As expected,
this framework
directly offers a momentum variable in the Feynman rules
and greatly simplifies the post-processing steps.
Moreover, manifest gauge invariance is retained 
by adopting symplectic perturbation theory.
In electromagnetism,
we provide an explicit check of this formalism
up to six points in the classical limit.
For Yang-Mills and gravity,
we derive the scalar Compton amplitudes
in \Sec{NA},
which utilizes
covariant color-kinematics duality \cite{cheung2021cck}.

Notably, this framework is exactly 
the approach taken 
by the current author
for computing the
gravitational Compton amplitudes
of the Kerr black hole
to all orders in spin:
\begin{align}
	\frac{
		\M_\text{Kerr}^{+-}
	}{
		\M_\text{scalar}^{+-}
	}
	\,=\,
	\,
	\bigg[\,{
		1 + \c + \c^2\mem
		\bigg(\,{
			\frac{\c{\mem+\mem}\b}{\a{\mem+\mem}\b}\,
			\frac{\mathe^{\a}{\mem-\,}\a{\mem-\,}1}{\a^2}
			- 
			\frac{\c{\mem-\mem}\a}{\a{\mem+\mem}\b}\,
			\frac{\mathe^{-\b}{\mem+\,}\b{\mem-\,}1}{\b^2}
		}\mem\bigg)
	}\,\bigg]\,
		\mathe^{-\a/2}\mem
		\mathe^{\b/2}
	\,.
\end{align}
We believe it is worth explicating
this technique first in the simpler setup of scalar particles.
On a related note,
the worldline effective theory of spinning black holes
\cite{Levi:2022rrq,Levi:2015msa,Levi:2018nxp,Porto:2016pyg,Porto:2008tb,Porto:2006bt,Porto:2005ac,Goldberger:2004jt,gmoov,ambikerr0,ambikerr1,Kim:2024grz,Ben-Shahar:2023djm,Scheopner:2023rzp}
can force one to employ noncanonical coordinates \cite{gmoov,ambikerr0,ambikerr1}.
However, it seems that
a dedicated analysis for symplectic perturbation theory
treating gauge theory and gravity systematically
has been lacking in the literature,
establishing another rationale for the analysis of this paper.

Lastly, although 
the key focus of this paper
will be exclusively put on
the original in-out formalism
\cite{Feynman:1948ur,Feynman:1950ir,Schwinger:1951nm,vH,polchinski1985worldlineformalism,Bern:1990cu,Bern:1991aq,Strassler:1992zr,Schubert:2001he,schubert2012lectures,schwartz2014qft,witten2015every},
our results in \Sec{General}
are universal and versatile
so that they can be applied to any type of a perturbative worldline framework,
such as the in-in formalism due to
\rrcite{Mogull:2020sak,Kopp:2022acm,Bohnenblust:2025gir,Comberiati:2022cpm,Wang:2022ntx,Haddad:2024ebn,Jakobsen:2021zvh,Shi:2021qsb,Kim:2024svw,Kim:2025hpn}.
To establish this point,
the implications of \Sec{General}
are showcased in \App{EXPL},
where we explore
the computation of impulse as a one-point function due to
Berends-Giele recursion,
a master formula for recoil operators \cite{Cheung:2023lnj-EMR,jordan2023,Cheung:2024byb},
and double copy structures.

\textit{Note added.}---%
During the final preparation of this paper,
it was brought to the present author's attention that
the relation between LSZ reduction and worldline topologies
has been pointed out also by
a recent work \cite{Bonezzi:2025iza},
in terms of an analysis involving $bc$ ghosts.
Although the conclusions coincide,
the approaches taken are different.
In our case, 
the noncompact topologies 
are discovered from a pursuit of gauge invariance
in phase space,
and an adaptation of the view \cite{Cheung:2022pdk} that LSZ reduction implements bulk-to-boundary propagation.

%	\newpage
	\section{Universal Features of Perturbation Theory in Phase Space}
\label{General}

In this section, we provide an overview of phase space worldline formalism from a bird's-eye view.
This top-down exposition essentially studies \textit{symplectic sigma models},
observing their universal features that are useful to remember.

\subsection{Hamiltonian Mechanics and Symplectic Geometry}
\label{General.mech}

A symplectic manifold
is an even-dimensional manifold $\P$
equipped with a closed non-degenerate two-form $\omega$ called the symplectic form.
For physicists, a symplectic manifold is the phase space
of a Hamiltonian system
for which an explicit action principle can be given locally.

A polarization
of a symplectic manifold
is a foliation by Lagrange submanifolds,
whose working definition is
a choice of a one-form $\theta$
whose exterior derivative is the symplectic form: $\omega = d\theta$.
We will refer to this one-form $\theta$ as symplectic potential.
Note that $\theta$ is not unique and is subject to an equivalence
$\theta \sim \theta + d\lambda$ for smooth functions $\lambda$ in $\P$.

A Hamiltonian system is fully specified by
a polarized symplectic manifold $(\P, \omega, \theta)$
and a function $H$, the Hamiltonian.
The Hamiltonian defines the system's time evolution.
While the pair $(\omega,H)$ dictates the bulk dynamics,
the polarization
is also an essential input for a complete definition,
albeit its roles are subtler.
In the classical theory, the polarization encodes the boundary condition for the variational principle.
In the quantum theory, the polarization is required for a concrete realization of the Hilbert space.

The master object that carries all these data
is the \textit{phase space action}: the physicist currency.
The phase space action takes the following form,
\begin{align}
	\label{ps-action}
	S[\z]
	\,=
		\int d\s\,\,\,
			\Big(\,\mem{
				\theta_i(\z(\s))\mem \dot{\z}^i(\s)
				- H(\z(\s))
			}\,\Big)
	\,,
\end{align}
from which the symplectic potential $\theta = \theta_i(\z)\mem d\z^i$ and the Hamiltonian $H(\zeta)$ are read off.
The symplectic form then follows by the exterior derivative: $\omega = d\theta$.
Here, we have supposed (local) coordinates $\z^i$.
The map $\z : \s \mapsto \z^i(\s)$ defines a sigma model from a one-dimensional domain to the symplectic manifold $\P$.

For a concrete example,
take a free scalar particle in flat $d$-dimensional spacetime
with Cartesian coordinates $x^\a$.
In a fixed parameterization,
its first-order action reads
\begin{align}
	\label{ps-action.free}
	S[x,p]
	\,=
		\int d\s\,\,\,
		\bigg(\,\mem{
			p_\a\mem \dot{x}^\a
			-\frac{1}{2}\mem (\hem p^2 + m^2)
		}\,\bigg)
	\,.
\end{align}
From \eqref{ps-action.free},
one deduces that
the phase space is the symplectic manifold
$\P = T^*\mathbb{R}^d$
with coordinates $\z^i = (x^\a,p_\a)$,
equipped with the symplectic form $\omega = dp_\a \swedge dx^\a$.
The polarization data is chosen as $\theta = p_\a dx^\a$,
foliating 
$\P = T^*\mathbb{R}^d$
with leaves of constant spacetime position $x^\a$.
Indeed,
the variational principle must set $\delta x^\a$ to zero at the boundary:
definite positions.
This implies that the on-shell action is a function of positions.
Similarly,
when one evaluates a path integral with the above action,
one obtains a transition amplitude between position eigenstates:
the position representation.
If the polarization data is instead chosen to be $\theta = -x^\a\mem dp_\a$, 
the derivative term in \eqref{ps-action.free}
is replaced with $-x^\a\mem \dot{p}_\a$,
leading to the momentum representation.

\subsection{Worldline Field Bases: Canonical Versus Noncanonical}
\label{General.kinkan}

Now let us formulate perturbation theory
in the above geometric framework for Hamiltonian mechanics.
The natural mathematical construct
is a one-parameter family of Hamiltonian systems
in the same phase space manifold $\P$,
the parameter of which describes a coupling constant.
This implements the physicist notion of ``free theory'' and ``interacting theory,''
where the former is realized by the zero coupling limit.
Again, this is concretely defined by a one-parameter family of phase space actions.

The free theory often exhibits a global symmetry, specifying a natural set of coordinates.
The interacting theory, however,
is subject to 
an ambiguity:
as any sigma model does,
the Hamiltonian system defined by the action in \eqref{ps-action}
suffers from, or perhaps enjoys, the freedom to choose coordinates
in the target manifold.
In particular, one can freely perform a coupling-dependent, generic diffeomorphism in the phase space---%
which is not necessarily a canonical transformation---%
that approaches to the identity in the zero coupling limit.
In a field-theoretic language,
this corresponds to worldline field redefinition ambiguities.

In short,
the question of field basis arises
in the symplectic sigma model.
Among all possible choices for the field basis, however,
two extreme cases stand out.

First of all,
the symplectic structure can be fixed to that of the free theory.
In this case, the Hamiltonian is coupling-dependent, solely encoding the interactions.

On the other hand,
one can also consider fixing the Hamiltonian to its free theory form,
stipulating that
the symplectic structure 
carries the sole dependency in the coupling constant
and fully encodes the interactions.
In this case,
the closure $d\omega = 0$ of the symplectic form, 
or equivalently the Jacobi identity of the Poisson bracket,
may impose a nontrivial consistency condition on the interactions,
as is observed by
Feynman \cite{dyson1990feynman} 
and
Souriau \cite{souriau1970structure}.

To illustrate the issue in a further explicit way,
let us suppose the phase space of the free theory
is described with \textit{canonical} coordinates
for simplicity,
the local existence of which is always guaranteed 
by the celebrated Darboux theorem \cite{Darboux}.
This means to work with canonical Poisson brackets in the free theory,
which is a reasonable consideration for many simple cases.
Freezing the symplectic structure means to retain the canonical Poisson brackets in the interacting theory as well.
However, freezing the Hamiltonian will make the Poisson brackets deform into a \textit{noncanonical} form
in the presence of interactions.

The former approach,
employing canonical coordinates (canonical field basis),
is familiar from textbooks.
However, we shall explicate that
the latter approach,
employing noncanonical coordinates (noncanonical field basis),
can also exhibit some charms.
There are at least two major motivations:
simplicity of Feynman rules
and
manifest gauge invariance.

In particular,
a paradigmatic example of the above dichotomy
between canonical and noncanonical coordinates
is found in the familiar 
relationship between
canonical and kinetic momenta.
Consider a charged particle in external electromagnetic fields.
In the very worldline formalism due to Feynman \cite{feynman1948space},
one employs
the configuration-space action,
\begin{align}
	\label{feyn-ex.x}
	S[x]
	\,=
	\int d\s\,\,\,
	\bigg(\,\mem{
		\frac{1}{2}\, \dot{x}^2
		- \frac{1}{2}\, m^2
		+\, qA_\a(x)\mem \dot{x}^\a
	}\,\bigg)
	\,,
\end{align}
where the one-form $A = A_\a(x)\mem dx^\a$ is the electromagnetic gauge connection.
Notably, the photon coupling is linear,
which is usually portrayed as the charm of the first-quantized approach.
Recall that the second-quantized approach
to scalar quantum electrodynamics
involves a ``spurious'' quadratic photon vertex
whose sole role is to ensure gauge invariance.

Now suppose one desires to switch to the phase space formulation.
Due to usual procedures,
one identifies from \eqref{feyn-ex.x} that the \textit{canonical momentum} is $P_\a = \dot{x}_\a + qA_\a(x)$,
deducing the following phase space action via Legendre transformation:
\begin{align}
	\label{feyn-ex.xP}
	S[x,P]
	\,=
	\int d\s\,\,\,
	\bigg(\,\mem{
		P_\a\mem \dot{x}^\a
		- \frac{1}{2}\mem \Big(\,{
			(P-qA(x))^2 {\mem+\,} m^2
		}\,\Big)
	}\,\bigg)
	\,.
\end{align}
This standard formulation, however, has several undesirable features.
First of all, the coupling to the photon is quadratic,
harming the charm of the worldline formalism.
Moreover, the canonical momentum is gauge dependent,
since it transforms as $P_\a \mapsto P_\a + q\mem \partial_\a\ve(x)$
for $A_\a(x) \mapsto A_\a(x) + \partial_\a\ve(x)$.

These problems are resolved at once by employing
the \textit{kinetic momentum}:
\begin{align}
	\label{kinkan}
	p_\a \,=\, P_\a - qA_\a(x)
	\,.
\end{align}
Amusingly,
\eqref{kinkan}
describes 
a worldline field redefinition
that cubicizes the Feynman rule,
which becomes trivialized in the $q{\,\to\,}0$ limit.
Namely, it reattains the linearity of the photon coupling:
the charm of the first-quantized formulation.
Moreover, the kinetic momentum $p_\a$
is gauge invariant and coincides with the physical velocity $\dot{x}_\a$ on equations of motion.
The resulting phase space action is
\begin{align}
	\label{feyn-ex.xp}
	S[x,p]
	\,=
	\int d\s\,\,\,
	\bigg(\,\mem{
		p_\a\mem \dot{x}^\a
		+ qA_\a(x)\mem \dot{x}^\a
		-\frac{1}{2}\mem (\hem p^2 + m^2)
	}\,\bigg)
	\,.
\end{align}

The action in \eqref{feyn-ex.xP} perturbs the Hamiltonian,
by working in a canonical worldline field basis $(x^\a,P_\a)$:
\begin{align}
	\label{feyn-ex.HPT}
	\theta \,=\, P_\a\mem dx^\a
	\,,\quad
	\omega \,=\, dP_\a \swedge dx^\a 
	\,,\quad
	H \,=\,
	\frac{1}{2}\mem \Big(\,{
		(P-qA(x))^2 {\mem+\,} m^2
	}\,\Big)
	\,.
\end{align}
In contrast, the action in \eqref{feyn-ex.xp} perturbs the symplectic structure,
by working in a noncanonical worldline field basis $(x^\a,p_\a)$:
\begin{align}
	\label{feyn-ex.SPT}
	\theta \,=\, p_\a\mem dx^\a + qA
	\,,\quad
	\omega \,=\, dp_\a \swedge dx^\a + qF
	\,,\quad
	H \,=\,
	\frac{1}{2}\mem (\hem p^2 + m^2)
	\,.
\end{align}
Here, $F = dA$ denotes the field strength two-form. 
Notably, the symplectic form in \eqref{feyn-ex.SPT} is gauge-invariantly split into 
the free theory part $dp_\a \swedge dx^\a$ and the interaction part $qF$,
so gauge invariance is manifested for bulk dynamics
\cite{woodhouse1997geometric,souriau1970structure,sternberg1978classical,guillemin1990symplectic}.
In fact, $p_\a$ shall be a physical variable
since the Hamiltonian, stating the mass-shell condition,
has retained its form.

The perturbation theories due to these two approaches
may be referred to as
Hamiltonian and symplectic perturbation theories, respectively.

In general,
we can mathematically formalize the perturbation theories in phase space
in the following form,
where
$\omega^\circ = d\theta^\circ$
and
$\omega' = d\theta'$:
\begin{align}
	\label{split}
	\theta \,=\, \theta^\circ + \theta'
	&\,,\quad
	\omega \,=\, \omega^\circ + \omega'
	\,,\quad
	H \,=\, H^\circ + H'
	\,.
\end{align}
Here, the accents $^\circ$ and $^\prime$ are employed
for denoting free and interaction parts.
Hamiltonian perturbation theory is when $H'$ solely encodes the interactions.
Symplectic perturbation theory is when $\theta'$ and $\omega'$ solely encode the interactions.
In particular,
\eqref{feyn-ex.HPT} describes
$H' = -qA(x) \mdot P + \frac{1}{2}\mem q^2 A^2(x)$,
while
\eqref{feyn-ex.SPT} describes
$\omega' = qF$.

\subsection{All-Order Variation of Phase Space Action}	
\label{General.variation}

Having provided a brief introduction to Hamiltonian mechanics
as a symplectic sigma model,
our goal now is to 
extract Feynman rules.
In accordance with \eqref{split},
we consider a generic scenario
with both Hamiltonian and symplectic perturbations,
although one would specialize in either one of the two perturbation schemes eventually.

Explicitly, the free and interacting Lagrangians are given as the following:
\begin{align}
		L^\circ[\zeta] &=
		\theta^\circ_i(\z)\mem \dot{\z}^i
		- H^\circ(\z)
		\,,\quad
		\label{psL}
		L[\zeta] =
		\theta_i(\z)\mem \dot{\z}^i
		- H(\z)
		\,.
\end{align}
The all-order variation of $L[\z]$ in \eqref{psL} is given by
\begin{align}
	\begin{split}
		\label{pf-2}
		L[\z{\,+\,}\dz] 
		\,=\,
		{}& L[\z] +
		L_\text{bd}[\z,\dz]
		-
		\sum_{n=1}^\infty\,
		\frac{1}{n!}\mem
			H_{,i_1i_2\cdots i_n\nem}\hnem(\z)\mem \dz^{i_1}\hem \dz^{i_2}\hem \cdots \dz^{i_n}
			\\[-0.15\baselineskip]
			&
			+
			\sum_{n=0}^\infty\,
			\frac{1}{(n\hem{\mem+\mem}1)!}\mem
				\Big(\mem{
					-\dot{\z}^k\mem
					\omega_{kj,i_1\cdots i_n}\hnem(\z)
					\mem \dz^j
					-n\mem \omega_{ji_1,i_2\cdots i_n}\hnem(\z)
					\mem \delta\dot{\z}^j
			}\mem\Big)\hhem
			\mem \dz^{i_1}{\hhnem\cdots\mem}\dz^{i_n}
			\,,
	\end{split}
\end{align}
where the boundary term is given by
\begin{align}
	\label{Lbd}
	L_\text{bd}[\z,\dz]
	\,=\,
	\frac{d}{d\s}\hem \bigg[\,\mem{
		\sum_{n=0}^\infty\,
		\frac{1}{(n\hem{\mem+\mem}1)!}\,
		\theta_{j,i_1\cdots i_n}\hnem(\z)
		\mem \dz^j
		\mem \dz^{i_1}{\cdots\mem}\dz^{i_n}
	}\mem\bigg]
	\,.
\end{align}
In obtaining \eqref{pf-2}, 
the following identity has been used
to maximally convert
the derivatives of the symplectic potential
to those of the symplectic form:
\begin{align}
	\label{max-convert}
	\begin{split}
		&{
			-n\mem \omega_{ji_1,i_2\cdots i_n}\hnem(\z)
			\mem \delta\dot{\z}^j
			\mem \dz^{i_1}{\hhnem\cdots\mem}\dz^{i_n}
		}
		\,,\\[0.2\baselineskip]
		&=\mem
		(n\hem{\mem+\mem}1)\mem
		\Big(\mem{
			\theta_{j,i_1\cdots i_n}\hnem(\z)
			-
			\theta_{(j,i_1\cdots i_n)}\hnem(\z)
		}\hem\Big)\hhem
		\mem \delta\dot{\z}^j
		\mem \dz^{i_1}{\hhnem\cdots\mem}\dz^{i_n}
		\,,\\[0.2\baselineskip]
		&=\mem
		\begin{aligned}[t]
			&
			(n\hem{\mem+\mem}1)\mem
			\theta_{j,i_1\cdots i_n}\hnem(\z)
			\mem \delta\dot{\z}^j
			\mem \dz^{i_1}{\hhnem\cdots\mem}\dz^{i_n}
			\\[-0.1\baselineskip]
			&
			+
			\dot{\z}^k\mem
			\theta_{j,ki_1\cdots i_n}\hnem(\z)
			\mem \dz^j
			\mem \dz^{i_1}{\hhnem\cdots\mem}\dz^{i_n}
			- \frac{d}{d\s}\hem \Big(\hem\hhem{
				\theta_{j,i_1\cdots i_n}\hnem(\z)
				\mem \dz^j
				\mem \dz^{i_1}{\cdots\mem}\dz^{i_n}
			}\Big)
			\,,
		\end{aligned}
		\\[0.2\baselineskip]
		&=\mem
		\begin{aligned}[t]
			&
			\Big(\hem{
				(n\hem{\mem+\mem}1)\mem
				\theta_{j,i_1\cdots i_n}\hnem(\z)
				\mem \delta\dot{\z}^j
				\mem \dz^{i_1}{\hhnem\cdots\mem}\dz^{i_n}
				+
				\dot{\z}^k\mem
				\theta_{k,ji_1\cdots i_n}\hnem(\z)
				\mem \dz^j
				\mem \dz^{i_1}{\hhnem\cdots\mem}\dz^{i_n}
			}\Big)
			\\[-0.1\baselineskip]
			&
			+
			\dot{\z}^k\mem
			\omega_{kj,i_1\cdots i_n}\hnem(\z)
			\mem \dz^j
			\mem \dz^{i_1}{\hhnem\cdots\mem}\dz^{i_n}
			- \frac{d}{d\s}\hem \Big(\hem\hhem{
				\theta_{j,i_1\cdots i_n}\hnem(\z)
				\mem \dz^j
				\mem \dz^{i_1}{\cdots\mem}\dz^{i_n}
			}\Big)
			\,.
		\end{aligned}
	\end{split}
\end{align}
Finally, after applying
the free-interaction split in \eqref{split},
\eqref{pf-2} boils down to
\begin{align}
	\begin{split}
		\label{Lparts}
		L[\z {\,+\,} \dz]
		\,=\,
		L^\circ[\z]
		&
		+ L_\text{bd}[\z,\dz]
		+ \dz^i\mem \Big(\,{
			\omega^\circ_{ij}(\z)\mem \dot{\z}^j
			- H^\circ_{,i}\hnem(\z)
		}\,\Big)
		\\
		&
		+ L_\text{prop}[\z,\dz]
		+ L_\text{sv}[\z,\dz]
		+ L_\text{hv}[\z,\dz]
		\,,
	\end{split}
\end{align}
where the bracketed term
vanishes 
when
the background trajectory is 
a saddle 
of the free theory,
say
$\s \mapsto \bmz^i\hnem(\s)$.

\newpage

Note that the Lagrangian may be endowed with extra boundary terms
if one wants to prescribe different polarizations for the in and out phase spaces.
In the scattering context,
it will suffice to examine the free theory limit of the boundary terms.

\subsection{Feynman rules}
\label{General.feyn}

We shall now explicitly spell out
each group of terms
in \eqref{Lparts},
$L_\text{prop}[\z,\dz]$, $L_\text{hv}[\z,\dz]$, and $L_\text{sv}[\z,\dz]$.
For simplicity,
we assume that the free theory is described 
in canonical coordinates,
so the components $\omega^\circ_{ij}$ 
of the free theory symplectic form
are constants.
The inverse of $\omega^\circ_{ij}$ as an antisymmetric matrix
will be denoted as $(\omega^\circ{}^{-1})^{ij}$,
which describes 
the free theory's Poisson bracket:
$\{ \z^i , \z^j \}^\circ = (\omega^\circ{}^{-1})^{ij}$.

Throughout this analysis,
we will observe some notable features in the Feynman rules
that are useful to keep in mind,
as is summarized below.
\begin{enumerate}
	\item 
		The bare propagator
		encodes the free theory's Poisson bracket.
	\item
		Vertices arise from Hamiltonian and symplectic perturbations.
		They are dubbed as Hamiltonian and symplectic vertices, respectively.
	\item 
		Vertices sourcing no worldline fluctuation
		compute the interaction action about the background trajectory.
	\item
		Vertices sourcing one worldline fluctuation
		arise from leading corrections to the Hamiltonian equations of motion.
		Especially, the symplectic vertex of valence one
		is quickly derived by
		a formula analogous to the Lorentz force.
	\item
		The symplectic vertices come in two types: regular and pinched.
\end{enumerate}

\subsubsection{Propagator}
\label{General.feyn.prop}

For constant $\omega^\circ_{ij}$,
$L_\text{prop}[\z,\dz]$
in \eqref{Lparts} is given by
\begin{align}
	\label{Lprop}
	L_\text{prop}[\z,\dz]
	\,=\, \frac{1}{2}\, \omega^\circ_{ij}\,
	\dz^i\mem
	\delta\dot{\z}^j
	\,,
\end{align}
which is quadratic in the worldline fluctuation $\dz^i$.
Ignoring the boundary terms for the moment,
\eqref{Lprop} derives the bare Green's function in the following form:
\begin{align}
	\label{generic.prop}
	G^{ij}\hnem(\s_1,\s_2)
	\,=\,
		(\omega^\circ{}^{-1})^{ij}\,
		\frac{1}{-\partial_{\s_1}}\mem
		\delta(\s_1,\s_2)
	\,,
\end{align}
where
we have denoted $\delta(\s_1,\s_2) := \delta(\s_1{\mem-\,}\s_2)$.
Notably, we observe that the propagator encodes the Poisson bracket of the free theory,
in terms of the tensor factor $(\omega^\circ{}^{-1})^{ij}$.

As usual,
a variety of choices
can be made for the causality prescription.
In particular, the antisymmetric one-dimensional Green's function is given by
\begin{align}
	\label{greens.sym}
	\Theta(\s_1,\s_2)
	\,=\,
	\bigg\{
	\begin{array}{ll}
		-1/2
		&
		\quad\text{if}\quad
		\s_1 > \s_2
		\,,\\
		+1/2
		&
		\quad\text{if}\quad
		\s_1 < \s_2
		\,,
	\end{array}
\end{align}
while the retarded and advanced Green's functions are
\begin{subequations}
	\label{greens.ra}
\begin{align}
	\label{greens.ret}
	\Theta_>(\s_1,\s_2)
	\,=\,
	\bigg\{
	\begin{array}{ll}
		-1
		&
		\quad\text{if}\quad
		\s_1 > \s_2
		\,,\\
		0
		&
		\quad\text{if}\quad
		\s_1 < \s_2
		\,,
	\end{array}
	\\
	\label{greens.adv}
	\Theta_<(\s_1,\s_2)
	\,=\,
	\bigg\{
	\begin{array}{ll}
		0
		&
		\quad\text{if}\quad
		\s_1 > \s_2
		\,,\\
		+1
		&
		\quad\text{if}\quad
		\s_1 < \s_2
		\,.
	\end{array}
\end{align}
\end{subequations}
Adopting \eqref{greens.sym}
for the anti-derivative in \eqref{generic.prop},
we obtain the symmetric propagator,
\begin{align}
	\label{prop.sym}
	\phantom{\delta \z^i\hnem(\t_1)}
	\adjustbox{valign=c}{\begin{tikzpicture}[]
			\node (o) at (0,0) {};
			\node (xshift) at (0.8,0) {};
			\node (i0) at ($(o)$) {};
			\node (i1) at ($(i0)+1*(xshift)$) {};
			\node (i2) at ($(i0)+2*(xshift)$) {};
			\node (v) at (i1) {};
			\node[t] (L) at ($(i0)$) {$\mathllap{
					\delta \z^i\hnem(\s_1)
				}$};
			\node[t] (R) at ($(i2)$) {$\mathrlap{
					\delta \z^j\hnem(\s_2)
				}$};
			\draw[linear] (R)--(L);
	\end{tikzpicture}}
	\kern-0.125em
	\phantom{{\delta \z^j\hnem(\s_2)}}
	\,\,&=\,\,\,
	(\omega^\circ{}^{-1})^{ij}\,
	\Theta(\s_1,\s_2)
	\,,
\end{align}
which is denoted with a line carrying no directionality.
On the other hand,
the retarded and advanced propagators, 
employing the retarded and advanced Green's functions in \eqref{greens.ra},
will be denoted with an arrow:
\begin{align}
	\label{prop.ra}
	\phantom{\delta \z^i\hnem(\t_1)}
	\adjustbox{valign=c}{\begin{tikzpicture}[]
			\node (o) at (0,0) {};
			\node (xshift) at (0.8,0) {};
			\node (i0) at ($(o)$) {};
			\node (i1) at ($(i0)+1*(xshift)$) {};
			\node (i2) at ($(i0)+2*(xshift)$) {};
			\node (v) at (i1) {};
			\node[t] (L) at ($(i0)$) {$\mathllap{
					\delta \z^i\hnem(\s_1)
				}$};
			\node[t] (R) at ($(i2)$) {$\mathrlap{
					\delta \z^j\hnem(\s_2)
				}$};
			\draw[qprop] (R)--(L);
	\end{tikzpicture}}
	\kern-0.125em
	\phantom{{\delta \z^j\hnem(\s_2)}}
	\,\,&=\,\,\,
	(\omega^\circ{}^{-1})^{ij}\,
	\Theta_>(\s_1,\s_2)
	\,\,\,=\,\,\,
	(\omega^\circ{}^{-1})^{ji}\,
	\Theta_<(\s_2,\s_1)
	\,.
\end{align}
Note that these propagators are related as
\begin{align}
	\label{prop.relation}
	\adjustbox{valign=c}{\begin{tikzpicture}[]
			\node (o) at (0,0) {};
			\node (xshift) at (0.8,0) {};
			\node (i0) at ($(o)$) {};
			\node (i1) at ($(i0)+1*(xshift)$) {};
			\node (i2) at ($(i0)+2*(xshift)$) {};
			\node (v) at (i1) {};
			\node (L) at ($(i0)$) {};
			\node (R) at ($(i2)$) {};
			\draw[linear] (R)--(L);
	\end{tikzpicture}}
	\,\,\,=\,\,\,
	\frac{1}{2}\,\Big(\,\,{
	\adjustbox{valign=c}{\begin{tikzpicture}[]
			\node (o) at (0,0) {};
			\node (xshift) at (0.8,0) {};
			\node (i0) at ($(o)$) {};
			\node (i1) at ($(i0)+1*(xshift)$) {};
			\node (i2) at ($(i0)+2*(xshift)$) {};
			\node (v) at (i1) {};
			\node (L) at ($(i0)$) {};
			\node (R) at ($(i2)$) {};
			\draw[qprop] (R)--(L);
	\end{tikzpicture}}
	\,\,+\,\,
	\adjustbox{valign=c}{\begin{tikzpicture}[]
			\node (o) at (0,0) {};
			\node (xshift) at (0.8,0) {};
			\node (i0) at ($(o)$) {};
			\node (i1) at ($(i0)+1*(xshift)$) {};
			\node (i2) at ($(i0)+2*(xshift)$) {};
			\node (v) at (i1) {};
			\node (L) at ($(i0)$) {};
			\node (R) at ($(i2)$) {};
			\draw[qprop] (L)--(R);
	\end{tikzpicture}}
	}\,\,\Big)
	\,.
\end{align}

The symmetric propagator in \eqref{prop.sym}
is relevant for computing time-ordered correlation functions of worldline operators.
In this sense, it serves as an analog of the Feynman propagator.
The retarded propagator in \eqref{prop.ra}
will be exclusively employed in an in-in formalism,
which completely specifies all phase space coordinates at the initial time.
Note that the propagators in \eqrefs{prop.sym}{prop.ra}
do not discriminate between the ``positions'' and the ``momenta,''
meaning that
they are completely ignorant of the polarization data.

This is to be contrasted with the setup in in-out formalisms,
where
either positions or momenta
will be specified
for each of the initial and final times
as per the boundary terms in $L_\text{bd}[\z,\dz]$.
As will be concretely seen in \Sec{QED.wlf},
the propagator
in this case
may not cleanly factorize into a tensor factor times a common time-domain function
and further treat coordinates in the phase space on an unequal footing.

Still, however, it remains true that 
the free theory Poisson bracket is encoded in
the propagator of the phase space sigma model.
This is because the commutator between $\dz^i$ and $\dz^j$ as worldline operators
can be computed 
from any Green's function
as
\begin{align}
	\label{pb-encode}
	\lim_{\e\to0^+}
	\Big(\,\hem{
		\expval{\mathrm{T}\{\mem
			\dz^i(\e)\mem \dz^j(0)
			-
			\dz^j(\e)\mem \dz^i(0)
		\mem\}}
		}\,\Big)
	\,=\,
	-i
	\lim_{\e\to0^+}
	\Big(\,{
		G^{ij}\hnem(\e,0)
		-
		G^{ji}\hnem(\e,0)
	}\,\Big)
	\,,
\end{align}
from which the Poisson bracket
$(\omega^\circ{}^{-1})^{ij}$
is extracted via Dirac's correspondence.

\subsubsection{Hamiltonian Vertices}
\label{General.feyn.hv}

Next, we spell out the group of terms denoted as
$L_\text{hv}[\z,\dz]$
in \eqref{Lparts}:
\begin{align}
	L_\text{hv}[\z,\dz]
	\,=\,
	- H'(\z)
	- H'_{,i}(\z)\mem \dz^i
	-
	\sum_{n=2}^\infty\,
	\frac{1}{n!}\mem
	H_{,i_1i_2\cdots i_n\nem}\hnem(\z)\mem \dz^{i_1}\hem \dz^{i_2}\hem \cdots \dz^{i_n}
	\,.
\end{align}
These terms give rise to vertices,
which we will refer to as ``Hamiltonian vertices.''

\newpage

The Hamiltonian vertex for $n {\,=\,} 0$ worldline fluctuations
is simply the interaction action due to the Hamiltonian perturbation:
\begin{align}
	\label{generic.H0}
	\adjustbox{valign=c}{\begin{tikzpicture}[]
			\node (o) at (0,0) {};
			\node (v) at (o) {};
			\node[gdot] (V) at ($(v)$) {};
	\end{tikzpicture}}\,\,\hem
	\,\,\,&=\,\,\,
	-\int d\s\,\,
	H'(\bmz(\s)\hnem)
	\,.
\end{align}
Here, we remind the reader that we are extracting Feynman rules around a free theory saddle $\s \mapsto \bmz(\s)$.
The Hamiltonian vertex sourcing $n {\,=\,} 1$ worldline fluctuation,
on the other hand,
is given by
the gradient of the interaction Hamiltonian:
\begin{align}
	\label{generic.H1}
	\adjustbox{valign=c}{\begin{tikzpicture}[]
			\node (o) at (0,0) {};
			\node (xshift) at (0.8,0) {};
			\node (i0) at ($(o)$) {};
			\node (i1) at ($(i0)+1*(xshift)$) {};
			\node (v) at (i1) {};
			\node[t] (L) at ($(i0)$) {$\mathllap{
					\delta\z^i\hnem(\s)
				}$};
			\draw[linear] (v)--(L);
			\node[poly1] (V) at ($(v)$) {};
	\end{tikzpicture}}\,\,
	\mem
	\kern-0.125em
	\,\,\,&=\,\,\,
	-H'_{,i}\hnem(\bmz(\s)\hnem)
	\,.
\end{align}
For $n{\,=\,}2$ worldline fluctuations, we have
\begin{align}
	\begin{split}
		\label{generic.H2}
		\phantom{{\delta \z^i\hnem(\s_1)}}
		\adjustbox{valign=c}{\begin{tikzpicture}[]
				\node (o) at (0,0) {};
				\node (xshift) at (0.8,0) {};
				\node (i0) at ($(o)$) {};
				\node (i1) at ($(i0)+1*(xshift)$) {};
				\node (i2) at ($(i0)+2*(xshift)$) {};
				\node[t] (L) at ($(i0)$) {$\mathllap{
						\delta \z^i\hnem(\s_1)
					}$};
				\node[t] (R) at ($(i2)$) {$\mathrlap{
						\delta \z^j\hnem(\s_2)
					}$};
				\draw[linear] (v)--(L);
				\draw[linear] (v)--(R);
				\node[poly2] (v) at (i1) {};
		\end{tikzpicture}}
		\kern-0.125em
		\phantom{\delta \z^j\hnem(\s_2)}
		\,&=\,\,\,
		-\delta(\s_1,\s_2)\,
		H_{,ij}\hnem(\bmz(\s_1)\hnem)
		\,.
	\end{split}
\end{align}
Similarly, for $n{\,\geq\,}2$, we find
\begin{align}
	\begin{split}
		\label{generic.Hn}
		\phantom{{\delta \z^{i_1}\hnem(\s_1)}}
		\adjustbox{valign=c}{\begin{tikzpicture}[]
				\node (o) at (0,0) {};
				\node (i-1) at ($(o)-(0-120:0.8)$) {};
				\node (i0) at ($(o)-(0:0.8)$) {};
				\node (i1) at ($(o)-(0+120:0.8)$) {};
				\node (v) at (o) {};
				\node[t] (I0) at (i0) {$\mathllap{
						\delta\z^{i_1}\hnem(\s_1)
					}$};
				\node[t] (I-1) at (i-1) {$\mathrlap{
						\delta\z^{i_n}\hnem(\s_n)
					}$};
				\node[t] (I1) at (i1) {$\mathrlap{
						\delta\z^{i_2}\hnem(\s_2)
					}$};
				\draw[linear] (o)--(I-1);
				\draw[linear] (o)--(I0);
				\draw[linear] (o)--(I1);
				\node[tdot] (dot2) at ($(o)-(0+120+30:0.35)$) {};
				\node[tdot] (dot3) at ($(o)-(0+120+30+30:0.35)$) {};
				\node[tdot] (dot4) at ($(o)-(0+120+30+30+30:0.35)$) {};
				\node[poly3] (V) at ($(v)$) {};
		\end{tikzpicture}}
		\kern-0.125em
		\phantom{\delta \z^{i_n}\hnem(\s_n)}
		\,&=\,\,\,
		-\delta^{(n-1)}(\s_1,\s_2,\cdots\hnem,\s_n)\,
		H_{,i_1i_2\cdots i_n\nem}\hnem(\bmz(\s_1)\hnem)
		\,,
	\end{split}
\end{align}
where 
$\delta^{(n-1)}(\s_1,\s_2,\cdots\hnem,\s_n) := 
	\delta(\s_1,\s_2)\,
	\delta(\s_1,\s_3)
	\,\cdots\,
	\delta(\s_1,\s_n)
$.
Note that we have chosen to not implement the free-interaction split in \eqrefs{generic.H2}{generic.Hn}.

\subsubsection{Symplectic Vertices}
\label{General.feyn.sv}

Finally, the group of terms denoted as $L_\text{sv}[\z,\dz]$ in \eqref{Lparts}
is found as
\begin{subequations}
\begin{align}
	L_\text{sv}[\z,\dz]
	\,&=\,
		L_\text{sv}^\text{regular}[\z,\dz]
		+
		L_\text{sv}^\text{pinched}[\z,\dz]
	\,,\\[0.25\baselineskip]
	\label{Lsv.regular}
	L_\text{sv}^\text{regular}[\z,\dz]
	\,&=\,
	\theta'_i\hnem(\z)\mem \dot{\z}^i
	-
	\sum_{n=1}^\infty\,
	\frac{1}{n!}\,
		\dot{\z}^j\mem
		\omega'_{ji_n,i_1\cdots i_{n-1}\hnem}\hnem(\z) 
		\, \dz^{i_1}{\hhnem\cdots\mem}\dz^{i_n}
	\,,\\
	\label{Lsv.pinched}
	L_\text{sv}^\text{pinched}[\z,\dz]
	\,&=\,
	-
	\sum_{n=2}^\infty\,
	\frac{n\hem{\mem-\mem}1}{n!}\,
	\omega'_{ji_1,i_2\cdots i_{n-1}\hnem}\hnem(\z)
	\, 
	\delta\dot{\z}^j\mem
	\dz^{i_1}{\hhnem\cdots\mem}\dz^{i_{n-1}}
	\,.
\end{align}
\end{subequations}
These terms describe vertices due to the perturbation in symplectic structure.
We will refer to them as ``symplectic vertices.''
They come in two types: regular and pinched.
The regular ones follow from \eqref{Lsv.regular},
while the pinched ones follow from \eqref{Lsv.pinched}.

The symplectic vertex with $n=0$ worldline fluctuations
is simply the interaction action
due to the symplectic perturbation:
\begin{align}
	\label{generic.int0}
	\adjustbox{valign=c}{\begin{tikzpicture}[]
			\node (o) at (0,0) {};
			\node (v) at (o) {};
			\node[odot] (V) at ($(v)$) {};
	\end{tikzpicture}}\,\,\hem
	\,\,\,&=\,\,\,
	\int d\s\,\,
	\theta'_i\hnem(\bmz(\s)\hnem)\mem \dot{\bmz}^i\hnem(\s)
	 \,.
\end{align}
The symplectic vertex with $n=1$ worldline fluctuation, 
on the other hand,
is given by
\begin{align}
	\label{generic.int1}
	\adjustbox{valign=c}{\begin{tikzpicture}[]
			\node (o) at (0,0) {};
			\node (xshift) at (0.8,0) {};
			\node (i0) at ($(o)$) {};
			\node (i1) at ($(i0)+1*(xshift)$) {};
			\node (v) at (i1) {};
			\node[t] (L) at ($(i0)$) {$\mathllap{
					\delta\z^i\hnem(\s)
				}$};
			\draw[linear] (v)--(L);
			\node[odot] (V) at ($(v)$) {};
	\end{tikzpicture}}\,\,
	\kern-0.125em
	\,\,\,&=\,\,\,
	\omega'_{ij}\hhnem(\bmz(\s)\hnem)\mem \dot{\bmz}^j\hnem(\s)
	 \,.
\end{align}
For $n=2$, the regular symplectic vertex is
\begin{align}
	\begin{split}
		\label{generic.int2}
		&
		\phantom{\delta\z^i\hnem(\t_1)}
		\adjustbox{valign=c}{\begin{tikzpicture}[]
				\node (o) at (0,0) {};
				\node (xshift) at (0.8,0) {};
				\node (i0) at ($(o)$) {};
				\node (i1) at ($(i0)+1*(xshift)$) {};
				\node (i2) at ($(i0)+2*(xshift)$) {};
				\node (v) at (i1) {};
				\node[t] (L) at ($(i0)$) {$\mathllap{
						\delta\z^i\hnem(\s_1)
					}$};
				\node[t] (R) at ($(i2)$) {$\mathrlap{
						\delta\z^j\hnem(\s_2)
					}$};
				\draw[linear] (v)--(L);
				\draw[linear] (v)--(R);
				\node[odot] (V) at ($(v)$) {};
		\end{tikzpicture}}
		\phantom{
			\delta\z^j\hnem(\t_2)
		}
		\kern-0.125em
		\,\,\,=\,\,\,
		-
		\dot{\bmz}^k\hnem(\s_1)
		\,
		\omega'_{\smash{k(i,j)}}\hnem(\bmz(\s_1)\hnem)
		\,
		\delta(\s_1,\s_2)
		\,,
	\end{split}
\end{align}
and similarly, for $n \geq 2$,
\begin{align}
	\begin{split}
		\label{generic.int3+}
		\phantom{{\delta\z^{i_1}\hnem(\s_1)}}
		\adjustbox{valign=c}{\begin{tikzpicture}[]
				\node (o) at (0,0) {};
				\node (i-1) at ($(o)-(0-120:0.8)$) {};
				\node (i0) at ($(o)-(0:0.8)$) {};
				\node (i1) at ($(o)-(0+120:0.8)$) {};
				\node (v) at (o) {};
				\node[t] (I0) at (i0) {$\mathllap{
						\delta\z^{i_1}\hnem(\s_1)
					}$};
				\node[t] (I-1) at (i-1) {$\mathrlap{
						\delta\z^{i_n}\hnem(\s_n)
					}$};
				\node[t] (I1) at (i1) {$\mathrlap{
						\delta\z^{i_2}\hnem(\s_2)
					}$};
				\draw[linear] (o)--(I-1);
				\draw[linear] (o)--(I0);
				\draw[linear] (o)--(I1);
				\node[tdot] (dot2) at ($(o)-(0+120+30:0.35)$) {};
				\node[tdot] (dot3) at ($(o)-(0+120+30+30:0.35)$) {};
				\node[tdot] (dot4) at ($(o)-(0+120+30+30+30:0.35)$) {};
				\node[odot] (V) at ($(v)$) {};
		\end{tikzpicture}}
		\kern-0.125em
		\phantom{{\delta\z^{i_n}(\s_n)}}
		\,\,\,=\,\,\,
		- 
		\dot{\bmz}^k\hnem(\s_1)\,
		\omega'_{k(i_1,i_2\cdots i_n)}\hnem(\bmz(\s_1)\hnem)
		\,
		\delta^{(n-1)}\hnem(\s_1,\s_2,\cdots\hnem,\s_n)
		\,.
	\end{split}
\end{align}

In the meantime,
the pinched symplectic vertices
are present for $n \geq 2$ worldline fluctuations.
The notations for these vertices
are as the following.
The pinched symplectic vertex for $n=2$
is defined as
\begin{align}
\begin{split}
	\label{generic.psv2}
	&
	\phantom{\delta\z^i\hnem(\t_1)}
	\adjustbox{valign=c}{\begin{tikzpicture}[]
			\node (o) at (0,0) {};
			\node (xshift) at (0.8,0) {};
			\node (i0) at ($(o)$) {};
			\node (i1) at ($(i0)+1*(xshift)$) {};
			\node (i2) at ($(i0)+2*(xshift)$) {};
			\node (v) at (i1) {};
			\node[t] (L) at ($(i0)$) {$\mathllap{
					\delta\z^i\hnem(\s_1)
				}$};
			\node[t] (R) at ($(i2)$) {$\mathrlap{
					\delta\z^j\hnem(\s_2)
				}$};
			\node (p) at ($(v)+(135:0.22)$) {\pin};
			\draw[linear] (v)--(L);
			\draw[linear] (v)--(R);
			\node[odot] (V) at ($(v)$) {};
	\end{tikzpicture}}
	\phantom{
		\delta\z^j\hnem(\t_2)
	}
	\kern-0.125em
	\,\,\,=\,\,\,
		\frac{1}{2}\,
		\omega'_{ij\vphantom{)}}\hhnem(\bmz(\s_2)\hnem)
		\,
		\frac{\partial}{\partial\s_1}\mem \delta(\s_1,\s_2)
	 \,.
\end{split}
\end{align}
Here, the check symbol (\pin) indicates the leg subject to the pinching.
Similarly,
the pinched symplectic vertex for $n=3$ is defined as
\begin{align}
	\begin{split}
		\label{generic.psv3}
			\phantom{{\delta\z^{i_1}\hnem(\s_1)}}
			\adjustbox{valign=c}{\begin{tikzpicture}[]
							\node (o) at (0,0) {};
							\node (i-1) at ($(o)-(0-120:0.8)$) {};
							\node (i0) at ($(o)-(0:0.8)$) {};
							\node (i1) at ($(o)-(0+120:0.8)$) {};
							\node (v) at (o) {};
							\node[t] (I0) at (i0) {$\mathllap{
											\delta\z^{i_1}\hnem(\s_1)
										}$};
							\node[t] (I-1) at (i-1) {$\mathrlap{
											\delta\z^{i_3}\hnem(\s_3)
										}$};
							\node[t] (I1) at (i1) {$\mathrlap{
											\delta\z^{i_2}\hnem(\s_2)
										}$};
							\draw[linear] (o)--(I-1);
							\draw[linear] (o)--(I0);
							\draw[linear] (o)--(I1);
							\node[odot] (V) at ($(v)$) {};
							\node (p) at ($(v)+(135:0.22)$) {\pin};
					\end{tikzpicture}}
			\kern-0.125em
			\phantom{{\delta\z^{i_n}\hnem(\s_n)}}
			=\,\,\,
			\frac{2}{3}\,
			\omega'_{i_1(i_2,i_3)}\hnem(\bmz(\s_2)\hnem)
			\,
			\frac{\partial}{\partial\s_1}\mem
			\delta^{(2)}\hnem(\s_1,\s_2,\s_3)
			\,.
		\end{split}
\end{align}
Finally,
the pinched symplectic vertices for 
for $n{\,\geq\,}2$
is defined as
\begin{align}
	\begin{split}
		\label{generic.psv3+}
		&
		\phantom{{\delta\z^{i_1}\hnem(\s_1)}}
		\adjustbox{valign=c}{\begin{tikzpicture}[]
				\node (o) at (0,0) {};
				\node (i-1) at ($(o)-(0-120:0.8)$) {};
				\node (i0) at ($(o)-(0:0.8)$) {};
				\node (i1) at ($(o)-(0+120:0.8)$) {};
				\node (v) at (o) {};
				\node[t] (I0) at (i0) {$\mathllap{
						\delta\z^{i_1}\hnem(\s_1)
					}$};
				\node[t] (I-1) at (i-1) {$\mathrlap{
						\delta\z^{i_n}\hnem(\s_n)
					}$};
				\node[t] (I1) at (i1) {$\mathrlap{
						\delta\z^{i_2}\hnem(\s_2)
					}$};
				\draw[linear] (o)--(I-1);
				\draw[linear] (o)--(I0);
				\draw[linear] (o)--(I1);
				\node[tdot] (dot2) at ($(o)-(0+120+30:0.35)$) {};
				\node[tdot] (dot3) at ($(o)-(0+120+30+30:0.35)$) {};
				\node[tdot] (dot4) at ($(o)-(0+120+30+30+30:0.35)$) {};
				\node[odot] (V) at ($(v)$) {};
				\node (p) at ($(v)+(135:0.22)$) {\pin};
		\end{tikzpicture}}
		\kern-0.125em
		\phantom{{\delta\z^{i_n}\hnem(\s_n)}}
		=\,\,\,
		\frac{n\hem{\mem-\mem}1}{n}\,
		\omega'_{i_1(i_2,i_3\cdots i_n)}\hnem(\bmz(\s_2)\hnem)
		\,
		\frac{\partial}{\partial\s_1}\mem
		\delta^{(n-1)}\hnem(\s_1,\s_2,\cdots\hnem,\s_n)
		\,.
	\end{split}
\end{align}
The rule is then to sum over 
the $n$ cyclic permutations of pinched leg
for each pinched symplectic vertex,
which is consistent with
the $1/n$ factor put in \eqref{generic.psv3+}.

Note that the factor $n{\mem-\,}1$ in \eqref{generic.psv3+}
corresponds to the number of cyclic permutations of
the symmetrized indices
$i_2,i_3,\cdots i_n$.
If one choose to remove this symmetrization
from the definition,
then the factor $n{\mem-\,}1$ can be eliminated:
\begin{align}
	(n\hem{\mem-\mem}1)\,
	\omega'_{i_1(i_2,i_3\cdots i_{n-1} i_n)}
	\,=\,
		\omega'_{i_1i_2,i_3\cdots i_{n-1} i_n}
		+
		\omega'_{i_1i_n,i_2\cdots i_{n-2} i_{n-1}}
		+
		\cdots
		+
		\omega'_{i_1i_3,i_4\cdots i_n i_2}
	\,.
\end{align}

The role of pinching in the diagrammatics
is the cancellation of propagators,
adjoining two vertices into a single contact interaction.

It should be clear from the derivation of \eqref{Lparts} in \Sec{General.variation}
that vertices of valence one
arise from the first-order variation of the Lagrangian,
which describes the leading perturbation in the classical equations of motion
(see \eqref{vari1}).
Thus, one can intuitively understand 
the valence-one vertices as the forces due to the interactions.
The Hamiltonian vertex in \eqref{generic.H1}
describes a \textit{gradient force} due to a scalar potential $H'$.
The symplectic vertex in \eqref{generic.int1}
describes a generalization of the \textit{Lorentz force}
in phase space,
which arises from
dotting the velocity vector $\dot{\bmz}$ to
a``magnetic field'' $\omega' = d\theta'$
due to a ``vector potential'' $\theta'$.
Indeed, 
the nonvanishing component of
$\omega'_{ij}\mem \dot{\z}^j$
exactly evaluates to the Lorentz force
$qF_{\m\n}\mem \dot{x}^\n$
for the example of electromagnetic interaction
given in \eqref{feyn-ex.SPT}.

\newpage

\section{Scattering Amplitudes from Worldline Formalism}
\label{WLF}

Despite the general applicability of the Feynman rules derived in \Sec{General.feyn},
we will specialize in
the classic worldline formalism
\cite{Feynman:1948ur,Feynman:1950ir,Schwinger:1951nm,vH,polchinski1985worldlineformalism,Bern:1990cu,Bern:1991aq,Strassler:1992zr,Schubert:2001he,schubert2012lectures,schwartz2014qft,witten2015every}
from now on.

The worldline formalism
is an in-out formalism.
Traditionally, it has obtained scattering amplitudes
by amputating off-shell propagators,
which takes LSZ reduction as an external input from field theory.
In this section, however, we will realize that
the LSZ reduction can be understood and implemented purely within the first-quantized picture,
in accordance with 
the boundary conditions available in phase space.
This results in 
an efficient variant of the worldline formalism
optimized for computing scattering amplitudes.

\subsection{Review of the Worldline Formalism}
\label{wlreview}

We shall
first provide a review of the original worldline formalism.
The story
begins with
the observation that
the free scalar propagator admits 
the following 
integral representation due to Schwinger \cite{Schwinger:1951nm}:
\begin{align}
	\label{schw1}
	\frac{-i}{
		p^2 + m^2 - i\e
	}
	\,=\mem
	\int_0^\infty dT\,\,
	\mathe^{-iT(p^2+m^2-i\e)}
	\,.
\end{align}
The interpretation of this representation becomes clear when one switches to the position space.
By utilizing elements in the quantum mechanics of a scalar particle,
one finds
\begin{align}
	\label{schw2}
	\Z^\circ(x_2,x_1)
	\,=\,
	\int \dbar^dp\,\,
	\mathe^{ipx_2}\,
	\frac{-i}{
		p^2 + m^2 - i\e
	}\, \mathe^{-ipx_1}
	\,=\mem
	\int_0^\infty dT\,\,
		\mathcal{K}^\circ(x_2,T \hem | x_1,0)
	\,,
\end{align}
where the integrand in the last expression
describes a transition amplitude from a position eigenstate to another:
\begin{align}
	\label{schwK}
	\mathcal{K}^\circ(x_2,T \hem | x_1,0)
	\,=\,
	\Bra{x_2}	
		\mathe^{-iT(\hat{p}^2+m^2-i\e)}
	\Ket{x_1}
	\,.
\end{align}
Notably, the transition amplitude in \eqref{schwK}
admits a path integral representation,
\begin{align}
	\label{schw3}
	\Z^\circ(x_2,x_1)
	\,=\mem
	\int_0^\infty dT\,\,
	\int^{x_2}_{x_1} \D{x} \D{p}\,\,
	\exp\bigg(\,{
		i
		\int_0^T d\s\,\,
		\bigg(\,{
			p_\a\mem \dot{x}^\a
			- (p^2+m^2)
		}\,\bigg)
	}\bigg)
	\,,
\end{align}
which is due to the correspondence
between operator and path integral formalisms.
The action in \eqref{schw3}
describes a sigma model from the finite interval $[0,T]$
to the phase space.
This is deduced from the fact that the operator $(\hat{p}^2 + m^2)$  in \eqref{schw2}
generates a time-$T$ evolution as a Hamiltonian.

In this sense, the parameter $T$ is referred to as the Schwinger proper time.
The partition function
in \eqref{schw3}
has implemented a complete sum over histories
for a particle travelling from $x_1$ to $x_2$
by allowing
all possible values of the proper time $T$.

The key physical message here is that
the transition amplitude in first quantization
yields the off-shell propagator in second quantization.
Crucially,
this correspondence continues to hold
in the presence of background fields.
For instance, the action of the particle in an electromagnetic background is given by
\begin{align}
	\label{pxA}
	\int_0^T d\s\,\,
	\bigg(\,{
		p_\a\mem \dot{x}^\a
		+ qA_\a(x)\mem \dot{x}^\a
		- (p^2+m^2)
	}\,\bigg)
	\,.
\end{align}
The partition function $\Z(x_2,x_1)$ due to 
this action
computes
the scalar Green's function in the electromagnetic background
\cite{Feynman:1948ur}:
\begin{align}
	\label{eq:greenA}
	\Big(\,{
		(\partial_2 {\,-\,} qA(x_2))^2 + m^2
	}\,\Big)\,
	\mathcal{Z}(x_2,x_1)
	\mem=\mem
	-i\mem \delta^{(d)}\hnem(x_2{\,-\,}x_1)
	\,.
\end{align}
In other words,
it reproduces
the two-point function in scalar quantum electrodynamics:
\begin{align}
	\label{eq:bridge}
	\expval{
		\bar{\phi}(x_2)
		\mem
		\phi(x_1)
	}_{\nem A}
	\,=\,
		\Z(x_2,x_1)
	\,.
\end{align}
This correspondence may be intuitively justified from the fact that the field $\phi(x)$
effectively
arises as the wavefunction in the quantum mechanics of the particle.
For a precise derivation, 
recall the worldline field redefinition in \eqref{kinkan}.

Provided this correspondence,
the standard steps for obtaining scattering amplitudes
proceed as the following.
First, one
takes the electromagnetic background to be a superposition of on-shell plane waves.
Next, one
extracts the part of the paritition function
that is
multilinear in the polarizations.
This derives an off-shell scalar propagator dressed with photons.
Finally,
amputating the massive external legs via LSZ reduction
yields
the higher-multiplicity analogs of the Compton amplitude.

Here, the LSZ reduction is simply taken as an instruction imported from 
the second-quantized framework.
Specifically, the LSZ reduction formula reads
\begin{align}
\begin{split}
	\label{LSZformula}
	&
	i\mem \A(p_2,p_1;\cdots)\,\hem
		\deltabar^{(d)}\hnem(-p_2{\mem+\,}p_1{\mem+\,}k)
	\\
	&
	\,=\mem
	\lim_{\substack{X_1\to0\\X_2\to0}}\,\,
		\int d^dx_2\, d^dx_1\,\,
			\mathe^{-ip_2x_2}\,\,
			iX_2
			\,\,
				\expval{
						\bar{\phi}(x_2)
						\mem
						\phi(x_1)
				}_{\nem A} \hem\Big|_\text{m.l.}
			\,\mem
			iX_1
			\,\,
			\mathe^{ip_1x_1}
	\,,
\end{split}
\end{align}
where we have assumed the plane wave background with total momentum $k$,
along with the extraction of the multilinear (m.l.) piece.
We have also employed shorthand notations,
\begin{align}
	\label{Xdef}
	X_2 \,:=\, {p_2}^2 {\mem+\,} m^2
	\,,\quad
	X_1 \,:=\, {p_1}^2 {\mem+\,} m^2
	\,,
\end{align}
and $\deltabar^{(d)}\hnem(p) := (2\pi)^d\mem \delta^{(d)}\hnem(p)$.
Clearly, the factors $iX_2$ and $iX_1$ in \eqref{LSZformula}
amputates the external propagators
while the limits put the external momenta on shell.
In this way, the scattering amplitude $\A(p_2,p_1;\cdots)$ is obtained from the off-shell propagator.

In practice,
however,
the post-processing described above
might be tedious to carry out.
The major bottlenecks are the two Fourier transforms,
which involve at least one Gaussian integral.
The multiplication by the factors $iX_2$, $iX_1$
and the limiting process
together
adds complexity as well.
Moreover, the partition function will be given in terms of an integral
over the finite domain $[0,T]$,
which may not be ideally simple.

To clarify, these are difficulties that are present when one's goal is to obtain on-shell scattering amplitudes in the momentum basis.
The tension lies in the fact that
the formalism's primary output is
an off-shell propagator in the position basis.

To be explicit,
let us plainly describe the process
for obtaining the simplest scattering amplitude
in scalar quantum electrodynamics:
the three-point tree amplitude.
To this end, one assumes the background of a single photon,
$A_\a(x) = e_\a\mem \mathe^{ikx}$.
The partition function in the interacting theory is given by
\begin{align}
	\label{shhA1}
	\Z(x_2,x_1)
	\,=\mem
	\int_0^\infty 
		dT\,\,
		\mathe^{-iTm^2}
	\int^{x_2}_{x_1} \D{x}\,\,
	\exp\bigg(\,{
		i
		\int_0^1 d\s\,\,
		\bigg(\,{
			\frac{1}{4T}\mem \dot{x}^2
			+ qA_\a(x)\mem \dot{x}^\a
		}\,\bigg)
	}\bigg)
	\,,
\end{align}
where we have reparametrized the worldline
such that the domain becomes $[0,1]$.
We have also integrated out $p$,
since the boundary conditions fix positions after all.
Plugging in $A_\a(x) = e_\a\mem \mathe^{ikx}$ to \eqref{shhA1}
and extracting the piece linear in the polarization $e$,
one finds
\begin{align}
	\begin{split}
		\label{ssch1}
		\int_0^\infty\nem \frac{dT}{(4\pi iT)^{d/2}}\,\,
		\exp\bigg(\,{
			\frac{i(x_2{\,-\mem}x_1)^2}{4T}
			- i\hem Tm^2
		}\,\bigg)
		\,\,
		\int_0^1 d\s\,\,
		iq\mem e\mdot (x_2{\,-\mem}x_1)\,
		\mathe^{ikx_1(1-\s)}\mem \mathe^{ikx_2\s}
		\,.
	\end{split}
\end{align}
This involves
expanding around the free theory saddle
$\bmx^\a(\s) = x_1^\a\mem (1{\mem-\mem}\s) + x_2^\a\mem \s$
and performing a Gaussian path integral
that derives the factor $(4\pi iT)^{-d/2}$.
To perform the Fourier transforms,
one can use the trick of promoting the polarization term
into an exponential at this order:
$e\mdot (x_2{\,-\mem}x_1) \to \oldexp(\hem e\mdot (x_2{\,-\mem}x_1))$.
Evaluating one Gaussian integral,
evaluating one ordinary Fourier integral,
re-extracting the piece linear in the polarization,
and finally imposing the on-shell conditions $k^2 = 0$ and $k\mdot e = 0$
then derives the result
\begin{align}
	\label{ssch2}
	i\mem
	\Big(\,{
		2q\mem e\mdot p_2
	}\,\Big)\mem
	\,\bigg[\,\hem{
		\int_0^\infty dT\,\, \mathe^{-iTX_2}\mem
		\int_0^1 d\s\,\, T\, \mathe^{iTK\s}
	}\,\bigg]\,
	\,
	\deltabar^{(d)}\nem( -p_2 {\mem+\,} p_1 {\mem+\,} k )
	\,,
\end{align}
where $K := 2k\mdot p_2$.
The square-bracketed term exactly describes the off-shell propagators to be amputated.
With a proper $i\e$ prescription, it evaluates to
\begin{align}
	\label{Ixx.eval}
	\int_0^\infty dT\,\, \mathe^{-iT(X_2 - i\e)}\mem
	\int_0^1 d\s\,\, T\, \mathe^{iTK\s}
	\,=\,
		\frac{-i}{X_2 {\mem-\,} i\e}\,
		\frac{-i}{X_1 {\mem-\,} i\e}
	\,,
\end{align}
where $X_2{\mem-\,}K = X_1$.
Therefore, LSZ reduction derives the scattering amplitude as
\begin{align}
	\label{3ptAmpQED}
	\A(p_2,p_1;(k,e))
	\,=\, 2q\mem e\mdot p_2
	\,.
\end{align}

Another clever method,
which could be helpful when exploring higher multiplicities,
is to recall a result from free theory that
\begin{align}
	\label{ssch2}
	\int_0^\infty\nem \frac{dT}{(4\pi iT)^{d/2}}\,\,
	\exp\bigg(\,{
		\frac{i(x_2{\,-\mem}x_1)^2}{4T}
		- i\hem Tm^2
	}\,\bigg)
	\,=\,
	\int \dbar^dp\,\,
	\mathe^{ip(x_2-x_1)}\,
	\frac{-i}{
		p^2 {\mem+\,} m^2 {\mem-\,} i\e
	}
	\,,
\end{align}
which 
follows from 
evaluating
\eqref{schw2}
in terms of a Gaussian path integral.
Using \eqref{ssch2},
the Fourier transform of \eqref{ssch1} can be obtained in the following form:
\begin{align}
	\label{ssch3}
	\int_0^1 d\s
	\int \dbar^dp\,\,\,
	\frac{-i}{
		p^2 {\mem+\,} m^2 {\mem-\,} i\e
	}\,\mem
	q\mem e\mdot \frac{\partial}{\partial p}\mem
			\deltabar^{(d)}\nem( -p {\mem+\,} p_2 {\mem-\,} k\s )
			\,\mem
			\deltabar^{(d)}\nem( -p_2 {\mem+\,} p_1 {\mem+\,} k )
	\,.
\end{align}
Integrating by parts,
one finds that the denominator structure in \eqref{Ixx.eval}
is reproduced in a Feynman parameterization:
\begin{align}
	\int_0^1 d\s\,\,
	\frac{1}{
		( X_2 {\mem-\,} K\s  {\mem-\,} i\e )^2
	}
	\,=\,
		\frac{1}{X_2 {\mem-\,} i\e}\,
		\frac{1}{X_2 {\mem-\,} K {\mem-\,} i\e}
	\,.
\end{align} 
Either way, however, is not completely trivial.

\newpage

Notably,
we will see in the following subsections that
the phase space approach could provide some shortcuts.
The momentum variable will be directly employed in the path integral and boundary conditions,
reducing the effort required for the Fourier transforms.
Moreover, LSZ reductions will be automated by extending the integral bounds to infinities.

\subsection{Alternative Boundary Conditions from Phase Space Worldline Formalism}
\label{Alter}

In \Sec{wlreview},
we have reviewed the original worldline formalism,
which is
a framework
rooted upon configuration space actions
and off-shell propagators.
The first key equation was
\begin{align}
	\label{ZK}
	\Z(x_2,x_1)
	\,=\mem
		\int_0^\infty dT\,\,
			\mathcal{K}(x_2,T \hem | x_1,0)
	\,,
\end{align}
which states the relation between
the partition function $\Z(x_2,x_1)$ due to a complete sum over histories
and the transition amplitude $\mathcal{K}(x_2,T \hem | x_1,0)$
that can be defined 
as an overlap between position eigenstates
in first quantization.
The second key equation was the path integral representation for the transition amplitude 
$\mathcal{K}(x_2,T \hem | x_1,0)$,
which is useful in its perturbative evaluation.
The last key equation was the LSZ reduction formula stated in \eqref{LSZformula},
taken as an external input from second quantization.
This describes a post-processing that extracts the scattering amplitude $\A(p_2,p_1,\cdots)$ 
out of the partition function $\Z(x_2,x_1)$.
In sum, the workflow can be represented as the following:
\begin{align}
	\label{workflow.xx}
	\mathcal{K}(x_2,T \hem | x_1,0)
		\quad\xrightarrow{\quad}\quad
	\Z(x_2,x_1)
		\quad\xrightarrow{\quad}\quad
	\A(p_2,p_1,\cdots)
	\,.
\end{align}

The idea in this subsection 
is to employ phase space actions 
to see if this workflow can be simplified.
For instance, let us revisit the free theory transition amplitude in \eqref{schwK},
but with the final state replaced with a momentum eigenstate:
\begin{align}
	\label{schwK.px}
	\mathcal{K}^\circ(p_2,T \hem | x_1,0)
	\,=\,
	\Bra{p_2}	
		\mathe^{-iT(\hat{p}^2+m^2)}
	\Ket{x_1}
	\,=\,
		\mathe^{-iT({p_2}^2 + m^2)}\,
		\mathe^{-ip_2x_1}
	\,.
\end{align}
The path integral implementation for $\mathcal{K}^\circ(p_2,T \hem | x_1,0)$ is
\begin{align}
	\label{Kpi.px}
	\int^{p_2}_{x_1} \D{x} \D{p}\,\,
	\exp\bigg(\,{
		-i\mem p_\a(T)\mem x^\a(T)
		+
		i
		\int_0^T d\s\,\,
		\bigg(\,{
			p_\a\mem \dot{x}^\a
			- (p^2+m^2)
		}\,\bigg)
	}\bigg)
	\,,
	\kern-0.1em
\end{align}
where the boundary term
originates from the Fourier kernel $\smash{\BraKet{p_2}{x_2}} = \oldexp(-ip_2x_2)$.
Indeed, variation of the action in \eqref{Kpi.px}
produces the boundary terms,
\begin{align}
	- \ddp_\a(T)\, x^\a(T)
	- p_\a(0)\, \dx^\a(0)
	\,,
\end{align}
which vanish for $\ddp_\a(T) = 0$ and $\dx^\a(0) = 0$.

Similarly, consider the transition amplitude between momentum eigenstates:
\begin{align}
	\label{schwK.pp}
	\mathcal{K}^\circ(p_2,T \hem | p_1,0)
	\,=\,
	\Bra{p_2}	
		\mathe^{-iT(\hat{p}^2+m^2)}
	\Ket{p_1}
	\,=\,
		\mathe^{-iT({p_2}^2 + m^2)}\,
		\deltabar^{(d)}\nem(p_2 {\mem-\mem} p_1)
	\,.
\end{align}
The path integral implementation for this object is
\begin{align}
	\label{Kpi.pp}
	\int^{p_2}_{p_1} \D{x} \D{p}\,\,
	\exp\bigg(\,{
		i
		\int_0^T d\s\,\,
		\bigg(\,{
			- \dot{p}_\a\mem x^\a
			- (p^2+m^2)
		}\,\bigg)
	}\bigg)
	\,,
\end{align}
whose action is given in the momentum polarization.
The boundary variation is
\begin{align}
	- \ddp_\a(T)\, x^\a(T)
	+ \ddp_\a(0)\, x^\a(0)
	\,,
\end{align}
which vanishes for $\ddp_\a(T) = 0 = \ddp_\a(0)$.

\newpage

Now let us add the interactions. The action in \eqref{Kpi.px} becomes
\begin{align}
	\label{mixy.xp}
	-p_\a(T)\mem x^\a(T)
	+
	\int_0^T d\s\,\,
	\bigg(\,{
		p_\a\mem \dot{x}^\a
		+ qA_\a(x)\mem \dot{x}^\a
		- (p^2+m^2)
	}\,\bigg)
	\,,
\end{align}
whose boundary variation is
\begin{align}
	\label{mixy.bd}
	- \ddp_\a(T)\mem x^\a(T)
	- p(0)\mem \dx^\a(0)
	+ qA_\a(x(T)\hnem)\mem \dx^\a(T)
	- qA_\a(x(0)\hnem)\mem \dx^\a(0)
	\,.
\end{align}
We learn that
$\ddp_\a(T) {\,=\,} 0$ and $\dx^\a(0) {\,=\,} 0$
no longer serve as a valid boundary condition,
due to the term $qA_\a(x(T)\hnem)\mem \dx^\a(T)$.
Since there is no condition ensuring the vanishing of $A_\a(x(T)\hnem)$,
it is 
not encouraged to 
describe the state at $\s {\;=\;} T$
in the momentum basis.

A resolution is to send the final time to far future: $T\to\infty$.
In the scattering context, the gauge potential vanishes in the asymptotically free regions: $A_\a(x(\infty)\hnem) \to 0$.
As a result, $\ddp_\a(\infty) = 0$ can indeed serve as a valid boundary condition.\footnote{
	The astute reader will sense a subtle connection between manifest gauge invariance and boundary conditions here.
	Suppose one insists to specify a momentum state in the bulk.
	When a gauge-invariant boundary action is prescribed at the bulk point,
	$p(0) \mdot x(0)$,
	the boundary variation there
	is $\ddp(0) \mdot x(0) - qA(x(0)\hnem) \mdot \dx(0)$.
	However, when a gauge-dependent boundary action is prescribed in terms of the canonical momentum,
	$P(0) \mdot x(0)$,
	the boundary variation
	is $\delta P(0) \mdot x(0)$.
	Hence the best one can do in the bulk
	is to specify the canonical momentum.
}
In this light,
the natural object to study is
the transition amplitude $\mathcal{K}(p_2,\infty \hem | x_1,0)$ for an infinite time lapse,
letting
the particle escape to far future in a definite momentum state:\footnote{
	Note that the path integral measure $\D{x}\D{p}$ develops no anomaly
	in noncanonical coordinates:
	consider the Liouville measure $\omega^d/d!$.
	The situation is different in gravity, however.
}
\begin{align}
		\label{jess.K.px}
		\int^{p_2}_{x_1} \D{x} \D{p}\,\,
		\exp\bigg(\,{
			-i\mem p(\infty)\mdot x^\a(\infty)
			+
			i
			\int_0^\infty\nem d\s\,\,
			\bigg(\,{
				p \mdot \dot{x}
				+ qA(x)\mdot \dot{x}
				- (p^2+m^2)
			}\,\bigg)
		}\bigg)
		\,.
		\kern-0.1em
\end{align}
Yet, there is an apparent problem.
In the free theory,
the on-shell action evaluates to
\begin{align}
	\label{myr.I}
	\lim_{\s_+ \to \infty}
	\Big(\,{
		- p_2 \mdot \bmx(\infty)
		+ 
		(\hem{
			p_2 \mdot \bmx(\infty)
			-	 p_2 \mdot x_1
		})
		- 
		({p_2}^2+m^2)\mem \s_+
	}\,\Big)
	\,,
\end{align}
which follows from plugging in the saddle trajectory,
\begin{align}
	\label{saddle.xp}
	\bmx(\s)
	\,=\, 
	x_1 + 2p_2\mem \s
	\,,\quad
	\bmp(\s)
	\,=\,
	p_2
	\,.
\end{align}
Crucially, the on-shell action in \eqref{myr.I}
is divergent unless the momentum escaped to the infinity,
$p_2$, is on-shell:
\begin{align}
	\label{myrX2}
	{p_2}^2 + m^2 \,=\, 0
	\,.
\end{align}
The same is true in the presence of interactions.
The implication is that
extending the worldline time bounds to infinities
will be accompanied by the impositions of on-shell conditions
for consistency.

Note that
this fact can be also approached from the operator formalism:
\begin{align}
	\begin{split}
		\label{myr4}
		\mathcal{K}^\circ(p_2,\infty \mem| x_1,0)
		\,&=\,
		\lim_{\s_+\to\infty}
		\Bra{p_2}
		\mathe^{-i\s_+(\hat{p}^2+m^2)}
		\Ket{x_1}
		\,,\\
		\,&=\,
		\lim_{\s_+\to\infty}
		\mathe^{-i\s_+({p_2}^2+m^2)}
		\BraKet{p_2}{x_1}
		\,=\,
		\BraKet{p_2}{x_1}
		\,=\,
		\mathe^{-ip_2x_1}
		\,.
	\end{split}
\end{align}
It should be clear that the transition amplitude is finite and nonzero
only if the escaped momentum is \textit{exactly} put on shell
(with no $i\e$).

\newpage

To have a concrete sense on what the transition amplitude in \eqref{jess.K.px} computes,
let us again consider the background of a single photon:
$A_\a(x) = e_\a\mem \mathe^{ikx}$.
The path integral
can be perturbatively computed by expanding around 
the free theory saddle in \eqref{saddle.xp}.
Extracting the piece linear in the polarization $e$,
one finds
\begin{align}
	\label{myr5}
	\mathe^{-ip_2x_1}\,
	\bigg(\,{
		\int_0^\infty d\s\,\,
			2iq\mem e\mdot p_2
			\, \mathe^{iK\s}
			\mem \mathe^{ikx_1}
	}\mem\bigg)
	\,.
\end{align}
Notably, the computation here is significantly simpler than that in \Sec{wlreview}.
One simply needs to multiply the free theory limit in \eqref{myr4}
with the interaction action due to the single-photon background.

Amusingly, when one
Fourier transforms from $x_1$ to $p_1$,
\eqref{myr5} becomes
\begin{align}
	\label{myr6}
	i\mem
	\Big(\,{
		2q\mem e\mdot p_2
	}\,\Big)\mem
	\,\bigg[\,\hem{
		\int_0^\infty d\s\,\, \mathe^{iK\s}
	}\,\bigg]\,
	\,
	\deltabar^{(d)}\nem( -p_2 {\mem+\,} p_1 {\mem+\,} k )
	\,.
\end{align}
With an adequate $i\e$ prescription, this evaluates to
\begin{align}
	\label{myr7}
	i\mem
	\Big(\,{
		2q\mem e\mdot p_2
	}\,\Big)\mem
	\,\bigg[\,\hem{
		\frac{-i}{X_2{\mem-\,}K-i\e}
	}\,\bigg]\,
	\,
	\deltabar^{(d)}\nem( -p_2 {\mem+\,} p_1 {\mem+\,} k )
	\,,
\end{align}
where we have reinstated $X_2 = {p_2}^2 + m^2$
while keeping in mind that 
we have been presuming $X_2 = 0$ from since \eqref{myrX2}.

Evidently, \eqref{myr7}
is the amplitude $\A(p_2,p_1;(k,e))$ derived in \eqref{3ptAmpQED}
times a single off-shell propagator with a momentum-conserving delta function.
This result seems to suggest that 
sending one end of the worldline to infinity
automatically implements a LSZ reduction,
so one simply needs to perform only one extra reduction by hand:
$X_1 = X_2 - K \to 0$.

With this observation, we can naturally anticipate that
sending two ends of the worldline to the infinities
will completely implement the LSZ reductions.
To compute the transition amplitude $\mathcal{K}(p_2,+\infty \mem | p_1,-\infty)$,
one may insert the resolution of the identity in the position basis
at an arbitrary finite time $\s_0$.
This is to circumvent the difficulty due to the absence of a saddle
that changes the free particle's momentum from $p_1$ to $p_2$.
The transition amplitude is then found as a sum of two terms:
\begin{align}
	2iq\mem 
	\int d^dx_0\,\,
	\bigg[\,{
			\int_{\s_0}^{+\infty} d\s\,\,
			e\mdot p_2
			\, \mathe^{iK\s}
		+
			\int_{-\infty}^{\s_0} d\s\,\,
			e\mdot p_1
			\, \mathe^{iK\s}
	}\,\bigg]
	\,\mem
	\mathe^{i(-p_2+k+p_1)x_0}
	\,.
\end{align}
On the support of the on-shell condition $k\mdot e = 0$,
the two integrals merge into one:
\begin{align}
	\label{myt6}
	i\mem
	\Big(\,{
		2q\mem e\mdot p_2
	}\,\Big)\mem
	\,\bigg[\,\hem{
		\int_{-\infty}^{+\infty} d\s\,\, \mathe^{iK\s}
	}\,\bigg]\,
	\,
	\deltabar^{(d)}\nem( -p_2 {\mem+\,} p_1 {\mem+\,} k )
	\,.
\end{align}
However, we should remind ourselves
that both of the two ends of the worldline 
have now escaped to the infinities,
so for the transition amplitude
to be finite and nonzero,
the two mass-shell conditions in \eqref{Xdef}
must be imposed together:
$X_2 = 0 = X_1$.
This implies that $K = 0$,
so what \eqref{myt6} really describes is
\begin{align}
	\label{myt7}
	i\mem
	\Big(\,{
		2q\mem e\mdot p_2
	}\,\Big)\mem
	\,\Big[\,\,\hem{
		\deltabar(0)
	}\,\mem\Big]\,
	\,
	\deltabar^{(d)}\nem( -p_2 {\mem+\,} p_1 {\mem+\,} k )
	\,.
\end{align}
Certainly, both of the LSZ reductions have been implemented,
yet leaving out a mysterious-looking 
infinite volume factor $\deltabar(0) = \int_{-\infty}^{+\infty} d\s$.

\newpage

To summarize, we have found that
the scattering amplitude 
can be derived in three different
versions of worldline computations.
The first is the traditional method in \Sec{wlreview},
which utilizes worldlines that
have the topology of a finite \textit{interval} $[0,1]$.
As reviewed in \eqref{ZK},
this involves a sum over the Schwinger proper time
and demands two LSZ reductions for the massive particle.
The second is to use worldlines that have
the topology of a \textit{half-line}, $\mathbb{R}_{\geq0} = [0,\infty)$.
This involves only one LSZ reduction for the post-processing.
The third is to use worldlines that have
the topology of a \textit{full line}, $\mathbb{R} = (-\infty,+\infty)$.
This involves 
a division by an infinite volume factor $\deltabar(0)$
and requires no LSZ reductions.

Throughout this analysis,
we have also observed 
an association between
worldline topology and boundary conditions in phase space.
The momentum is naturally an asymptotic variable
that characterizes states at infinities.
We have also concretely seen that
the momenta registering the infinities
shall be exactly on-shell
for consistency.

The astute reader will point out that
we have not yet introduced the notion of \textit{partition function} $\Z$
for the half-line and infinite line topologies:
so far, we have only discussed \textit{transition amplitudes} $\mathcal{K}$.
For the interval approach,
the relation between the partition function and the transition amplitude
is given in \eqref{ZK},
which we reproduce below:
\begin{align}
	\label{ZK.1}
	\text{Interval}:\quad
	&
	\Z(x_2,x_1)
	\,=\mem
		\int_0^\infty dT\,\,
		\mathcal{K}(x_2,T \hem | x_1,0)
	\,.
\end{align}
In \Sec{Moduli},
we will derive that
this relation is given for the half-line topology as
\begin{align}
	\label{ZK.2}
	\text{Half-Line}:\quad
	&
	\Z(p_2,x_1)
	\,=\mem
		\mathcal{K}(p_2,\infty \hem | x_1,0)
	\,.
\end{align}
That is, they are exactly equal.
For the full line, it follows that
\begin{align}
	\label{ZK.3}
	\text{Full Line}:\quad
	&
	\Z(p_2,p_1)
	\,=\mem
		\frac{1}{\deltabar(0)}\,\hem
		\mathcal{K}(p_2,+\infty \hem | p_1,-\infty)
	\,,
\end{align}
which explains the $\deltabar(0)$ factor.
Namely, discarding the $\deltabar(0)$ factor in \eqref{myt7}
is indeed the correct prescription for extracting the scattering amplitude.

Among these three options,
the half-line stands out.
The workflow in \eqref{workflow.xx} greatly simplifies
as the partition function is directly equated with the transition amplitude
while only one LSZ reduction is required:
\begin{align}
	\label{workflow.px}
	\mathcal{K}(p_2,T \hem | x_1,0)
	\,=\,
	\Z(p_2,x_1)
	\quad\xrightarrow{\quad}\quad
	\A(p_2,p_1,\cdots)
	\,.
\end{align}
The full line approach may also be promising,
but its technical complexity due to the absence a free theory saddle
can persist.

In the following subsections, 
it will be established that
the single unified equation
that systematically generates Eqs.\,(\ref{ZK.1}), (\ref{ZK.2}), and (\ref{ZK.3})
is
\begin{align}
	\label{ZKU}
	\Z
	\,=\mem
	\int\nem \frac{\D{\kappa}}{\vol(\Gauge)}\mem
	\int \D{x} \D{p}\,\,
	\mathe^{i\S[x,p;\k]}
	\,.
\end{align}
This is \textit{the} definition for the partition function of a relativistic particle,
implementing a complete sum over histories
with the einbein $\k$.

\subsection{LSZ Reduction, Worldline Topology, and Boundary Conditions}
\label{Tripos}

\begin{figure}
	\centering
	\begin{align*}
			{\renewcommand{\arraystretch}{1.5}
				\begin{array}{ccc}
						\includegraphics[valign=c,scale=0.6]{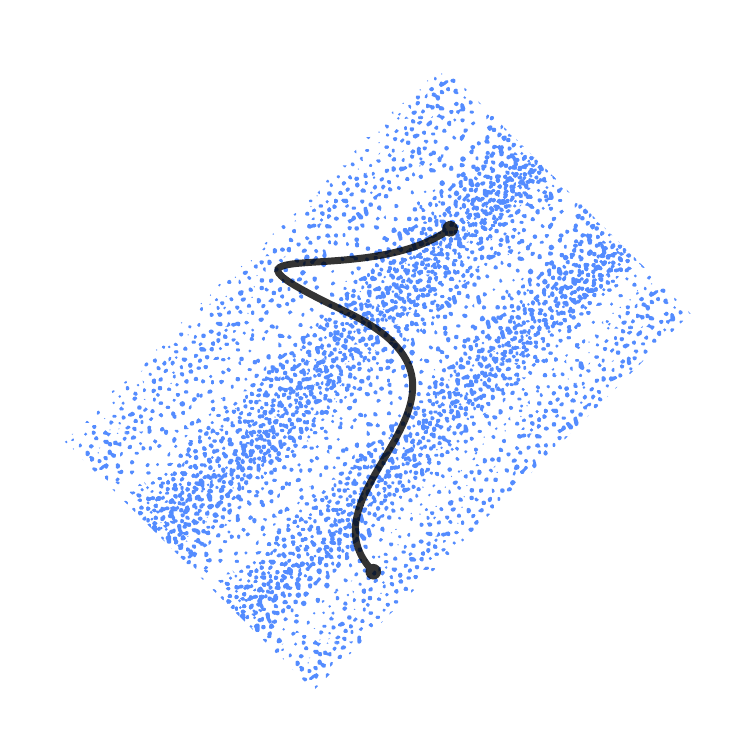}
						&
						\includegraphics[valign=c,scale=0.6]{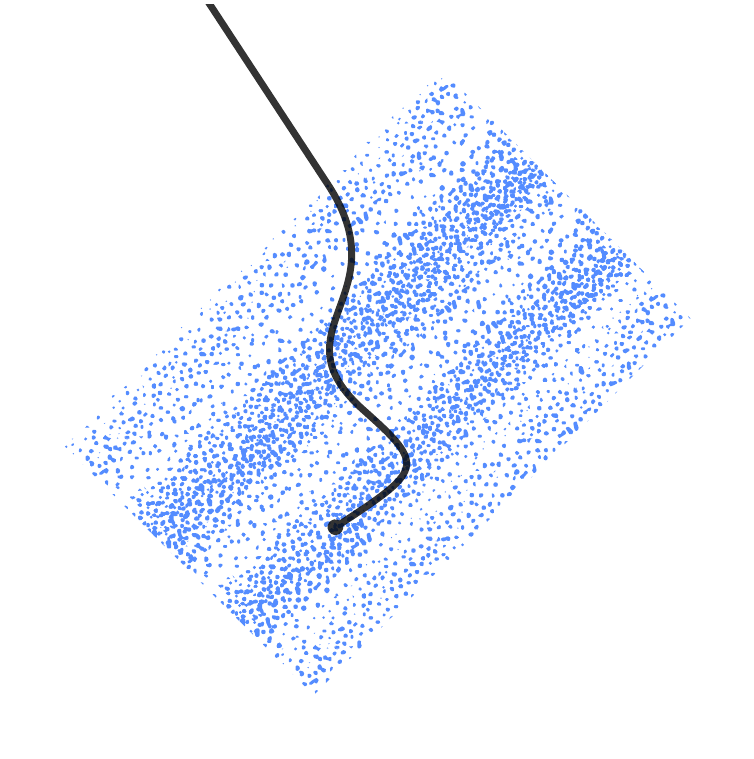}
						&
						\includegraphics[valign=c,scale=0.6]{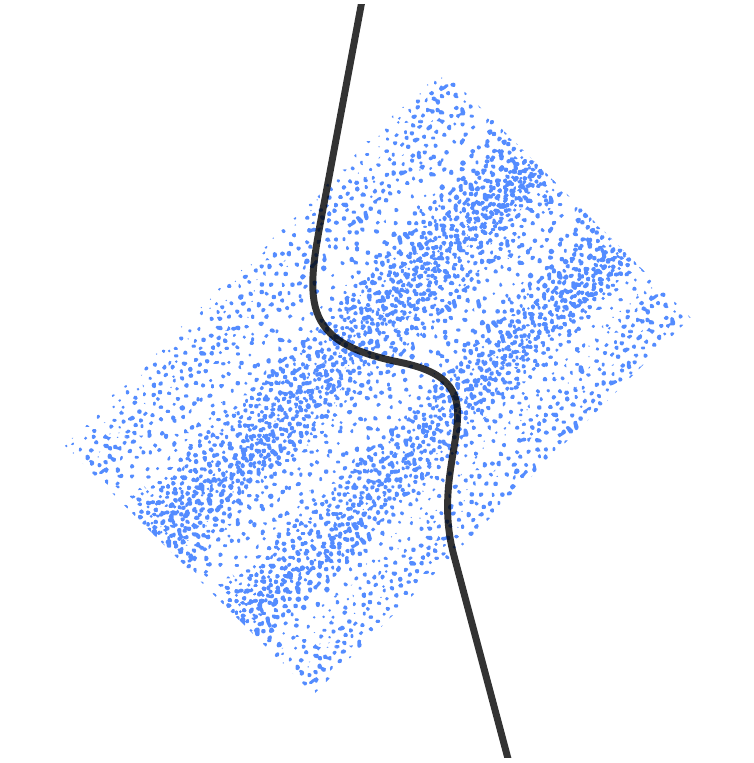}
						\\
						\text{\footnotesize
								(a)
								Interval
							}
						&
						\text{\footnotesize
								(b)
								Half-Line
							}
						&
						\text{\footnotesize
								(c)
								Full Line
							}
					\end{array}}
		\end{align*}
	\caption{
			Various worldline topologies for
			particle propagation in a sandwich geometry.
		}
	\label{fig:sand}
\end{figure}

Suppose a massive particle propagating in a ``sandwich geometry,''
where the support of external fields
separates 
the $d$-dimensional spacetime
into in and out regions
as depicted in \fref{fig:sand}.
Since our objective in this paper is
obtaining scattering amplitudes,
this is an essential stipulation:
the notion of 
asymptotically free states
should exist.

The scattering amplitudes
are obtained by LSZ-reducing the off-shell propagators.
To reiterate,
the role of LSZ reduction is to
cut the external legs and put them \textit{on shell},
which is crucial for the very qualification of 
scattering amplitudes
as asymptotic observables.
The slogan here is that
objects that are
measured, anchored, and defined
at the asymptotic infinities---i.e., the boundaries---%
can only be physical.

Indeed,
an insight has been that
the geometrical content of LSZ reduction
is the \textit{bulk-to-boundary propagation};
see \rcite{Cheung:2022pdk}, for instance.
What have escaped to the boundary are on-shell states.
Off-shell propagators, however, 
describe a virtual, bulk-to-bulk process.
LSZ reduction propagates the bulk endpoints of the off-shell propagator
to the boundaries,
making the particle register the infinities at far past and future.

In accordance with this 
notion,
we have visualized 
successive LSZ reductions of
an off-shell propagator
in \fref{fig:sand}.
Crucially,
this is a process that changes the \textit{worldline topology}.
The bulk-to-bulk propagation in \fref{fig:sand}\,(a)
describes an interval in the bulk,
which is topologically a real segment $[0,1]$.
The bulk-to-boundary propagation in \fref{fig:sand}\,(b)
describes a half-line,
which is topologically $\mathbb{R}_{\geq0} = [0,\infty)$.
The boundary-to-boundary propagation in \fref{fig:sand}\,(c)
describes a full line,
which is topologically $\mathbb{R} = (-\infty,+\infty)$.

When 
embraced
in the worldline context,
this insight implies that the LSZ reductions
could be implemented purely
within the 
first-quantized framework.
Instead of
performing the LSZ reductions \textit{by hand}
as a post-processing rule
imported from
field theory,
one may directly compute the 
half-line and full line propagators
in \fref{fig:sand}\,(b,c)
by literally 
first-quantizing a half-line and a full line.
In this way, the LSZ reductions are implemented from the very beginning.

In the phase space worldline formalism,
we have seen that
such an idea is naturally 
approached
in terms of the available \textit{boundary conditions}.
A state of definite position ($x$)
is bound to the bulk.
A state of definite momentum ($p$),
in contrast,
escapes to infinities.
While $p$ serves as a well-behaving asymptotic observable,
$x$ does not.
In this context,
the boundary conditions available in phase space
are naturally associated with the three topologies illustrated in \fref{fig:sand}.
The interval (bulk-to-bulk) topology
would be prescribed with a boundary condition 
fixing initial $x$ and final $x$.
The half-line (bulk-to-boundary) topology
would be prescribed with a boundary condition
fixing initial $x$ and final $p$.
Similarly,
the full line (boundary-to-boundary) topology
would be prescribed with a boundary condition
fixing initial $p$ and final $p$.

From this top-down exposition
on LSZ reduction, worldline topology, and boundary conditions,
it becomes clear that
scattering amplitude 
is a synonym for
boundary-to-boundary propagator.
Similarly, off-shell propagator
is just another name for bulk-to-bulk propagator.
The slight contextual difference is 
due to the usual basis in which these objects are represented:
momentum, position, etc.
The bulk-to-boundary propagator 
is a partly on-shell (boundary) object
with one off-shell (bulk) end.
This discussion should clarify
the reasons behind the structures observed in \Sec{Alter}.
Especially, the prescriptions for the LSZ reductions
are necessitated by the fact that
Eqs.\,(\ref{ZK.1}), (\ref{ZK.2}), and (\ref{ZK.3})
have computed the bulk-to-bulk, 
bulk-to-boundary,
and boundary-to-boundary
propagators in \fref{fig:sand}.

\begin{table}[t]
	\begin{align*}
		{\renewcommand{\arraystretch}{1.5}
			\begin{array}{ccc}
				\quad\qquad\qquad\mathclap{\text{%
						Worldline Topology%
				}}\qquad\qquad\quad
				&
				\quad\qquad\qquad\mathclap{\text{%
						Moduli Space%
				}}\qquad\qquad\quad
				&
				\quad\qquad\qquad\mathclap{\text{%
						Residual Gauge%
				}}\qquad\qquad\quad
				\\\hline
				\quad\qquad\qquad\mathclap{\text{%
						Interval%
				}}\qquad\qquad\quad
				&
				\quad\qquad\qquad\mathclap{\text{%
						$T \in (0,\infty)$
				}}\qquad\qquad\quad
				&
				\quad\qquad\qquad\mathclap{\text{%
						$\{\, \adjustbox{scale=0.75,valign=c}{$\bullet$} \,\}$%
				}}\qquad\qquad\quad
				\\
				\quad\qquad\qquad\mathclap{\text{%
						Half-Line%
				}}\qquad\qquad\quad
				&
				\quad\qquad\qquad\mathclap{\text{%
						$\{\, \adjustbox{scale=0.75,valign=c}{$\bullet$} \,\}$%
				}}\qquad\qquad\quad
				&
				\quad\qquad\qquad\mathclap{\text{%
						$\{\, \adjustbox{scale=0.75,valign=c}{$\bullet$} \,\}$%
				}}\qquad\qquad\quad
				\\
				\quad\qquad\qquad\mathclap{\text{%
						Full Line%
				}}\qquad\qquad\quad
				&
				\quad\qquad\qquad\mathclap{\text{%
						$\{\, \adjustbox{scale=0.75,valign=c}{$\bullet$} \,\}$%
				}}\qquad\qquad\quad
				&
				\quad\qquad\qquad\mathclap{\text{%
						$\mathbb{R}$%
				}}\qquad\qquad\quad
		\end{array}}
	\end{align*}
	\caption{%
		A summary of worldline topologies, their associated moduli spaces, and residual equivalences.
	}
	\label{topologies-table}
\end{table}

\subsection{Moduli Spaces for One-Dimensional Geometries}
\label{Moduli}

Now we provide the top-down derivation of Eqs.\,(\ref{ZK.1}), (\ref{ZK.2}), and (\ref{ZK.3}).
We show that they
are the faithful evaluations of the partition function of a relativistic particle
for the three topologies
as sketched in \eqref{ZKU}.
This analysis
necessitates correct identifications of the
\textit{moduli spaces} 
associated with the worldline topologies,
summarized in Table~\ref{topologies-table}.

The worldline of a relativistic particle is endowed with a Lagrange multiplier $\k$ imposing the mass-shell condition,
known as the einbein.
The einbein defines a \textit{dynamical} metric in one dimension
and is subject to the one-dimensional diffeomorphism---i.e., reparameterization---redundancy.

To define the path integral measure for $\k$, one has to properly implement the quotient along the gauge orbit.
Eventually,
the quotiented measure will be realized as 
an ordinary integral over the \textit{moduli space}
of one-dimensional metrics
on the worldline.

For the interval topology,
it is well-known that the Schwinger proper time $T$ serves as the moduli parameter.
This is the total length of the worldline up to a customary factor:
$2T = \int_0^1 \k(\s)\mem d\s$.
The total length is invariant under reparametrizations.
Intervals with different $T$ values cannot be mapped to each other via diffeomorphisms.
Therefore,
the metric geometries of an interval are invariantly classified by the total length $T \in (0,\infty)$.

Accordingly,
the quotiented path integral measure is realized as \cite{polchinski1985worldlineformalism}\footnote{
	Here, we are ignoring the Faddeev-Popov determinant.
}
\begin{align}
	\label{recipe.xx}
	\int
	\frac{\D{\kappa}}{\vol(\Gauge)}
	\quad\longrightarrow\quad
	\int_0^\infty\nem dT\,
	\int
	\D{\kappa}\,\,
	\delta[\k {\mem-\,} 2T]
	\,.
\end{align}
Here, $\delta[\k {\mem-\,} 2T]$ is a delta functional
that sets $\k(\s) = 2T$.
Namely, the recipe is to
perform an ordinary integral over moduli space
after gauge-fixing the einbein to a constant value $2T$.

For the half-line topology, however,
one finds that the moduli space collapses to a single point.
All half-lines are diffeomorphic to each other,
and there is nothing left after one performs the quotient due to this equivalence.
This yet assumes that
we sum over complete metrics,
for which the total length of the half-line is infinite.
Technically,
this makes the einbein asymptote to a nonzero constant $\k_\infty$.

With this understanding,
the quotiented measure for the half-line topology is
\begin{align}
	\label{recipe.px}
	\int
	\frac{\D{\kappa}}{\vol(\Gauge)}
	\quad\longrightarrow\quad
	\int
	\D{\kappa}\,\,
		\delta[\k {\mem-\,} \k_\infty]
	\,,
\end{align}
which gauge-fixes the einbein to a nonzero constant $\k_\infty$.
Note that the specific value chosen for $\k_\infty$ is irrelevant by the very gauge invariance:
one can rescale it away to any other value while affecting the value of the partition function.
Hence,
the replacement in \eqref{recipe.px} effectively drops the einbein path integral entirely.

Lastly, for the full line topology,
the moduli space 
of complete metrics
is again trivial:
all full lines are diffeomorphic to each other.
However, 
it should be understood that
there remains a residual gauge redundancy.
This is due to diffeomorphisms
that asymptote to a common constant value at both ends.
The space of such gauge redundancies is $\mathbb{R}$,
provided the assumption that the two ends of the full line are distinguishable.

In this case, the recipe for the einbein path integral is
\begin{align}
	\label{recipe.pp}
	\int
	\frac{\D{\kappa}}{\vol(\Gauge)}
	\quad\longrightarrow\quad
	\frac{1}{\deltabar(0)}\,
	\int
		\D{\kappa}\,\,
		\delta[\k {\mem-\,} \k_\infty]
	\,,
\end{align}
where $\deltabar(0) = \int_{\mathbb{R}} d\s = \vol(\mathbb{R})$
is the volume of the space of constant translations.
To say, the gauge quotient is bigger than the integral itself.
Again, the result is ignorant of the specific value chosen for the constant $\k_\infty$
by construction.
Hence the replacement in \eqref{recipe.px}
effectively drops the path integral
while discarding
an infinite volume factor
$\deltabar(0)$
as well.

A summary of these identifications
is given in Table~\ref{topologies-table}.
The interval is the only topology 
whose moduli space is nontrivial.
The full line exhibits residual gauge redundancies.

A more quantitative derivation may be given,
although the essence is simply 
the analysis of residual freedom
after fixing the einbein to a constant.
The infinitesimal gauge transformations
are given by
$\delta_\xi\hem \k(\s) = d(\hem \k(\s)\mem \xi(\s))/d\s$,
where $\xi(\s)$ describes a vector field in the one dimension.
For non-compact topologies,
one can take
the gauge-fixing functional
as $\chi[\k](\s) = \k(\s) - \k_\infty$
for a nonzero constant $\k_\infty$.
This removes the freedom to rescale the worldline parameter, for instance.
Residual gauge transformations are 
defined as
gauge transformations that preserve the gauge slice
$\chi[\k](\s) = 0$.
Evaluating $\delta_\xi\hem \chi[\k](\s)$
on $\chi[\k](\s) = 0$,
one finds that they
should be constant translations: $\dot{\xi}(\s) = 0$.
For the half-line, the boundary condition $\xi(0) = 0$
fixes that $\xi(\s) = 0$.
For the full line, 
there is no such restriction,
so
constant translations 
do
serve as residual gauge transformations.

It should be clear that
adopting the recipes in 
Eqs.\,(\ref{recipe.xx}), (\ref{recipe.px}), and (\ref{recipe.pp})
exactly reproduces 
Eqs.\,(\ref{ZK.1}), (\ref{ZK.2}), and (\ref{ZK.3})
from the very definition of the partition function
stated in \eqref{ZKU}
when specialized in the respective boundary conditions.
To be explicit,
the action $\S[x,p;\k]$ in \eqref{ZKU} is given by
\begin{subequations}
\begin{align}
	\label{S-QED.1}
	\text{Interval}:\quad
	&
		\int^1_0 d\s\,\,
		\bigg(\,{
			p_\a\mem \dot{x}^\a
			+ qA_\a(x)\mem \dot{x}^\a
			- \frac{\k}{2}\mem (p^2+m^2)
		}\,\bigg)
	\,,\\
	\label{S-QED.2}
	\text{Half-Line}:\quad
	&
	-p_\a(\infty)\mem x^\a(\infty)
	\,+
	\int^\infty_0\nem d\s\,\,
	\bigg(\,{
		p_\a\mem \dot{x}^\a
		+ qA_\a(x)\mem \dot{x}^\a
		- \frac{\k}{2}\mem (p^2+m^2)
	}\,\bigg)
	\,,\\
	\label{S-QED.3}
	\text{Full Line}:\quad
	&
	\int^{+\infty}_{-\infty}\nem d\s\,\,
	\bigg(\,{
		-\dot{p}_\a\mem x^\a
		+ qA_\a(x)\mem \dot{x}^\a
		- \frac{\k}{2}\mem (p^2+m^2)
	}\,\bigg)
	\,.
\end{align}
\end{subequations}

\subsection{Comparison and Relation Between Topologies}

Finally, it is instructive to revisit the integrals obtained
in \Secs{wlreview}{Alter}.
They can be summarized in terms of the following prototype integrals:
\begin{subequations}
	\label{pre}
	\begin{alignat}{3}
		\label{pre-xx}
		\I_\text{interval}
		&\,=\,
			\int_0^\infty dT\,\, \mathe^{-iTX_2}\mem
			\int_0^1 d\s\,\, T\, \mathe^{iTK\s}
		\,,\\
		\label{pre-xp}
		\I_\text{half-line}
		&\,=\,
			\int_0^\infty d\s\,\, \mathe^{iK\s}
		\,,\\
		\label{pre-pp}
		\I_\text{full\,line}
		&\,=\,
		\frac{1}{\deltabar(0)}\mem
			\int_{-\infty}^\infty d\s\,\, \mathe^{iK\s}
		\,.
	\end{alignat}
\end{subequations}
We recall that,
with proper $i\e$ prescriptions,
the full evaluation of these integrals gives
\begin{subequations}
\begin{align}
	\I_\text{interval}
	\,&=\,
	\frac{-i}{X_2 {\mem-\,} i\e}\mem \frac{-i}{X_2 {\mem-\,} K {\mem-\,} i\e}
	\,,\\
	\I_\text{half-line}
	\,&=\,
	\frac{-i}{-K {\mem-\,} i\e}
	\,,\\
	\I_\text{full\,line}
	\,&=\,
	\frac{1}{\deltabar(0)}\,
	\deltabar(K)
	\,,
\end{align}
\end{subequations}
encoding
propagators 
awaiting for two, one, and zero LSZ reductions, respectively.

All of these three integrals reduce to the same value, $1$.
Take the half-line integral in \eqref{pre-xp}.
Physically, its infinite integral bound already anticipates that $X_2$ is put on shell:
the leg for $X_2$ has escaped to the boundary.
Note that this was argued from a concrete point of view in \Sec{Alter}.
As a result, the other amputation, $X_2 {\mem-\,} K \to 0$, in the guise of $K {\,\to\,} 0$, only remains.
Similarly, for the full line integral in \eqref{pre-pp},
the correct interpretation should be that
both $X_2 = 0$ and $X_2 {\mem-\,} K = 0$
have been demanded by the open ends at $\s \to \pm\infty$.
Hence $\deltabar(K) = \deltabar(0)$
and $\I_\text{full\,line} = 1$.

While
this verifies the first-quantized implementation of LSZ reduction from worldline topologies,
it is also possible to establish through
the field-theoretic LSZ reduction
that the half-line integral representation in \eqref{pre-xp}
follows from the Schwinger proper time representation in \eqref{pre-xx}:
\begin{align}
	\begin{split}
		\label{interval-to-hline-3}
		\lim_{X_2\to0}\,
		\Big(\,{
			iX_2\,
			\I_\text{interval}
		}\,\Big)
		\,&=\,
		\lim_{X_2\to0}\,\,
			\int_0^\infty dT\,\, 
			\bigg(\mem{
				- \frac{\partial}{\partial T}\,\mathe^{-iTX_2}
			}\,\bigg)\mem
		\int_0^1 d\s\,\, T\, \mathe^{iTK\s}
		\,,\\
		\,&=\,
		\lim_{X_2\to0}\,\,
			\int_0^\infty dT\,\, \mathe^{-iTX_2}\mem
				\frac{\partial}{\partial T}
				\int_0^T d\t\,\, \mathe^{iK\t}
		\,,\\
		\,&=\,
			\int_0^\infty dT\,
				\frac{\partial}{\partial T}\nem
				\int_0^T d\t\,\, \mathe^{iK\t}
		\,,\\
		\,&=\,
			\int_0^\infty d\t\,\, 
				\mathe^{iK\t}
		\,=\, 
		\I_\text{half-line}
		\,.
	\end{split}
\end{align}
More generally, 
the LSZ reduction from the interval to the half-line
is can be shown by
the following identity,
which holds for $\Omega(0)=0$:
\begin{align}
\begin{split}
	\label{interval-to-hline}
	\lim_{X_2\to0}\,
	iX_2 
	\int_0^\infty dT\,\, 
	\mathe^{-iT(X_2-i\e)}\,\hem
	\Omega(T)
	\,=\,
	\Omega(\infty)
	\,.
\end{split}
\end{align}
The calculus in \eqrefs{interval-to-hline-3}{interval-to-hline} has been known in the literature \cite{Mogull:2020sak}.
However, its implication
that the field-theoretic LSZ reduction is equivalent to the worldline topology-based LSZ reduction
seems to have been not 
widely realized.

For the full line representation in \eqref{pre-pp},
we comment that
the \textit{first-quantized} S-matrix element
computed in a constant einbein gauge
exactly produces the same infinite line integral,
but without the $\deltabar(0)$ quotient.
By first-quantized S-matrix element,
we mean the S-matrix element 
that one defines within the framework of quantum mechanics
by means of Born approximation.
Works \cite{Mogull:2020sak,Kopp:2022acm}
have pointed out a relationship between 
the first-quantized and second-quantized amplitudes
(in the context of classical/eikonal limit).
The observation has been that they
seem to differ only by a factor of the form $\deltabar(K)$
encoding the difference between two on-shell conditions put for the two massive legs.
In our view, however,
the correct interpretation should be that
$K \to 0$
has been already required for using the full line topology registering both past and future infinities,
so the factor $\deltabar(K)$ is canceled by the very $\deltabar(0)$ quotient
mandated
by the analysis on residual reparametrization redundancies.
This verifies and provides a strict derivation of the observation
that the first-quantized and second-quantized amplitudes
differ by a single delta function factor.
To reiterate, it should be understood that dropping the delta function factor
is due to the faithful realization of the measure $\mathcal{D}\k/\nem\vol(\Gauge)$.

\section{Application to Electromagnetism}
\label{QED}

Having established the foundations,
we now specialize in specific examples
to initiate concrete demonstrations.
In this section, we develop the phase space worldline formalism
for scalar quantum electrodynamics.
Our concrete goal is to obtain 
higher-multiplicity Compton amplitudes
in the classical, eikonalized limit,
meaning that the electromagnetic backgrounds
consist of plane waves carrying macroscopic \textit{wavenumbers} instead of momenta.
Note, however, that our framework itself is applicable to any situation
and is capable of computing  quantum corrections.

\subsection{Worldline Formalism}
\label{QED.wlf}

Let us employ
the half-line topology
with the mixed (position-momentum) boundary condition.
As remarked around \eqref{workflow.px},
this setup is seems to be ideal
for an efficient extraction of scattering amplitudes.

As established in \Sec{WLF},
the bulk-to-boundary propagator is computed 
from the first-quantized framework
as the partition function
for the half-line topology,
\begin{align}
	\label{zpx}
	\Z(p_2,x_1)
	\,&=\mem
		\int\nem \frac{\D{\kappa}}{\vol(\Gauge)}\mem
		\int^{p_2}_{x_1} \D{x} \D{p}\,\,
		\mathe^{i\S[x,p;\k]}
	\,=\mem
		\int^{p_2}_{x_1} \D{x} \D{p}\,\,
		\mathe^{i\S[x,p]}
	\,,
\end{align}
where $\S[x,p]$ is
the gauge-fixed action:
\begin{align}
	\S[x,p]
	\label{S-QED.fixed}
	\,&=\mem
	-p_\a(\infty)\mem x^\a(\infty)
	+
	\int_0^\infty d\s\,\,
	\bigg(\,{
		p_\a\mem \dot{x}^\a
		+ qA_\a(x)\mem \dot{x}^\a
		- (p^2+m^2)
	}\,\bigg)
	\,.
\end{align}
Here, we have applied the recipe in \eqref{recipe.px}
and eliminated the arbitrary constant $\k_\infty$ by
a rescaling $\s \mapsto (2/\k_\infty)\mem \s$.

We expand around the free theory saddle:
\begin{align}
\begin{split}
	\label{eq:sad.xp}
		x^\a(\s)
		\,&=\,
		\bmx^\a(\s) + \dx^\a(\s)
		\,=\,
		x_1^\a + 2p_2^\a\mem \s + \dx^\a(\s)
		\,,\\
		p_\a(\s)
		\,&=\,
		\bmp_\a(\s) + \ddp_\a(\s)
		\hem\hhem\hhem
		\,=\,
		p_2{}_\a + \ddp_\a(\s)
		\,.
\end{split}
\end{align}
The on-shell action in the free theory is $-p_2\mdot x_1$,
given that the external momentum $p_2$ is on-shell.
As a result, the partition function in \eqref{zpx}
boils down to
\begin{align}
	\label{zxval}
	\Z(p_2,x_1)
	\,=\,
	\mathe^{-ip_2x_1}
	\int \D{\dx} \D{\ddp}\,\,
	\exp\bigg(\,{
		i
		\int_0^\infty d\s\,\,
			\Big(\,{
				\ddp_\a\hem \d\dot{x}^\a
				- \ddp^2
			}\,\Big)
	}\mem\bigg)
	\,\,\mathe^{iS'[x,p]}
	\,,
\end{align}
where the boundary condition for the path integral is
$\ddp_\a(\infty) = 0$, $\dx^\a(0) = 0$.
According to the procedure taken in \Sec{General.variation},
the interaction action in \eqref{zxval}
can be massaged as
\begin{align}
	S'[x,p]
	\,=\,
	S_\text{sv}[x,p]
	\,\hem+ 
	\int_0^\infty\nem d\s\,\,
	\frac{d}{d\s}\hem \bigg[\,\mem{
		\sum_{n=0}^\infty\,
		\frac{1}{(n\hem{\mem+\mem}1)!}\,
		A_{\b,\a_1\cdots \a_n}\hnem(x)
		\mem \dx^\b
		\mem \dx^{\a_1}{\cdots\mem}\dx^{\a_n}
	}\mem\bigg]
	\,,
\end{align}
so $S_\text{sv}[x,p]$ encodes the symplectic vertices.
The boundary term vanishes exactly by virtue of our boundary conditions
$A_\a(x(\infty)\hnem) {\,\to\,} 0$,
$\dx^\a(0) = 0$.
With this understanding, the partition function in \eqref{zxval}
is evaluated as
\begin{align}
	\begin{split}
		\label{Zfeyn}
		\mathcal{Z}(p_2,x_1)
		\,&=\,
		\mathe^{-ip_2x_1}
		\exp\Big(\,\mem{ i\mem (\text{%
				Connected Feynman graphs%
			})
		}\hem\,\Big)
		\,,
	\end{split}
\end{align}
with respect to the Feynman rules in \Sec{General.feyn}.

\newpage

The propagator needs to be elaborated on.
The boundary condition of the path integral in \eqref{zxval}
stipulates that
fluctuations in $\ddp$ always propagate toward the past,
whereas
fluctuations in $\dx$ always propagate toward the future.
Therefore, when the phase space coordinates are grouped as $\z^i = (x^\m , p_\m)$,
the bare Green's function is given by
\begin{align}
	\label{Gxp}
	G^{ij}\hnem(\s_1,\s_2)
	\,=\,
	\begin{pmatrix}
			\mem 0 & \delta^\m{}_\n\mem \Theta_>(\s_1,\s_2) \,
			\\
			-\delta_\m{}^\n\mem \Theta_<(\s_1,\s_2)  & 0 \mem
	\end{pmatrix}
	\,,
\end{align}
which is symmetric as
$G^{ij}\hnem(\s_1,\s_2) = G^{ji}\hnem(\s_2,\s_1)$
and satisfies the condition stated in \eqref{pb-encode}:
\begin{align}
	\s_1 > \s_2
	\qiq
	{-i\mem}
	\Big(\,{
		G^{ij}\hnem(\s_1,\s_2)
		-
		G^{ji}\hnem(\s_1,\s_2)
	}\,\Big)
	\,=\,
	\begin{pmatrix}
		\,\,i\mem\{ x^\m , x^\n \}^\circ & \,i\mem\{ x^\m , p_\n \}^\circ\,
		\\
		\,\,i\mem\{ p_\m , x^\n \}^\circ & \,i\mem\{ p_\m , p_\n \}^\circ\,
	\end{pmatrix}
	\,.
\end{align}

To extract the scattering amplitudes,
one assumes the background of $(N{\mem-\,}2)$ plane waves,
with wavenumbers $k_I$ for $I = 3,4,\cdots, N$, say.
In this background,
each vertex functional takes the form
\begin{align}
	\label{Vv}
	V_I[\dx,\ddp]
	\,=\,
		v_I[\dx,\ddp]\,
			\mathe^{ik_Ix_1}
	\,.
\end{align}
Therefore, the part of the partition function in \eqref{Zfeyn}
that is multilinear in the plane waves 
is given in the form
\begin{align}
	\label{zxval.ml}
	\Z(p_2,x_1)\mem\Big|_\text{m.l.}
	\, \mathe^{ip_1x_1}
	\,=\,
		\mathe^{-ip_2x_1}\,
		\BB{i\mem \F_N\, \mathe^{i(k_3+\cdots+k_N)x_1}}
		\,
		\mathe^{ip_1x_1}
	\,,
\end{align}
where $\F_N$ is the sum of all relevant Feynman diagrams
computed with the reduced vertex functional $v_I[\dx,\ddp]$ in \eqref{Vv}.
Clearly, the Fourier transform from $x_1$ to $p_1$ yields the momentum-conserving delta function.
Therefore, the amplitude is found by LSZ reducing $\F_N$ by once.
In the eikonal limit, it is given by
\begin{align}
	\label{amplitudeN}
	\A(p_2,p_1;3,4,\cdots,N)
	\,=\,	
	\lim_{(K_3{\mem+\,} \cdots {\mem+\,}K_N) \to 0\vph}
	\,
		-i\hem(K_3{\mem+\,} \cdots {\mem+\,}K_N)
		\, \F_N
	\,,
\end{align}
since the off-shell propagator for the leg $1$ 
on $X_2 = {p_2}^2 + m^2 = 0$
is
\begin{align}
\label{eikonalized-propagator}
	\lim_{X_2\to0\vph}\,
	\frac{-i\hbar}{
		\bigbig{
			p_2 - \hbar\mem (k_3 {\mem+\,} \cdots {\mem+\,} k_N)
		}^{\nem2}
		+ m^2
	}
	\,=\,
	\frac{i}{
		(K_3{\mem+\,} \cdots {\mem+\,}K_N)
	}
	+ \mathcal{O}(\hbar)
	\,.
\end{align}
Here, we have 
restored $\hbar$ and denoted
\begin{align}
	K_I
	\,:=\,
		2k_I \mdot p_2
	\,.
\end{align}
\eqref{amplitudeN} is the very ultimate formula 
of our framework
that produces the amplitudes.

\subsection{Feynman Rules}
\label{QED.feyn}

Let us now explicitly spell out the Feynman rules
by applying the result from \Sec{General.feyn}.
From \eqref{S-QED.fixed}, the elements of the symplectic geometry are identified as
\begin{align}
\begin{split}
	\label{qed-cheatsheet}
	\omega^\circ \,=\,
		dp_\a \swedge dx^\a
	&\,,\quad
	H^\circ \,=\,
		p^2 + m^2
	\,,\\
	\theta' \,=\,
		A_\a(x)\mem dx^\a
	&\,,\quad
	\omega' \,=\,
		\frac{1}{2}\, F_{\a\b}(x)\mem dx^\a \swedge dx^\b
	\,.
\end{split}
\end{align}

Firstly,
the bare propagator 
is found as
\begin{align}
	\label{scalar-prop}
	\adjustbox{valign=c}{\begin{tikzpicture}[]
			\node (o) at (0,0) {};
			\node (xshift) at (0.8,0) {};
			\node (i0) at ($(o)$) {};
			\node (i1) at ($(i0)+1*(xshift)$) {};
			\node (i2) at ($(i0)+2*(xshift)$) {};
			\node[t] (x) at ($(i0)$) {$\mathllap{
					\delta x^\a\hnem(\s_1)
				}$};
			\node[t] (p) at ($(i2)$) {$\mathrlap{
					\delta p_\b\hnem(\s_2)
				}$};
			\draw[prop] (p)--(x);
	\end{tikzpicture}}
	\kern-0.125em\phantom{\small
		\delta p_\n\hnem(\s_2)
	}
	\,\,\,&=\,\,\,
	\delta^\a{}_\b
	\,
	\Theta_{>}(\s_1,\s_2)
	\,,
\end{align}
which nicely displays the structure of the free theory's Poisson bracket:
$\{ x^\a , p_\b \}^\circ = \delta^\a{}_\b$.
Here, the arrow signifies the \textit{species}:
the worldline fields $\z^i(\s)$ have split into two groups,
$x^\a(\s)$ and $p_\a(\s)$.
Notably,
our mixed boundary condition
aligns the directionality of this arrow with the causality of the Green's function $\Theta_>(\s_1,\s_2)$,
so there is no need for introducing another arrow.
Amusingly,
the species
$\delta x^\a$ and $\delta p_\a$
serve as
``up and down indices''
in this diagrammatics.

Secondly, the only Hamiltonian vertex is
\begin{align}
	\label{QED.hv}
	\phantom{
		\delta p_\n\hnem(\s_2)
	}\kern-0.125em
	\adjustbox{valign=c}{\begin{tikzpicture}[]
			\node (o) at (0,0) {};
			\node (r1) at ($(o)$) {};
			\node (l1) at ($(o)+(165:1.3)$) {};
			\node (l2) at ($(o)+(195:1.3)$) {};
			\node (ynudge) at (0,0.01265) {};
			\node[cdot] (v) at (r1) {};
			\node[t] (p1) at ($(l1)$) {$\mathllap{
					\delta p_\a\hnem(\s_1)
				}$};
			\node[t] (p2) at ($(l2)$) {$\mathllap{
					\delta p_\b\hnem(\s_2)
				}$};
			\draw[prop] ($(v)+(ynudge)$)--(p1);
			\draw[prop] ($(v)-(ynudge)$)--(p2);
	\end{tikzpicture}}
	\,
	\,\,\,&=\,\,\,
	-2\eta^{\a\b}
	\,
	\delta(\s_1,\s_2)
	\,.
\end{align}

It remains to obtain the symplectic vertices.
To this end, consider the electromagnetic background of $(N{\mem-\,}2)$ plane waves:
\begin{align}
	\label{AFwaves}
	A_\a\hnem(x)
	\,=\,
	\sum_{I=3}^N\,
		e_I{}_\a\,
		\mathe^{ik_Ix}
	\,,\quad
	F_{\a\b}(x)
	\,=\,
	\sum_{I=3}^N\,
		i\mem \varphi_I{}_{\a\b}\,
		\mathe^{ik_Ix}
	\,.
\end{align}
Here, we have employed antisymmetric tensor (bivector) polarizations $\varphi_I{}_{\a\b}$,
which are \textit{gauge-invariant}
and satisfy $k_I^\a\mem \varphi_I{}_{\a\b} = 0$ and $(k_I \mwedge \varphi_I)_{\c\a\b} = 0$.
The vector polarizations satisfy $(k_I \mwedge e_I)_{\a\b} = \varphi_I{}_{\a\b}$.
On the saddle trajectory, \eqref{AFwaves} evaluates to
\begin{align}
	\label{AFwaves-on}
	A_\a\hnem(\bmx(\s)\hnem)
	\,=\,
	\sum_{I=3}^N\,
		e_I{}_\a\,
		\mathe^{iK_I\s}\,
		\mathe^{ik_Ix_1}
	\,,\quad
	F_{\a\b}(\bmx(\s)\hnem)
	\,=\,
	\sum_{I=3}^N\,
		i\mem \varphi_I{}_{\a\b}\,
		\mathe^{iK_I\s}\,
		\mathe^{ik_Ix_1}
	\,.
\end{align}

To extract the reduced vertices 
for computing $\F_N$ in \eqref{amplitudeN},
one drops the factors $\mathe^{ik_Ix_1}$ in \eqref{AFwaves-on}.
The regular symplectic vertex of worldline fluctuation valence $n=0$ is
\begin{align}
	\label{QED.sv0}
	\adjustbox{valign=c,raise=-1.2em}{\begin{tikzpicture}[]
			\node (o) at (0,0) {};
			\node (yshift) at (0,-0.9) {};
			\node (i3) at ($(o)+(yshift)$) {};
			\node[cdot] (v) at ($(o)$) {};
			\node[T] (A) at ($(i3)$) {};
			\draw[wiggly] (v)--(A);
	\end{tikzpicture}}
	\,\,
	\,\,\,&=\,\,\,
	\BB{2 e\mdot p}
	\,
	\frac{i}{K}
	\,.
\end{align}
The regular symplectic vertex of valence $n=1$ is
\begin{align}
	\label{QED.sv1}
	\adjustbox{valign=c,raise=-0.9em}{\begin{tikzpicture}[]
			\node (o) at (0,0) {};
			\node (xshift) at (1.3,0) {};
			\node (yshift) at (0,-0.9) {};
			\node (i1) at ($(o)-(xshift)$) {};
			\node (i2) at ($(o)+(xshift)$) {};
			\node (i3) at ($(o)+(yshift)$) {};
			\node[cdot] (v) at ($(o)$) {};
			\node[T] (A) at ($(i3)$) {};
			\node[t] (x) at ($(i2)$) {$\mathrlap{
					\delta x^\a\hnem(\s)
				}$};
			\draw[wiggly] (v)--(A);
			\draw[prop] (x)--(v);
	\end{tikzpicture}}
	\kern-0.125em\phantom{\small
		\delta x^\a\hnem(\s)
	}
	\,\,\,&=\,\,\,
	\BB{
		2i\hem \varphi_{\a\b}\mem p^\b
	}
	\,
	\mathe^{i K\s}
	\,.
\end{align}
The regular symplectic vertex of valence $n=2$ is
\begin{align}
	\label{QED.sv2}
	\adjustbox{valign=c,raise=-0.65em}{\begin{tikzpicture}[]
			\node (o) at (0,0) {};
			\node (xshift) at (1.0,0) {};
			\node (yshift) at (0,-0.9) {};
			\node (i1) at ($(o)-(xshift)$) {};
			\node (x1) at ($(o)+(15:1.3)$) {};
			\node (x2) at ($(o)+(-15:1.3)$) {};
			\node (a) at ($(o)+(yshift)$) {};
			\node[cdot] (V) at ($(o)$) {};
			\node[T] (A) at ($(a)$) {};
			\node[t] (X1) at ($(x1)$) {$\mathrlap{
					\delta x^{\a_1}\hnem(\s_1)
				}$};
			\node[t] (X2) at ($(x2)$) {$\mathrlap{
					\delta x^{\a_2}\hnem(\s_2)
			}$};
			\draw[wiggly] (V)--(A);
			\draw[prop] (X1)--(V);
			\draw[prop] (X2)--(V);
	\end{tikzpicture}}
	\kern-0.125em\phantom{\small
		\delta x^{\a_2}\hnem(\s_2)
	}
	\,\,\,&=\,\,\,
	\BB{
		2i^2\hem k_\wrap{(\a_1} \varphi_\wrap{\a_2)\b}\mem p^\b
		\hem}
	\,
	\mathe^{i K\s_1}\,
	\delta(\s_1,\s_2)
	\,.
\end{align}
The regular symplectic vertex of valence $n\geq2$ is
\begin{align}
	\label{QED.svn}
	\adjustbox{valign=c,raise=-0.425em}{\begin{tikzpicture}[]
			\node (o) at (0,0) {};
			\node (xshift) at (1.0,0) {};
			\node (yshift) at (0,-0.9) {};
			\node (i1) at ($(o)-(xshift)$) {};
			\node (x1) at ($(o)+(22.5:1.3)$) {};
			\node (x2) at ($(o)+(-22.5:1.3)$) {};
			\node (a) at ($(o)+(yshift)$) {};
			\node[cdot] (V) at ($(o)$) {};
			\node[T] (A) at ($(a)$) {};
			\node[t] (X1) at ($(x1)$) {$\mathrlap{
					\delta x^{\a_1}\hnem(\s_1)
				}$};
			\node[t] (X2) at ($(x2)$) {$\mathrlap{
					\delta x^{\a_n}\hnem(\s_n)
				}$};
			\draw[wiggly] (V)--(A);
			\draw[prop] (X1)--(V);
			\draw[prop] (X2)--(V);
			\node[tdot] (dot2) at ($(o)-(0+180-10:0.55)$) {};
			\node[tdot] (dot3) at ($(o)-(0+180:0.55)$) {};
			\node[tdot] (dot4) at ($(o)-(0+180+10:0.55)$) {};
	\end{tikzpicture}}
	\kern-0.125em\phantom{\small
		\delta x^{\a_2}\hnem(\s_2)
	}
	\,\,\,&=\,\,\,
	\BB{
		2\hem i^n\hem k_\wrap{(\a_1} \nem{\cdots}\mem k_\wrap{\a_{n-1}} \varphi_\wrap{\a_n)\b}\mem p^\b
	}
	\,
	\mathe^{i K\s_1}
	\,
	\delta^{(n-1)}\hnem(\s_1,\cdots,\s_n)
	\,.
\end{align}

\newpage

When listing these formulae, we have omitted the photon label $I$.
Also, the coupling $q$ has been peeled off.
Note that all of our vertices are linear in the photon by construction.
Moreover, we have further simplified our notation by abbreviating $p_2$ as $p$,
hoping that there arises no confusion with the dynamical worldline degrees of freedom $p(\s)$.
The rationale is that the difference between $p_2$ and $p_1$ is of order $\O(\hbar^1)$
in the eikonal limit,
so the classical amplitude could be effectively viewed as a function of 
a single momentum variable $p$
and the photon variables.

With this understanding,
the pinched symplectic vertex of valence $n=2$ is
\begin{align}
	\label{QED.psv2}
	\adjustbox{valign=c,raise=-0.65em}{\begin{tikzpicture}[]
			\node (o) at (0,0) {};
			\node (xshift) at (1.0,0) {};
			\node (yshift) at (0,-0.9) {};
			\node (i1) at ($(o)-(xshift)$) {};
			\node (x1) at ($(o)+(15:1.3)$) {};
			\node (x2) at ($(o)+(-15:1.3)$) {};
			\node (a) at ($(o)+(yshift)$) {};
			\node[cdot] (V) at ($(o)$) {};
			\node[T] (A) at ($(a)$) {};
			\node[t] (X1) at ($(x1)$) {$\mathrlap{
					\delta x^{\a_1}\hnem(\s_1)
				}$};
			\node[t] (X2) at ($(x2)$) {$\mathrlap{
					\delta x^{\a_2}\hnem(\s_2)
				}$};
			\node (p) at ($(V)+(43:0.33)$) {\pin};
			\draw[wiggly] (V)--(A);
			\draw[prop] (X1)--(V);
			\draw[prop] (X2)--(V);
	\end{tikzpicture}}
	\kern-0.125em\phantom{\small
		\delta x^{\a_2}\hnem(\s_2)
	}
	\,\,\,&=\,\,\,
	\BB{
		\tfrac{1}{2}\,
			i^2\hem \varphi_{\a_1\a_2}
	}
	\,
	\mathe^{i K\s_2}\,
		\frac{1}{i}
		\frac{\partial}{\partial\s_1}\,
		\delta(\s_1,\s_2)
	\,.
\end{align}
The pinched symplectic vertex of valence $n=3$ is
\begin{align}
	\label{QED.psvn}
	\adjustbox{valign=c,raise=-0.425em}{\begin{tikzpicture}[]
			\node (o) at (0,0) {};
			\node (xshift) at (1.0,0) {};
			\node (yshift) at (0,-0.9) {};
			\node (i1) at ($(o)-(xshift)$) {};
			\node (x1) at ($(o)+(22.5:1.3)$) {};
			\node (x2) at ($(o)+(0:1.3)$) {};
			\node (x3) at ($(o)+(-22.5:1.3)$) {};
			\node (a) at ($(o)+(yshift)$) {};
			\node[cdot] (V) at ($(o)$) {};
			\node[T] (A) at ($(a)$) {};
			\node[t] (X1) at ($(x1)$) {$\mathrlap{
					\delta x^{\a_1}\hnem(\s_1)
				}$};
			\node[t] (X2) at ($(x2)$) {$\mathrlap{
					\delta x^{\a_2}\hnem(\s_2)
				}$};
			\node[t] (X3) at ($(x3)$) {$\mathrlap{
					\delta x^{\a_3}\hnem(\s_3)
			}$};
			\node (p) at ($(V)+(52:0.33)$) {\pin};
			\draw[wiggly] (V)--(A);
			\draw[prop] (X1)--(V);
			\draw[prop] (X2)--(V);
			\draw[prop] (X3)--(V);
	\end{tikzpicture}}
	\kern-0.125em\phantom{\small
		\delta x^{\a_2}\hnem(\s_2)
	}
	\,\,\,&=\,\,\,
	\BB{
		\tfrac{2}{3}\,
		i^3\hem \varphi_{\a_1(\a_2} k_{\a_3)}
	}
	\,
	\mathe^{i K\s_2}\,
	\frac{1}{i}
	\frac{\partial}{\partial\s_1}\,
	\delta^{(2)}\hnem(\s_1,\s_2,\s_3)
	\,.
\end{align}
The pinched symplectic vertex of valence $n\geq2$ is
\begin{align}
\label{QED.psvn}
\kern-0.8em
\adjustbox{valign=c,raise=-0.425em}{\begin{tikzpicture}[]
		\node (o) at (0,0) {};
		\node (xshift) at (1.0,0) {};
		\node (yshift) at (0,-0.9) {};
		\node (i1) at ($(o)-(xshift)$) {};
		\node (x1) at ($(o)+(22.5:1.3)$) {};
		\node (x2) at ($(o)+(-22.5:1.3)$) {};
		\node (a) at ($(o)+(yshift)$) {};
		\node[cdot] (V) at ($(o)$) {};
		\node[T] (A) at ($(a)$) {};
		\node[t] (X1) at ($(x1)$) {$\mathrlap{
				\delta x^{\a_1}\hnem(\s_1)
			}$};
		\node[t] (X2) at ($(x2)$) {$\mathrlap{
				\delta x^{\a_n}\hnem(\s_n)
			}$};
		\node (p) at ($(V)+(52:0.33)$) {\pin};
		\draw[wiggly] (V)--(A);
		\draw[prop] (X1)--(V);
		\draw[prop] (X2)--(V);
		\node[tdot] (dot2) at ($(o)-(0+180-10:0.55)$) {};
		\node[tdot] (dot3) at ($(o)-(0+180:0.55)$) {};
		\node[tdot] (dot4) at ($(o)-(0+180+10:0.55)$) {};
\end{tikzpicture}}
\kern-0.125em\phantom{\small
	\delta x^{\a_2}\hnem(\s_2)
}
\,\,&=\,\,\,
	\BB{
		\tfrac{n-1}{n}\,
		i^n\hem \varphi_{\a_1(\a_2} k_{\a_3} {\cdots}\hem k_{\a_n)}
	}
	\,
	\mathe^{i K\s_2}\,
	\frac{1}{i}
	\frac{\partial}{\partial\s_1}\,
	\delta^{(n-1)}\hnem(\s_1,\cdots,\s_n)
\,.
\kern-0.2em
\end{align}

Finally, note that the fat propagator 
due to the Hamiltonian vertex
is
\begin{align}
	\label{Gxp.fat}
	\begin{pmatrix}
		\,-2\eta^{\a\b}\mem \Delta(\s_1,\s_2)\, & \,\delta^\m{}_\n\mem \Theta_>(\s_1,\s_2)\,
		\\
		\,-\delta_\m{}^\n\mem \Theta_<(\s_1,\s_2)\,  & 0 \mem
	\end{pmatrix}
	\,,
\end{align}
which only differs from the bare propagator in \eqref{scalar-prop}
by a single component:
\begin{align}
	\label{QED.hv}
	\phantom{
		\delta p_\n\hnem(\s_2)
	}\kern-0.125em
	\adjustbox{valign=c}{\begin{tikzpicture}[]
			\node (o) at (0,0) {};
			\node (r1) at ($(o)$) {};
			\node (l1) at ($(o)+(165:1.3)$) {};
			\node (l2) at ($(o)+(195:1.3)$) {};
			\node (ynudge) at (0,0.01265) {};
			\node[cdot] (v) at (r1) {};
			\node[t] (p1) at ($(l1)$) {$\mathllap{
					\dx^\a\hnem(\s_1)
				}$};
			\node[t] (p2) at ($(l2)$) {$\mathllap{
					\dx^\b\hnem(\s_2)
				}$};
			\draw[prop] ($(v)+(ynudge)$)--(p1);
			\draw[prop] ($(v)-(ynudge)$)--(p2);
	\end{tikzpicture}}
	\,
	\,\,\,=\,\,\,
	-2\eta^{\a\b}
	\,
	\Delta(\s_1,\s_2)
	\,.
\end{align}
Here, $\Delta(\s_1,\s_2)$ denotes a symmetric second-order propagator:
\begin{align}
	-\frac{\partial^2}{{\partial \s_1}^2}\,
	\Delta(\s_1,\s_2) \,=\, 
	\delta(\s_1,\s_2)
	\,,\quad
	\Delta(\s_1,\s_2) \,=\, \Delta(\s_2,\s_1)
	\,.
\end{align}
Explicitly, it is given by
\begin{align}
	\Delta(\s_1,\s_2)
	\,=\,
	-
	\frac{
		|\s_1 {\mem-\,} \s_2|
	}{2}
	\,.
\end{align}
As a shorthand notation,
we might desire to denote
\begin{align}
	\label{scalar-prop2}
	\adjustbox{valign=c}{\begin{tikzpicture}[]
			\node (o) at (0,0) {};
			\node (xshift) at (0.8,0) {};
			\node (i0) at ($(o)$) {};
			\node (i1) at ($(i0)+1*(xshift)$) {};
			\node (i2) at ($(i0)+2*(xshift)$) {};
			\node[t] (x1) at ($(i0)$) {$\mathllap{
					\delta x^\a\hnem(\s_1)
				}$};
			\node[t] (x2) at ($(i2)$) {$\mathrlap{
					\delta x^\b\hnem(\s_2)
				}$};
			\draw[linear] (x1)--(x2);
	\end{tikzpicture}}
	\kern-0.125em\phantom{\small
		\delta x^\b\hnem(\s_2)
	}
	\,\,\,&=\,\,\,
		-2\eta^{\a\b}
		\,
		\Delta(\s_1,\s_2)
	\,,
\end{align}
hoping that no confusion would arise
as the $x$-$x$ propagator is the only channel without a species change.
Note also that the full propagator in \eqref{Gxp.fat}
also satisfies the condition stated in \eqref{pb-encode},
as $\lim_{\s_1 \to \s_2} \Delta(\s_1,\s_2) = 0$.

\subsection{Compton Amplitudes}

Finally, we demonstrate how the classical $n$-Compton amplitudes in the eikonalized regime
are derived in our formalism.
By an $n$-Compton amplitude,
we mean the analog of the Compton amplitude for photon multiplicity $n$:
a massive particle scattering off $n$ photons.

\subsubsection{Multiplicity 1}
\label{Compton.EM1}

The $1$-Compton amplitude is simply the three-point amplitude.
The relevant diagram is
\begin{align}
	\F_3
	\quad=\quad
	\adjustbox{valign=c,raise=-1.2em}{\begin{tikzpicture}[]
			\node (o) at (0,0) {};
			\node (yshift) at (0,-0.9) {};
			\node (i3) at ($(o)+(yshift)$) {};
			\node[cdot] (v) at ($(o)$) {};
			\node[T] (A) at ($(i3)$) {$3$};
			\draw[wiggly] (v)--(A);
	\end{tikzpicture}}
\quad=\quad
	\BB{2 e_3\mdot p}
	\,
	\frac{i}{K_3}
	\,.
\end{align}
Using the formula in \eqref{amplitudeN},
we find
\begin{align}
	\label{QED.A3}
	\A_3
	\,=\,
		\lim_{K_3\to0\vph}
			-iK_3\, \F_3
	\,=\,
		2 e_3\mdot p
	\,.
\end{align}
Surely, this amplitude is gauge invariant:
on $K_3 = 2k_3\mdot p = 0$,
the shift 
$e_3 \mapsto e_3 + \lambda\mem k_3$
changes its value by zero.

\subsubsection{Multiplicity 2}
\label{Compton.EM2}

The $2$-Compton amplitude is simply the Compton amplitude.
We find
\begin{align}
	\label{QED.F4}
	\F_4
	\quad=\quad
	\adjustbox{valign=c,raise=-1.35em}{\begin{tikzpicture}[]
			\node (o) at (0,0) {};
			\node (xshift) at (0.8,0) {};
			\node (yshift) at (0,-0.9) {};
			\node (i3) at ($(o)$) {};
			\node (i4) at ($(o)+1*(xshift)$) {};;
			\node[cdot] (v3) at ($(i3)$) {};
			\node[cdot] (v4) at ($(i4)$) {};
			\node[T] (A3) at ($(i3)+(yshift)$) {$3$};
			\node[T] (A4) at ($(i4)+(yshift)$) {$4$};
			\draw[wiggly] (i3)--(A3);
			\draw[wiggly] (i4)--(A4);
			\draw[linear] (i3)--(i4);
	\end{tikzpicture}}
	\quad=\quad
		\BB{-8\mem p\hhem\varphi_3\varphi_4p}
		\, 
		\int d\s_3 d\s_4\,\,
			\mathe^{iK_3\s_3}\, \mathe^{iK_4\s_4}\,
			\Delta(\s_3,\s_4)
	\,,
\end{align}
where we have employed the shorthand notation in \eqref{scalar-prop2}.
The integral in \eqref{QED.F4} 
is conveniently evaluated in the time domain:
\begin{align}
	\label{INT4}
	\int d\s_3 d\s_4\,\,
		\mathe^{iK_3\s_3}\, \mathe^{iK_4\s_4}\,
		\Delta(\s_3,\s_4)
	\,=\,
	\int d\s\,\,
		\mathe^{iK_3\s}\mem
		\frac{1}{-{\partial_\s}^2}\,
		\mathe^{iK_4\s}
	\,=\,
		\frac{i}{K_3{\mem+\,}K_4\vph}\,
		\frac{1}{{K_4}^2\vph}
	\,.
\end{align}
Here, 
the integral bounds are $[0,\infty)$
unless otherwise stated.
The amplitude is found as
\begin{align}
	\label{QED.A4}
	\A_4
	\,=\,
		\lim_{K_3+K_4\to0\vph}
		-i\hem(K_3{\mem+\,}K_4)\, \F_4
	\,=\,
		8\mem p\hhem\varphi_3\varphi_4p\,
		\frac{1}{{K_3}{K_4}\vph}
	\,,
\end{align}
on the support of $K_3{\mem+\,}K_4 = 0$.
This correctly exhibits the $s$- and $u$-channel poles
(represented as $1/K_3$ and $1/K_4$ in the eikonal limit)
and is symmetric under $3\mlra4$ exchange.

Notably, a charm of this derivation for the Compton amplitude
is that gauge invariance is manifested
at all intermediate steps.
This is by virtue of symplectic perturbation theory,
where the symplectic vertices 
emitting at least one worldline fluctuation
are directly formulated in terms of the field strength.
In particular, the numerator of \eqref{QED.A4}
describes the dot product between two ``Lorentz forces''
that the 
photons $3$ and $4$ exert on the background trajectory,
$\varphi_3p$ and $\varphi_4p$.

\subsubsection{Multiplicity 3}
\label{Compton.EM3}

For the $3$-Compton amplitude,
we encounter two classes of diagrams:
\begin{align}
	\label{QED.F5}
	\F_{345}^\text{(r)}
	\,\,\,=\,\,\,
	\adjustbox{valign=c,raise=-1.07em}{\begin{tikzpicture}[]
			\node (o) at (0,0) {};
			\node (xshift) at (0.8,0) {};
			\node (yshift) at (0,-0.9) {};
			\node (i3) at ($(o)$) {};
			\node (i4) at ($(o)+1*(xshift)$) {};
			\node (i5) at ($(o)+2*(xshift)$) {};
			\node[cdot] (v3) at ($(i3)$) {};
			\node[cdot] (v4) at ($(i4)$) {};
			\node[cdot] (v5) at ($(i5)$) {};
			\node[T] (A3) at ($(i3)+(yshift)$) {$3$};
			\node[T] (A4) at ($(i4)+(yshift)$) {$4$};
			\node[T] (A5) at ($(i5)+(yshift)$) {$5$};
			\node[t] (x) at ($(i2)$) {};
			\node (p) at ($(v4)+(135:0.22)$) {\vpin};
			\draw[wiggly] (i3)--(A3);
			\draw[wiggly] (i4)--(A4);
			\draw[wiggly] (i5)--(A5);
			\draw[linear] (i3)--(i4)--(i5);
	\end{tikzpicture}}
	\quad
	\,,\quad
	\F_{345}^\text{(p)}
	\,\,\,=\,\,\,
	\adjustbox{valign=c,raise=-1.07em}{\begin{tikzpicture}[]
			\node (o) at (0,0) {};
			\node (xshift) at (0.8,0) {};
			\node (yshift) at (0,-0.9) {};
			\node (i3) at ($(o)$) {};
			\node (i4) at ($(o)+1*(xshift)$) {};
			\node (i5) at ($(o)+2*(xshift)$) {};
			\node[cdot] (v3) at ($(i3)$) {};
			\node[cdot] (v4) at ($(i4)$) {};
			\node[cdot] (v5) at ($(i5)$) {};
			\node[T] (A3) at ($(i3)+(yshift)$) {$3$};
			\node[T] (A4) at ($(i4)+(yshift)$) {$4$};
			\node[T] (A5) at ($(i5)+(yshift)$) {$5$};
			\node[t] (x) at ($(i2)$) {};
			\node (p) at ($(v4)+(135:0.22)$) {\pin};
			\draw[wiggly] (i3)--(A3);
			\draw[wiggly] (i4)--(A4);
			\draw[wiggly] (i5)--(A5);
			\draw[linear] (i3)--(i4)--(i5);
	\end{tikzpicture}}
	\quad
	\,.
\end{align}
Here, we have used the shorthand notation in \eqref{scalar-prop2}.

It is convenient to
denote the ``Lorentz force'' factors as 
\begin{align}
	f_I{}_\a
	\,:=\,
	(\varphi_I p)_\a
	\,=\,
	\varphi_I{}_{\a\b}\mem p^\b
	\,.
\end{align}
Due to the magnetic field equation (Jacobi identity) of Maxwell fields,
it holds that
\begin{align}
	\label{magphi}
	(k_I \mwedge \varphi_I)_{\c\a\b}
	\,=\, 0
	\qiq
	\varphi_I{}_{\a\b}
	\,=\,
	-\frac{2}{K_I}\,
	( k_I \mwedge f_I )_{\a\b}
	\,.
\end{align}
In terms of these variables, scalar invariants are formed as
\begin{align}
	\label{kf-invariants}
	k_I \cdot k_J
	\,=:\,
		\kk{IJ}
	\,,\quad
	k_I \cdot\nem f_J 
	\,=:\,
		\kf{IJ}
	\,,\quad
	f_I \cdot\hnem f_J
	\,=:\,
		\ff{IJ}
	\,.
\end{align}

For the regular diagram $\F_{345}^\text{(r)}$,
the tensor factor is
\begin{align}
	16\,
		f_3^\a\,
		( k_4 \modot f_4 )_{\a\b}\,
		f_5^\b
	\,=\,
	16\,
		( f_3 ( k_4 \modot f_4 )f_5 )
	\,,
\end{align}
where $\odot$ is the symmetrized tensor product.
The integral factor is
\begin{align}
	\label{QED.I5r}
	\int d\s\,\,
		\mathe^{iK_3\s}\,
			\frac{1}{-{\partial_\s}^2}\,
		\mathe^{iK_4\s}\,
			\frac{1}{-{\partial_\s}^2}\,
		\mathe^{iK_5\s}
	\,=\,
		\frac{i}{K_3{\mem+\,}K_4{\mem+\,}K_5\vph}\,
		\frac{1}{{(K_4{\mem+\,}K_5)}^2\vph}\mem
		\frac{1}{{K_5}^2\vph}\,
	\,.
\end{align}
Thus, the contribution to the amplitude is
\begin{align}
	\label{QED.A5r}
	\A_{345}^\text{(r)}
	\,=\,
	16\,
		( f_3 ( k_4 \modot f_4 )f_5 )
	\,
	\frac{1}{{K_3}^2{K_5}^2\vph}
	\,,
\end{align}
on the support of $K_3{\mem+\,}K_4{\mem+\,}K_5 = 0$.

For the pinched diagram $\F_{345}^\text{(p)}$,
the tensor factor is
\begin{align}
		8\,
		f_3^\a\,
		\varphi_4{}_{\a\b}\,
		f_5^\b
	\,=\,
		-16\,
			(f_3 ( k_4 \mwedge f_4 ) f_5)
		\,
		\frac{1}{K_4}
	\,,
\end{align}
where we have used \eqref{magphi}.
The integral factor is
\begin{align}
	\label{QED.I5p}
	\int d\s\,\,
	\mathe^{iK_3\s}\,
	\frac{1}{-{\partial_\s}^2}\,
	\frac{1}{i}\mem
	\partial_\s\,
	\mathe^{iK_4\s}\,
	\frac{1}{-{\partial_\s}^2}\,
	\mathe^{iK_5\s}
	\,=\,
		\frac{i}{K_3{\mem+\,}K_4{\mem+\,}K_5\vph}\,
		\frac{1}{K_4{\mem+\,}K_5\vph}\mem
		\frac{1}{{K_5}^2\vph}
	\,.
\end{align}
Thus, the contribution to the amplitude is
\begin{align}
	\label{QED.A5p}
	\A_{345}^\text{(p)}
	\,=\,
	16\,
	( f_3 ( k_4 \mwedge f_4 )f_5 )
		\,
		\frac{1}{K_3 K_4 {K_5}^2\vph}
		\,
	\,,
\end{align}
on the support of $K_3{\mem+\,}K_4{\mem+\,}K_5 = 0$.

\newpage

Therefore, from symmetry factor considerations,
the final amplitude is found as
\begin{align}
\begin{split}
	\label{A5.sf}
		\A_5
		\,&=\,
		\BB{
			\A_{345}^\text{(r)}
			+
			\A_{345}^\text{(p)}
			+
			\A_{543}^\text{(p)}
		}
		+ (\text{cyc.})_3
		\,,\\
		\,&=\,
		16\,
		\bb{
			( f_3 ( k_4 \modot f_4 )f_5 )
			\,
			\frac{1}{{K_3}^2{K_5}^2\vph}
			+
			( f_3 ( k_4 \mwedge f_4 )f_5 )
			\,
			\frac{K_3 {\mem-\,} K_5}{{K_3}^2{K_4}{K_5}^2\vph}
		}\mem
		+ (\text{cyc.})_3
		\,,\\
		\,&=\,
		16\,
		\bb{
			( f_3 ( k_4 \mtensor f_4 )f_5 )
			\mem
			\bb{
				\frac{1}{{K_3}^2{K_5}^2\vph}
				+ \frac{K_3 {\mem-\,} K_5}{{K_3}^2K_4{K_5}^2\vph}
			}
		}\mem
		+ (\text{perm.})_6
		\,,\\
		\,&=\,
		-32\,
		\bb{
			( f_3 ( k_4 \mtensor f_4 )f_5 )
			\,
			\frac{1}{{K_3}^2 K_4 K_5\vph}
		}\mem
		+ (\text{perm.})_6
		\,,
\end{split}
\end{align}
where we have summed over
the three cyclic permutations (cyc.)
or the six permutations (perm.)
of $3,4,5$.

Massaging the above expression a bit more,
we find
\begin{align}
	\label{QED.A5}
	\A_5
	\,=\,
		-32\,
			\kf{35}\mem \ff{34}
	\,
	\frac{1}{K_3 K_4 {K_5}^2\vph}
	+ (\text{perm.})_6
	\,.
\end{align}
We have verified that \eqref{QED.A5}
is indeed equal to
the classical limit of the $3$-Compton amplitude
computed in the second-quantized framework.
To see this quickly,
adopt an axial gauge
with respect to the second-leg massive momentum $p_2$,
which configures the polarization vectors as
\begin{align}
	\label{axial-p2}
	e_I{}_\a
	\mem=\mem
		-\frac{2}{K_I}\,
			\varphi_I{}_{\a\b}\mem p_2^\b
	\quad\iff\quad
	k_I \mdot e_I \mem=\mem 0
	\,,\quad
	p_2 \mdot e_I \mem=\mem 0
	\,,\quad
	(k_I \mwedge e_I)_{\a\b}
	\mem=\mem
		\varphi_I{}_{\a\b}
	\,.
\end{align}
This immediately eliminates all ladder diagrams,
while also imposing that
the second leg must connect to a quartic vertex.
Consider the planar diagram
where the second leg meets photons $3$ and $4$
while the first leg meets photon $5$.
Applying the Feynman rules of scalar quantum electrodynamics, 
one finds
\begin{align}
	\label{2ndQED.A5diagram}
	\BB{
		-2\mem e_3 \mdot e_4
	}
	\,
	\frac{1}{
		X_2 {\,-\,} \hbar(K_3{\mem+\mem}K_4)
		+\O(\hbar^2)
	}
	\,
	\BB{
		2\mem e_5 \mdot (p_2{\mem-\mem}\hbar k_3{\mem-\mem}\hbar k_4)
	}
	\,,
\end{align}
whose $\hbar\to0$ limit
is nicely finite
as $X_2 = {p_2}^2 {\mem+\mem} m^2$ and $e_5 \mdot p_2$ are put to zero.
Straightforward algebra then equates
\eqref{2ndQED.A5diagram}
with
\begin{align}
	\label{2ndQED.A5diagram.eval}
	-32\,
	\ff{34}\,
	\frac{1}{K_3 K_4 {K_5}^2\vph}
	\,
	\BB{
		\kf{35}
		+
		\kf{45}
	}\hem
\,,
\end{align}
by plugging in \eqref{axial-p2}.
Considering the other two diagrams
achieves \eqref{QED.A5}.

Note that the process of handling the integral factors
in \eqrefs{QED.I5r}{QED.I5p}
may be expedited as the following,
which presumes $K_3 \mplus K_4 \mplus K_5 = 0$
from the beginning:
\begin{subequations}
	\begin{align}
		\den^\text{(Ir)}_{345}
		\,&=\,
		\frac{1}{\deltabar(0)}
		\int d\s\,\,
		\mathe^{iK_3\s}\,
		\frac{1}{-{\partial_\s}^2}\,
		\mathe^{iK_4\s}\,
		\frac{1}{-{\partial_\s}^2}\,
		\mathe^{iK_5\s}
		\,=\,
		\frac{1}{{(K_4 \mplus K_5)}^2\vph}
		\frac{1}{{K_5}^2\vph}
		\,,\\
		\den^\text{(Ip-a)}_{3456}
		\,&=\,
		\frac{1}{\deltabar(0)}
		\int d\s\,\,
		\mathe^{iK_3\s}\,
		\frac{i}{{\partial_\s}}\,
		\mathe^{iK_4\s}\,
		\frac{1}{-{\partial_\s}^2}\,
		\mathe^{iK_5\s}
		\,=\,
		\frac{1}{{K_4 \mplus K_5}\vph}
		\frac{1}{{K_5}^2\vph}
		\,.
	\end{align}
\end{subequations}
This simplified approach
evaluates the full line integral to
directly extract the part that contributes to the amplitude,
automating the LSZ reduction.

\subsubsection{Multiplicity 4}
\label{Compton.EM4}

Finally, we challenge 
one order higher.
The relevant diagrams now come in
five classes
falling in two topologies.
For what will be called the I-topology,
the diagrams are
\begin{align}
\begin{split}
	\label{QED.F6.Itop}
	\F_{3456}^\text{(Ir)}
	\,\,\,=\,\,\,
	\adjustbox{valign=c,raise=-1.07em}{\begin{tikzpicture}[]
			\node (o) at (0,0) {};
			\node (xshift) at (0.8,0) {};
			\node (yshift) at (0,-0.9) {};
			\node (i3) at ($(o)$) {};
			\node (i4) at ($(o)+1*(xshift)$) {};
			\node (i5) at ($(o)+2*(xshift)$) {};
			\node (i6) at ($(o)+3*(xshift)$) {};
			\node[cdot] (v3) at ($(i3)$) {};
			\node[cdot] (v4) at ($(i4)$) {};
			\node[cdot] (v5) at ($(i5)$) {};
			\node[cdot] (v6) at ($(i6)$) {};
			\node[T] (A3) at ($(i3)+(yshift)$) {$3$};
			\node[T] (A4) at ($(i4)+(yshift)$) {$4$};
			\node[T] (A5) at ($(i5)+(yshift)$) {$5$};
			\node[T] (A6) at ($(i6)+(yshift)$) {$6$};
			\node[t] (x) at ($(i2)$) {};
			\node (p) at ($(v4)+(135:0.22)$) {\vpin};
			\draw[wiggly] (i3)--(A3);
			\draw[wiggly] (i4)--(A4);
			\draw[wiggly] (i5)--(A5);
			\draw[wiggly] (i6)--(A6);
			\draw[linear] (i3)--(i4)--(i5)--(i6);
	\end{tikzpicture}}
	\quad
	&\,,\quad
	\F_{3456}^\text{(Ipp-a)}
	\,\,\,=\,\,\,
	\adjustbox{valign=c,raise=-1.07em}{\begin{tikzpicture}[]
			\node (o) at (0,0) {};
			\node (xshift) at (0.8,0) {};
			\node (yshift) at (0,-0.9) {};
			\node (i3) at ($(o)$) {};
			\node (i4) at ($(o)+1*(xshift)$) {};
			\node (i5) at ($(o)+2*(xshift)$) {};
			\node (i6) at ($(o)+3*(xshift)$) {};
			\node[cdot] (v3) at ($(i3)$) {};
			\node[cdot] (v4) at ($(i4)$) {};
			\node[cdot] (v5) at ($(i5)$) {};
			\node[cdot] (v6) at ($(i6)$) {};
			\node[T] (A3) at ($(i3)+(yshift)$) {$3$};
			\node[T] (A4) at ($(i4)+(yshift)$) {$4$};
			\node[T] (A5) at ($(i5)+(yshift)$) {$5$};
			\node[T] (A6) at ($(i6)+(yshift)$) {$6$};
			\node[t] (x) at ($(i2)$) {};
			\node (p) at ($(v4)+(135:0.22)$) {\pin};
			\node (p) at ($(v5)+(135:0.22)$) {\pin};
			\draw[wiggly] (i3)--(A3);
			\draw[wiggly] (i4)--(A4);
			\draw[wiggly] (i5)--(A5);
			\draw[wiggly] (i6)--(A6);
			\draw[linear] (i3)--(i4)--(i5)--(i6);
	\end{tikzpicture}}
	\quad
	\,,\\
	\F_{3456}^\text{(Ip-a)}
	\,\,\,=\,\,\,
	\adjustbox{valign=c,raise=-1.07em}{\begin{tikzpicture}[]
			\node (o) at (0,0) {};
			\node (xshift) at (0.8,0) {};
			\node (yshift) at (0,-0.9) {};
			\node (i3) at ($(o)$) {};
			\node (i4) at ($(o)+1*(xshift)$) {};
			\node (i5) at ($(o)+2*(xshift)$) {};
			\node (i6) at ($(o)+3*(xshift)$) {};
			\node[cdot] (v3) at ($(i3)$) {};
			\node[cdot] (v4) at ($(i4)$) {};
			\node[cdot] (v5) at ($(i5)$) {};
			\node[cdot] (v6) at ($(i6)$) {};
			\node[T] (A3) at ($(i3)+(yshift)$) {$3$};
			\node[T] (A4) at ($(i4)+(yshift)$) {$4$};
			\node[T] (A5) at ($(i5)+(yshift)$) {$5$};
			\node[T] (A6) at ($(i6)+(yshift)$) {$6$};
			\node[t] (x) at ($(i2)$) {};
			\node (p) at ($(v4)+(135:0.22)$) {\pin};
			\draw[wiggly] (i3)--(A3);
			\draw[wiggly] (i4)--(A4);
			\draw[wiggly] (i5)--(A5);
			\draw[wiggly] (i6)--(A6);
			\draw[linear] (i3)--(i4)--(i5)--(i6);
	\end{tikzpicture}}
	\quad
	&\,,\quad
	\F_{3456}^\text{(Ipp-b)}
	\,\,\,=\,\,\,
	\adjustbox{valign=c,raise=-1.07em}{\begin{tikzpicture}[]
			\node (o) at (0,0) {};
			\node (xshift) at (0.8,0) {};
			\node (yshift) at (0,-0.9) {};
			\node (i3) at ($(o)$) {};
			\node (i4) at ($(o)+1*(xshift)$) {};
			\node (i5) at ($(o)+2*(xshift)$) {};
			\node (i6) at ($(o)+3*(xshift)$) {};
			\node[cdot] (v3) at ($(i3)$) {};
			\node[cdot] (v4) at ($(i4)$) {};
			\node[cdot] (v5) at ($(i5)$) {};
			\node[cdot] (v6) at ($(i6)$) {};
			\node[T] (A3) at ($(i3)+(yshift)$) {$3$};
			\node[T] (A4) at ($(i4)+(yshift)$) {$4$};
			\node[T] (A5) at ($(i5)+(yshift)$) {$5$};
			\node[T] (A6) at ($(i6)+(yshift)$) {$6$};
			\node[t] (x) at ($(i2)$) {};
			\node (p) at ($(v4)+(135:0.22)$) {\pin};
			\node (p) at ($(v5)+(45:0.22)$) {\pin};
			\draw[wiggly] (i3)--(A3);
			\draw[wiggly] (i4)--(A4);
			\draw[wiggly] (i5)--(A5);
			\draw[wiggly] (i6)--(A6);
			\draw[linear] (i3)--(i4)--(i5)--(i6);
	\end{tikzpicture}}
	\quad
	\,,\\
	\F_{3456}^\text{(Ip-b)}
	\,\,\,=\,\,\,
	\adjustbox{valign=c,raise=-1.07em}{\begin{tikzpicture}[]
			\node (o) at (0,0) {};
			\node (xshift) at (0.8,0) {};
			\node (yshift) at (0,-0.9) {};
			\node (i3) at ($(o)$) {};
			\node (i4) at ($(o)+1*(xshift)$) {};
			\node (i5) at ($(o)+2*(xshift)$) {};
			\node (i6) at ($(o)+3*(xshift)$) {};
			\node[cdot] (v3) at ($(i3)$) {};
			\node[cdot] (v4) at ($(i4)$) {};
			\node[cdot] (v5) at ($(i5)$) {};
			\node[cdot] (v6) at ($(i6)$) {};
			\node[T] (A3) at ($(i3)+(yshift)$) {$3$};
			\node[T] (A4) at ($(i4)+(yshift)$) {$4$};
			\node[T] (A5) at ($(i5)+(yshift)$) {$5$};
			\node[T] (A6) at ($(i6)+(yshift)$) {$6$};
			\node[t] (x) at ($(i2)$) {};
			\node (p) at ($(v5)+(135:0.22)$) {\pin};
			\draw[wiggly] (i3)--(A3);
			\draw[wiggly] (i4)--(A4);
			\draw[wiggly] (i5)--(A5);
			\draw[wiggly] (i6)--(A6);
			\draw[linear] (i3)--(i4)--(i5)--(i6);
	\end{tikzpicture}}
	\quad
	&\,,\quad
	\F_{3456}^\text{(Ipp-c)}
	\,\,\,=\,\,\,
	\adjustbox{valign=c,raise=-1.07em}{\begin{tikzpicture}[]
			\node (o) at (0,0) {};
			\node (xshift) at (0.8,0) {};
			\node (yshift) at (0,-0.9) {};
			\node (i3) at ($(o)$) {};
			\node (i4) at ($(o)+1*(xshift)$) {};
			\node (i5) at ($(o)+2*(xshift)$) {};
			\node (i6) at ($(o)+3*(xshift)$) {};
			\node[cdot] (v3) at ($(i3)$) {};
			\node[cdot] (v4) at ($(i4)$) {};
			\node[cdot] (v5) at ($(i5)$) {};
			\node[cdot] (v6) at ($(i6)$) {};
			\node[T] (A3) at ($(i3)+(yshift)$) {$3$};
			\node[T] (A4) at ($(i4)+(yshift)$) {$4$};
			\node[T] (A5) at ($(i5)+(yshift)$) {$5$};
			\node[T] (A6) at ($(i6)+(yshift)$) {$6$};
			\node[t] (x) at ($(i2)$) {};
			\node (p) at ($(v4)+(45:0.22)$) {\pin};
			\node (p) at ($(v5)+(135:0.22)$) {\pin};
			\draw[wiggly] (i3)--(A3);
			\draw[wiggly] (i4)--(A4);
			\draw[wiggly] (i5)--(A5);
			\draw[wiggly] (i6)--(A6);
			\draw[linear] (i3)--(i4)--(i5)--(i6);
	\end{tikzpicture}}
	\quad
	\,.
\end{split}
\end{align}
For what will be called the Y-topology,
the diagrams are
\begin{align}
	\label{QED.F6.Ytop}
	\F_{3456}^\text{(Yr)}
	\,\,\,&=\,\,\,
	\adjustbox{valign=c,raise=-1.07em}{\begin{tikzpicture}[]
			\node (o) at (0,0) {};
			\node (xshift) at (0.8,0) {};
			\node (yshift) at (0,-0.9) {};
			\node (i3) at ($(o)$) {};
			\node (i4) at ($(o)+1*(xshift)$) {};
			\node (i5) at ($(i4)+(0.8,0.6)$) {};
			\node (i6) at ($(i4)+(0.8,-0.6)$) {};
			\node[cdot] (v3) at ($(i3)$) {};
			\node[cdot] (v4) at ($(i4)$) {};
			\node[cdot] (v5) at ($(i5)$) {};
			\node[cdot] (v6) at ($(i6)$) {};
			\node[T] (A3) at ($(i3)+(yshift)$) {$3$};
			\node[T] (A4) at ($(i4)+(yshift)$) {$4$};
			\node[T] (A5) at ($(i5)+(yshift)$) {$5$};
			\node[T] (A6) at ($(i6)+(yshift)$) {$6$};
			\node[t] (x) at ($(i2)$) {};
			\draw[wiggly] (i3)--(A3);
			\draw[wiggly] (i4)--(A4);
			\draw[wiggly] (i5)--(A5);
			\draw[wiggly] (i6)--(A6);
			\draw[linear] (i3)--(i4);
			\draw[linear] (i4)--(i5);
			\draw[linear] (i4)--(i6);
	\end{tikzpicture}}
	\quad
	\,,\quad
	\F_{3456}^\text{(Yp)}
	\,\,\,=\,\,\,
	\adjustbox{valign=c,raise=-1.07em}{\begin{tikzpicture}[]
			\node (o) at (0,0) {};
			\node (xshift) at (0.8,0) {};
			\node (yshift) at (0,-0.9) {};
			\node (i3) at ($(o)$) {};
			\node (i4) at ($(o)+1*(xshift)$) {};
			\node (i5) at ($(i4)+(0.8,0.6)$) {};
			\node (i6) at ($(i4)+(0.8,-0.6)$) {};
			\node[cdot] (v3) at ($(i3)$) {};
			\node[cdot] (v4) at ($(i4)$) {};
			\node[cdot] (v5) at ($(i5)$) {};
			\node[cdot] (v6) at ($(i6)$) {};
			\node[T] (A3) at ($(i3)+(yshift)$) {$3$};
			\node[T] (A4) at ($(i4)+(yshift)$) {$4$};
			\node[T] (A5) at ($(i5)+(yshift)$) {$5$};
			\node[T] (A6) at ($(i6)+(yshift)$) {$6$};
			\node[t] (x) at ($(i2)$) {};
			\node (p) at ($(v4)+(135:0.22)$) {\pin};
			\draw[wiggly] (i3)--(A3);
			\draw[wiggly] (i4)--(A4);
			\draw[wiggly] (i5)--(A5);
			\draw[wiggly] (i6)--(A6);
			\draw[linear] (i3)--(i4);
			\draw[linear] (i4)--(i5);
			\draw[linear] (i4)--(i6);
	\end{tikzpicture}}
	\quad
	\,.
\end{align}

We begin with the I-topology.
The tensor factors are
\begin{subequations}
\begin{align}
	\ten^\text{(Ir)}_{3456}
	\,&=\,
	+32\,
	(f_3
		\hem( k_4 \modot f_4 )\hem( k_5 \modot f_5 )\hem
	f_6)
	\,,\\[0.15\baselineskip]
	\ten^\text{(Ip-a)}_{3456}
	\,&=\,
	-32\,
		(f_3
			\hem( k_4 \mwedge f_4 )\hem( k_5 \modot f_5 )\hem
		f_6)
		\,\frac{1}{K_4}
	\,,\\
	\ten^\text{(Ip-b)}_{3456}
	\,&=\,
	-32\,
		(f_3
			\hem( k_4 \modot f_4 )\hem( k_5 \mwedge f_5 )\hem
		\hem f_6)
		\,\frac{1}{K_5}
	\,,\\
\begin{split}
	\ten^\text{(Ipp-a)}_{3456}
	\,&=\,
	+32\,
		(f_3
		\hem( k_4 \mwedge f_4 )\hem( k_5 \mwedge f_5 )\hem
		\hem f_6)
		\,\frac{1}{K_4K_5}
	\,=\,
		-\ten^\text{(Ipp-b)}_{3456}
	\,=\,
		-\ten^\text{(Ipp-c)}_{3456}
	\,.
\end{split}
\end{align}
\end{subequations}
The integral factors 
after the LSZ reduction
are
\begin{subequations}
\begin{align}
		\den^\text{(Ir)}_{3456}
		\,&=\,
			\frac{1}{{(K_4 \mplus K_5 \mplus K_6)}^2\vph}
			\frac{1}{{(K_5 \mplus K_6)}^2\vph}
			\frac{1}{{K_6}^2\vph}
		\,,\\
		\den^\text{(Ip-a)}_{3456}
		\,&=\,
			\frac{1}{{K_4 \mplus K_5 \mplus K_6}\vph}
			\frac{1}{{(K_5 \mplus K_6)}^2\vph}
			\frac{1}{{K_6}^2\vph}
		\,,\\
		\den^\text{(Ip-b)}_{3456}
		\,&=\,
			\frac{1}{{(K_4 \mplus K_5 \mplus K_6)^2}\vph}
			\frac{1}{{K_5 \mplus K_6}\vph}
			\frac{1}{{K_6}^2\vph}
		\,,\\
		\den^\text{(Ipp-a)}_{3456}
		\,&=\,
			\frac{1}{{K_4 \mplus K_5 \mplus K_6}\vph}
			\frac{1}{{K_5 \mplus K_6}\vph}
			\frac{1}{{K_6}^2\vph}
		\,,\\
		\den^\text{(Ipp-b)}_{3456}
		\,&=\,
			\frac{1}{{K_4 \mplus K_5 \mplus K_6}\vph}
			\frac{1}{{(K_5 \mplus K_6)^2}\vph}
			\frac{1}{{K_3 \mplus K_4 \mplus K_5}\vph}
			\,,\\
		\den^\text{(Ipp-c)}_{3456}
		\,&=\,
			\frac{1}{{(K_4 \mplus K_5 \mplus K_6)^2}\vph}
			(K_3 \mplus K_4)(K_5 \mplus K_6)
			\frac{1}{{(K_5 \mplus K_6)^2}\vph}
			\frac{1}{{K_6}^2\vph}
		\,,
\end{align}
\end{subequations}
which can be directly identified from the diagrams in \eqref{QED.F6.Itop},
in fact.

\newpage

Employing a similar algebra as in \eqref{A5.sf},
the part of the amplitude
due to
the I-topology diagrams is found 
from symmetry factor considerations
as
\begin{align}
	\label{A6I}
	\A_6^\text{(I)}
	\,&=\,
	\bb{
		\frac{1}{2}\,
		\A_{3456}^\text{(Ir)}
		+
		\A_{3456}^\text{(Ip-a)}
		+
		\A_{3456}^\text{(Ip-b)}
		+
		\A_{3456}^\text{(Ipp-a)}
		+
		\frac{1}{2}\,
		\A_{3456}^\text{(Ipp-b)}
		+
		\frac{1}{2}\,
		\A_{3456}^\text{(Ipp-c)}
	}
	+ (\text{perm.})_{24}
	\nonumber
	\,,\\
	&\,=\,
\begin{aligned}[t]
	&
		\frac{128}{\K{3}\K{4}\K{5}\K{6}}\,
		\bb{
			\frac{
				\ff{34}\mem\kf{45}\mem\kf{56}
			}{
				(\K{3}\mplus\K{4})\mem\K{6}
			}
			-
			\frac{1}{2}\hem
			\frac{
				\kf{43}\mem\ff{45}\mem\kf{56}
			}{
				\K{3}\mem\K{6}
			}
			-
			\frac{1}{2}\hem
			\frac{
				\ff{34}\mem\kk{45}\mem\ff{56}
			}{
				(\K{3}\mplus\K{4})
				(\K{5}\mplus\K{6})
			}
		}
	\\
	& + (\text{perm.})_{24}
	\,,
\end{aligned}
\end{align}
where
$\A_{3456}^\text{(Ir)} = \ten_{3456}^\text{(Ir)}\mem \den_{3456}^\text{(Ir)}$,
etc.

For the Y-topology, 
on the other hand,
the tensor factors are
\begin{subequations}
	\begin{align}
		\label{ten6-Yr}
		\ten_{3456}^\text{(Yr)}
		\,&=\,
		128\mem \ff{34}\mem\kf{45}\mem\kf{46}
		\,,\\[0.15\baselineskip]
		\label{ten6-Yp}
		\ten_{3456}^\text{(Yp)}
		\,&=\,
		-\frac{128}{3}\,\BB{
			\nem
			\BB{
				\ff{45}\mem\kf{43}\mem\kf{46}
				-
				\ff{43}\mem\kf{45}\mem\kf{46}
			}
			+ (5\mlra6)
		}
		\,\frac{1}{K_4}
		\,.
	\end{align}
\end{subequations}
The integral factors are
\begin{align}
			\den^\text{(Yr)}_{3456}
			\,=\,
			\frac{1}{{(K_4 \mplus K_5 \mplus K_6)}^2\vph}
			\frac{1}{{K_5}^2\vph}
			\frac{1}{{K_6}^2\vph}
			\,,\quad
			\den^\text{(Yp)}_{3456}
			\,&=\,
			\frac{1}{{K_4 \mplus K_5 \mplus K_6}\vph}
			\frac{1}{{K_5}^2\vph}
			\frac{1}{{K_6}^2\vph}
			\,.
\end{align}
From symmetry factor considerations,
the part of the amplitude 
due to the Y-topology diagrams 
is found as
\begin{align}
\begin{split}
	\label{A6Y}
	\A_6^\text{(Y)}
	\,&=\,
	\bb{
		\frac{1}{6}\,
		\A_{3456}^\text{(Yr)}
		+
		\frac{1}{2}\,
		\A_{3456}^\text{(Yp)}
	}
	+ (\text{perm.})_{24}
	\,=\,
	- \frac{64\mem
		\ff{34}\mem\kf{35}\mem\kf{36}
	}{
		\K{3}\K{4}\K{5}^2\K{6}^2
	}
	+ (\text{perm.})_{24}
	\,.
\end{split}
\end{align}

Finally,
the total amplitude 
is the sum of \eqrefs{A6I}{A6Y}:
\begin{align}
	\label{A6}
	\A_6
	\,=\,
	\A_6^\text{(I)}
	+
	\A_6^\text{(Y)}
	\,.
\end{align}
By taking the same approach as in \Sec{Compton.EM3},
one can readily check that
$\A_6$ in \eqref{A6}
indeed matches with
the classical limit of the $4$-Compton amplitude 
computed in the second-quantized framework.
Note that 
the on-shell condition for the massive leg $1$
imposes
not only
$\K{3}\mplus\K{4}\mplus\K{5}\mplus\K{6} = 0$
at $\O(\hbar^1)$
but 
also
a six-term identity at $\O(\hbar^2)$:
\begin{align}
	\label{6kk=0}
	\kk{34}
	+
	\kk{35}
	+
	\kk{36}
	+
	\kk{45}
	+
	\kk{46}
	+
	\kk{56}
	\,=\,
	0
	\,.
\end{align}
\eqref{6kk=0} can be used for 
an alternative representation of
$(-\kk{45}/(\K{3}\mplus\K{4})(\K{5}\mplus\K{6})) + (\text{perm.})_{24}$
as
$(\kk{34}/2(\K{3}\mplus\K{4})(\K{5}\mplus\K{6})) + (\text{perm.})_{24}$.

This concludes our 
test of the phase space worldline formalism
in electrodynamics.
We have shown that the formalism
manifests gauge invariance
whenever two or more photons are involved in the process.
We have also verified that
the results agree with the computations in the second-quantized formalism.

Throughout, we have 
always used the identity in \eqref{magphi}
to break down the tensor factors
into products of
the scalar invariants in \eqref{kf-invariants}.
This
helped the comparison with the second-quantized formalism,
for instance.
However, note that one may also choose to not use the identity in \eqref{magphi},
in which case 
one will encounter longer sequences of the linearized field strengths
such as
$p\hem \varphi_3\varphi_4\varphi_5p$.
Finally, we remark again that our formalism
is capable of computing quantum contributions,
although we have only considered the classical limit here
for simplicity.

\section{Application to Yang-Mills and Gravity}
\label{NA}

Finally,
we would like to offer a glimpse
into nonabelian and gravitational interactions.
In this section, we compute
the Compton amplitudes of multiplicity two
by scattering a scalar particle in
the backgrounds of nonlinearly superposed plane waves
in Yang-Mills theory and general relativity.
We will observe how
the phase space formalism
draws a parallel between gauge theory and gravity,
based on the universality of the Feynman rules in \Sec{General.feyn}.

\subsection{Phase Space}
\label{NA.ps}

We begin with gauge theory.
The equations of motion of a particle in an external nonabelian gauge field
trace back to Wong \cite{wong1970field}.
The corresponding Hamiltonian formulation is viable
with various choices for the microscopic implementation of the particle's nonabelian charge.

In this paper,
we adopt a purely bosonic implementation, for simplicity.
Suppose the gauge Lie algebra $\g = \su(N)$
with structure constants $f^a{}_{bc}$.
The phase space of the particle can be concretely given as
$\P = T^*\mathbb{R}^d \mtimes T^*\g$,
with coordinates
$x^\a$, $p_\a$, $\phi^a$, and $\tphi_a$.
In the free theory,
the symplectic structure is given as
\begin{align}	
\begin{split}
	\label{colored}
	\theta^\circ \,&=\,
		p_\a\mem dx^\a
		+ \tphi_a\mem d\phi^a
	\,,\\
	\omega^\circ \,&=\,
		dp_\a \wedge dx^\a
		+ d\tphi_a \wedge d\phi^a
	\,,
\end{split}
\end{align}
featuring canonical Poisson brackets
$\{ x^\a , p_\b \}^\circ = \delta^\a{}_\b$,
$\{ \phi^a , \tphi_b \}^\circ = \delta^a{}_b$.
The nonabelian charge is implemented as a composite variable,
\begin{align}
	\label{qq}
	q_a \,:=\, \tphi_b\, f^b{}_{ac}\, \phi^c
	\qiq
	\{{q_a},{q_b}\}^\circ
	\,=\,
		q_c\, f^c{}_{ab}
	\,,
\end{align}
whose Hamiltonian vector field defines an $\mathrm{SU}(N)$ group action
in the phase space.

As is well-known,
the particle is minimally coupled to a nonabelian gauge connection $A^a = A^a{}_\a(x)\mem dx^\a$
by covariantizing the symplectic potential in \eqref{colored}:
\begin{align}
	\theta \,=\,
		p_\a\mem dx^\a
		+ \tphi_a\mem D\phi^a
	\qquad\text{where}\qquad
	D\phi^a \,=\,
		d\phi^a 
		+ f^a{}_{cb}\mem A^c\mem \phi^b
	\,.
\end{align}
Here, $D$ denotes the covariant exterior derivative
such that $D\phi^a = d\phi^a + f^a{}_{cb}\mem A^c\mem \phi^b$.
The symplectic form then follows as $\omega = d\theta$.
As a result, the modification on the symplectic structure is given as
\begin{align}
\begin{split}
	\label{YM.sp}
	\theta'
	\,&=\,
		q_a\mem A^a
	\,,\quad
	\omega'
	\,=\,
		dq_a \wedge A^a
		+ \frac{1}{2}\, q_a\mem dA^a
	\,,
\end{split}
\end{align}
while the Hamiltonian is left as the same.

Let us now move on to gravitational coupling.
In the conventional description,
a scalar particle
is minimally coupled to gravity 
via the metric field.
In the Hamiltonian description,
this means to have
\begin{align}
	\label{Grav.HPT}
	\theta
	\,=\,
		P_\m\mem dx^\m
	\,,\quad
	\omega
	\,=\,
		dP_\m \swedge dx^\m
	\,,\quad
	H \,=\,
	\frac{1}{2}\mem 
		(g^{-1}\hnem(x)\hnem)^{\m\n}\,
		P_\m\hem P_\n
	\,,
\end{align}
where $(g^{-1}\hnem(x)\hnem)^{\m\n}$ denotes the inverse metric.

\newpage

Evidently, \eqref{Grav.HPT} describes a Hamiltonian perturbation theory,
paralleling 
\eqref{feyn-ex.HPT}.
To achieve a symplectic perturbation theory, we perform a worldline field redefinition
in terms of an orthonormal frame $e^\a{}_\m(x)$:
\begin{align}
	\label{Grav.kinkan}
	P_\m
	\,=\,
		p_\a\mem e^\a{}_\m(x)
	\qquad\text{where}\qquad
		\eta_{\a\b}\mem e^\a{}_\m(x)\mem e^\b{}_\n(x)
		\,=\,
			g_{\m\n}(x)
	\,.
\end{align}
Interestingly,
\eqref{Grav.kinkan} 
identifies the orthonormal-frame momentum $p_\a$ as the kinetic momentum
and establishes its relationship with the canonical momentum $P_\m$,
in parallel with \eqref{kinkan}.
As a result, the Hamiltonian is brought back to its free theory form,
while the symplectic structure is modified:
\begin{align}
	\label{Grav.SPT}
	\theta
	\,=\,
		p_\a\mem e^\a
	\,,\quad
	\omega
	\,=\,
		dp_\a\mem e^\a
		+ p_\a\mem de^\a
	\,,\quad
	H \,=\,
	\frac{1}{2}\mem 
		\eta^{\a\b}\mem p_\a\mem p_\b
	\,=\,
	\frac{1}{2}\mem
		p^2
	\,.
\end{align}
In particular,
when the tetrad one-form is expanded around a flat background as
$e^\a = e^\a{}_\m(x)\mem dx^\m = \delta^\a{}_\m\mem dx^\m + h^\a{}_\m(x)\mem dx^\m$,
the symplectic perturbations are
\begin{align}
	\label{Grav.sp}
	\theta' \,=\,
		p_\a\mem h^\a
	\,,\quad
	\omega' \,=\,
		dp_\a \wedge h^\a
		+ \frac{1}{2}\mem p_\a\mem dh^\a
	\,.
\end{align}

Amusingly,
\eqref{Grav.sp}
is isomorphic to
\eqref{YM.sp}
via the following correspondences:
\begin{align}
\begin{split}
	\label{isom.YM-Grav}
	q_a
	\quad\text{(color charge)}
	&\qquad\xleftrightarrow{\,}\qquad
	p_\a
	\quad\text{(kinetic momentum)}
	\,,\\
	A^a
	\quad\text{(gauge field)}
	&\qquad\xleftrightarrow{\,}\qquad
	h^\a
	\quad\text{(tetrad perturbation)}
	\,.
\end{split}
\end{align}
Therefore, we identify an isomorphism at the level of particle symplectic structure.
Note that both \eqref{YM.sp} and \eqref{Grav.sp}
achieves cubic coupling on the worldline
via symplectic perturbation theory.

Meanwhile, note that \eqref{YM.sp} exhibits a crucial difference
when compared to \eqref{qed-cheatsheet}.
Namely, the symplectic form $\omega$ is not gauge-invariantly split into free and interacting parts.
A gauge-invariant split would instead be
\begin{align}
	\label{YM.covsplit}
	\omega
	\,=\,
		\BB{
			Dp_\a \swedge dx^\a
			+ D\tphi_a \swedge D\phi^a
		} + \BB{
			q_a\mem F^a
		}
	\,,
\end{align}
where
$F^a = dA^a + \frac{1}{2}\, f^a{}_{bc}\mem A^b \mwedge A^c$
is the field strength two-form.
This means that,
although we have employed the kinetic momentum $p_\a$
that coincides with the physical velocity $\dot{x}_\a$ on equations of motion
in \eqref{YM.sp},
the perturbation theory will not manifest gauge covariance in general.
The issue is whether to expand around the ``flat'' background
in \eqref{colored}
or the covariantized background in \eqref{YM.covsplit}.
In this paper,
we stick to the former approach.
A more satisfactory treatment following the latter
shall be presented in a future work.

For the gravitational coupling,
we remark that a gauge-invariant split of the symplectic form in \eqref{Grav.SPT}
is viable by employing 
any affine connection:
\begin{align}
	\label{Grav.covsplit}
	\omega
	\,=\,
		\BB{
			Dp_\a \swedge e^\a
		} + \BB{
			p_\a\mem T^\a
		}
	\,.
\end{align}
This parallels \eqref{YM.covsplit},
where
$T^\a = De^\a$
is the torsion two-form.
Amusingly, \eqref{Grav.covsplit}
identifies that the torsion $T^\a$ is the ``gravitational field strength,''
given the identification of the momentum $p_\a$ as the ``gravitational charge'' due to \eqref{isom.YM-Grav}.\footnote{
	In fact,
	it is not difficult to see that
	this correspondence can be physically understood in the gravitoelectromagnetism
	\cite{wald2010general}
	sense.
	In particular, the torsion encodes the difference between the auxiliary connection $D$ and the Levi-Civita connection,
	which is a tensor.
}

\subsection{Worldline Formalism}
\label{NA.wlf}

A key lesson to be learned from \Secs{General}{QED}
is that 
the symplectic geometry of the particle
in the backgrounds of superposed plane waves
encodes all the data
for deriving the Compton amplitudes.
Meanwhile,
in \Sec{NA.ps},
we have identified
an isomorphism between 
gauge theory and gravity couplings
at the level of
symplectic structures.
Thus,
we might
expect that
this structural similarity will be inherited down to the Compton amplitudes.

With this anticipation,
we mechanically implement
our phase space formalism established in \Sec{QED}
for the nonabelian interactions.

First, we identify the saddle trajectory for the particle.
Explicitly, we take
\begin{align}
\begin{split}
	\label{eq:sad.xpcolor}
		x^\a(\s)
		\,&=\,
		\bmx^\a(\s) + \dx^\a(\s)
		\,=\,
		x_1^\a + 2p_2^\a\mem \s + \dx^\a(\s)
	\,,\\
		p_\a(\s)
		\,&=\,
		\bmp_\a(\s) + \ddp_\a(\s)
		\hem\hhem\hhem
		\,=\,
		p_2{}_\a + \ddp_\a(\s)
	\,,\\
		\phi^a(\s)
		\,&=\,
		\bm{\phi}^a(\s) + \dphi^a(\s)
		\hem\hhem\hhem
		\,=\,
		\phi_1^\a + \dphi^a(\s)
	\,,\\
		\smash{\tphi}_a(\s)
		\,&=\,
		\smash{\bm{\tphi}}_a(\s) + \smash{\dtphi}_a(\s)
		\hem\hhem\hhem
		\,=\,
		\smash{\tphi}_2{}_\a + \smash{\dtphi}_a(\s)
	\,,
\end{split}
\end{align}
which will compute the transition amplitude
from a definite-$\phi$ state to a definite-$\tphi$ state.
In the quantum computation, one would peel off factors of $\phi^a_1$ and $\tphi_2{}_a$
to examine the result in each tensor representation;
see \cite{Corradini:2016czo,Edwards:2017nvs,Bastianelli:2015iba,Ahmadiniaz:2015xoa}, for instance.
However, since our goal here is to study the classical limit,
we can simply take $\phi_1^a$ and $\tphi_2{}_a$
as classical parameters
defining a classical nonabelian charge
$q_a = \tphi_2{}_b\mem f^b{}_{ac}\mem \phi^c_1$.

Next, we identify the background field configuration,
on which the particle scatters off.
We take
the nonlinear superposition of two plane waves
with classical values of wavenumbers.
In Yang-Mills theory,
for instance,
such a nonlinear superposition is given as
\begin{align}
	\label{A-nonlinear}
	A^a{}_\a(x)
	\,=\,
		g\, c_3^a\, e_3{}_\a\, \mathe^{ik_3x}
		+ g\, c_4^a\, e_4{}_\a\, \mathe^{ik_4x}
		+ \O(g^2)
	\,.
\end{align}
The $\O(g^1)$ part in \eqref{A-nonlinear}
is a simple sum of two plane waves.
The nonlinear parts from $\O(g^2)$
arise due to
interactions between them,
dictated by Yang-Mills equations of motion.
Together, they construct an on-shell configuration
up to each perturbative order.

As is well-known,
the nonlinear completion of a linearized solution
is perturbatively obtained by running Berends-Giele recursion \cite{BerendsGiele}.
To this end,
we adopt a formulation of Yang-Mills equations
put forward by Cheung and Mangan \cite{cheung2021cck},
together with an axial gauge condition
$A^a{}_\a(x)\mem \eta^\a= 0$
for a fixed reference vector $\eta^\a$.
Notably, this formulation 
features a covariant analog of
color-kinematics duality
based on the Lorentz algebra
in $d$ dimensions,
$\tg = \so(1,d-1)$,
by viewing the Yang-Mills field strength $F^a{}_{\a\b}(x)$ as a fundamental object.
As a result,
it naturally organizes the expressions
in terms of the bivector polarizations $\varphi_I{}_{\a\b}$.
Details are left to \App{CK}.

To be explicit,
the solution representing the nonlinear superposition of two plane waves in Yang-Mills theory
is found as
\begin{align}
\begin{split}
	\label{F34}
	F^{a}{}^{\a\b}(x)
	\,=\,
	\begin{aligned}[t]
		&
		ig\,
			c_3^a\, \varphi_3^{\a\b}\,
				\mathe^{ik_3x}
		+
		ig\,
			c_4^a\, \varphi_4^{\a\b}\,
				\mathe^{ik_4x}
		\\
		&{}
		+
		2g^2\,
			c_{34}^a\mem
			\bb{
				\varphi_{34}^{\a\b}
				- \varphi_3^{\a\b}\,
					\frac{
						\eta\hhem\varphi_4\hnem k_3
					}{
						k_4\eta
					}
				+ \varphi_4^{\a\b}\,
					\frac{
						\eta\hhem\varphi_3\hnem k_4
					}{
						k_3\eta
					}
			}\mem
				\frac{1}{t_{34}}\,
				\mathe^{i(k_3+k_4)x}
		+ \O(g^3)
		\,,
	\end{aligned}
\end{split}
\end{align}
where we have omitted contracted indices
such as 
$\eta\hhem\varphi_4\hnem k_3 = \eta^\a \varphi_4{}_{\a\b} k_3^\b$
and defined
\begin{align}
	c_{34}^a := f^a{}_{bc}\mem c_3^b\mem c_4^c
	\,,\quad
	(\varphi_{34}{})^a{}_\b := 
		(\varphi_3)^\a{}_\c\mem (\varphi_4)^\c{}_\b
		-
		(\varphi_4)^\a{}_\c\mem (\varphi_3)^\c{}_\b
	\,,\quad
	t_{34} := -2\mem k_3\mdot k_4
	\,.
\end{align}
Note how the bivector polarizations 
serve
as ``color'' factors
valued in the Lorentz algebra.

For gravity,
we take a similar approach 
by utilizing the Penrose wave equation
\cite{Penrose:1960eq,Ryan:1974nt}.
This allows us to treat
the Riemann curvature 
$R_{\a\b\c\d}(x) = R_\wrap{[\a\b][\c\d]}(x) = R_\wrap{[\c\d][\a\b]}(x)$
as a fundamental object,
viewed as a symmetric bi-adjoint field 
for the Lorentz algebra.
The details are left to \App{CK}.

\subsection{Feynman Rules}

To extract the Feynman rules,
we evaluate
the symplectic perturbations
in the plane-wave backgrounds.
For multiplicity two,
the relevant elements
are simply
the differential forms
$\theta'$ and $\omega'$,
which take the following form:
\begin{align}
\begin{split}
	\theta' 
	\,&=\,
		g\mem \theta'{}^3
		+ g\mem \theta'{}^4
		+ g^2\mem \theta'{}^{34}
		+ \O(g^3)
	\,,\\
	\omega' 
	\,&=\,
		g\mem \omega'{}^3
		+ g\mem \omega'{}^4
		+ \O(g^2)
	\,.
\end{split}
\end{align}
In particular,
one needs to evaluate $\omega'^3$, $\omega'^4$, and $\theta'^{34}$.

For the particle coupled to Yang-Mills theory,
we find from the ``Lorentz force law'' in \eqref{generic.int1}
that the valence-one vertex is
\begin{align}
	\label{YMcalc.v1}
	\dz^i\mem \omega'^3_{ij}(\bmz)\mem \dot{\bmz}^j
	\,=\,
	\bbsq{
		\delta q_a\,
		\bb{
			4\mem
				c_3^a\,
				\frac{
					\eta\hhem\varphi_3p
				}{
					\a
				}
		}
		+
		\delta x^\a\,
		\bb{
			2i\,
				qc_3\,
				(\varphi_3 p)_\a
		}
	}\,
		\mathe^{ik_3x}
		\mem\mathe^{iK_3\s}
	\,.
\end{align}
Again, $q_a$, $x^\a$, and $p_\a$ 
in this equation should really be
$\tphi_2{}_\b\mem f^b{}_{ac}\mem \phi_1^\a$,
$x_1^\a$,
and
$p_2{}_\a$.
With the understanding that we are specializing in the classical limit,
we have dropped the subscripts
to simplify the notation.
For the valence-zero vertex,
we find 
\begin{align}
	\label{YMcalc.v0}
	\theta'_i(\bmz)\mem \dot{\bmz}^i
	\,=\,
		\bb{
			-
			\frac{
				8i\mem qc_{34}
				\,
				p\hhem\varphi_3\varphi_4p
			}{K_3t_{34}}
		}
		\bb{
			1 + \frac{\d}{\a\b}
		}
	\,,
\end{align}
which exploits some algebra
regarding the Jacobi identities
$(k_3 \swedge \varphi_3)_{\c\a\b} = 0$
and
$(k_4 \swedge \varphi_4)_{\c\a\b} = 0$.
Here, we have employed several shorthand notations:
\begin{align}
	\label{abdparam}
	\a \,:=\, 2k_3\eta
	\,,\quad
	\b \,:=\, 2k_4\eta
	\,,\quad
	\d := 
		-
		\frac{
			2t_{34}
			(\eta\hhem\varphi_3p)
			(\eta\hhem\varphi_4p)
		}{
			p\hhem\varphi_3\varphi_4p
		}
	\,.
\end{align}
Diagrammatically,
\eqref{YMcalc.v1}
describes the following vertices:
\begin{subequations}
\begin{align}
	\label{YM.sv1(x)}
	\adjustbox{valign=c,raise=-1.2em}{\begin{tikzpicture}[]
			\node (o) at (0,0) {};
			\node (xshift) at (1.3,0) {};
			\node (yshift) at (0,-0.9) {};
			\node (i1) at ($(o)-(xshift)$) {};
			\node (i2) at ($(o)+(xshift)$) {};
			\node (i3) at ($(o)+(yshift)$) {};
			\node[cdot] (v) at ($(o)$) {};
			\node[T] (A) at ($(i3)$) {$3$};
			\node[t] (x) at ($(i2)$) {$\mathrlap{
					\delta x^\a\hnem(\s)
				}$};
			\draw[wiggly] (v)--(A);
			\draw[prop] (x)--(v);
	\end{tikzpicture}}
	\kern-0.125em\phantom{\small
		\delta x^\a\hnem(\s)
	}
	\,\,\,&=\,\,\,
	\BB{
		2i\mem qc_3\mem (\varphi_3p)_\a
	}
	\,
	\mathe^{i K_3\s}
	\,,\\
	\label{YM.sv1(phi)}
	\adjustbox{valign=c,raise=-1.2em}{\begin{tikzpicture}[]
			\node (o) at (0,0) {};
			\node (xshift) at (1.3,0) {};
			\node (yshift) at (0,-0.9) {};
			\node (i1) at ($(o)-(xshift)$) {};
			\node (i2) at ($(o)+(xshift)$) {};
			\node (i3) at ($(o)+(yshift)$) {};
			\node[cdot] (v) at ($(o)$) {};
			\node[T] (A) at ($(i3)$) {$3$};
			\node[t] (x) at ($(i2)$) {$\mathrlap{
					\delta \phi^a\hnem(\s)
				}$};
			\draw[wiggly] (v)--(A);
			\draw[dprop] (x)--(v);
	\end{tikzpicture}}
	\kern-0.125em\phantom{\small
		\delta \phi^a\hnem(\s)
	}
	\,\,\,&=\,\,\,
	\bb{
		4\mem \tphi_b\hem f^b{}_{ca}\mem c_3^c
		\,\frac{
			\eta\hhem\varphi_3p
		}{\a}
	}
	\,
	\mathe^{i K_3\s}
	\,,\\
	\label{YM.sv1(tphi)}
	\adjustbox{valign=c,raise=-1.2em}{\begin{tikzpicture}[]
			\node (o) at (0,0) {};
			\node (xshift) at (-1.3,0) {};
			\node (yshift) at (0,-0.9) {};
			\node (i2) at ($(o)+(xshift)$) {};
			\node (i3) at ($(o)+(yshift)$) {};
			\node[cdot] (v) at ($(o)$) {};
			\node[T] (A) at ($(i3)$) {$4$};
			\node[t] (x) at ($(i2)$) {$\mathllap{
					\delta \tphi_a\hnem(\s)
				}$};
			\draw[wiggly] (v)--(A);
			\draw[dprop] (v)--(x);
	\end{tikzpicture}}
	\kern-0.125em\phantom{\small
		\delta \tphi_a\hnem(\s)
	}
	\,\,\,&=\,\,\,
	\bb{
		4\mem f^a{}_{cb}\mem c_4^c\mem \phi^b
		\,\frac{
			\eta\hhem\varphi_4p
		}{\b}
	}
	\,
	\mathe^{i K_4\s}
	\,.
\end{align}
\end{subequations}
On the other hand, \eqref{YMcalc.v0}
describes a contact vertex:
\begin{align}
	\label{YM.sv0}
	\adjustbox{valign=c,raise=-1.5em}{\begin{tikzpicture}[]
			\node (o) at (0,0) {};
			\node (xshift) at (0.5,0) {};
			\node (yshift) at (0,-0.9) {};
			\node (i3) at ($(o)+(yshift)-(xshift)$) {};
			\node (i4) at ($(o)+(yshift)+(xshift)$) {};
			\node[cdot] (v) at ($(o)$) {};
			\node[T] (A) at ($(i3)$) {$3$};
			\node[T] (B) at ($(i4)$) {$4$};
			\draw[wiggly] (v)--(A);
			\draw[wiggly] (v)--(B);
	\end{tikzpicture}}
	\,\,\,&=\,\,\,
	\,\,
	\bb{
		-
		\frac{
			8i\mem qc_{34}
			\,
			p\hhem\varphi_3\varphi_4p
		}{K_3t_{34}\vph}
	}
	\bb{
		1 + \frac{\d}{\a\b}
	}
	\, \frac{i}{K_3 \mplus K_4}
	\,.
\end{align}

In \eqrefs{YM.sv1(phi)}{YM.sv1(tphi)},
we have denoted assigned a dotted line for the color degrees of freedom.
Note that the color propagator is given by
\begin{align}
	\label{scalar-prop}
	\adjustbox{valign=c}{\begin{tikzpicture}[]
			\node (o) at (0,0) {};
			\node (xshift) at (0.8,0) {};
			\node (i0) at ($(o)$) {};
			\node (i1) at ($(i0)+1*(xshift)$) {};
			\node (i2) at ($(i0)+2*(xshift)$) {};
			\node[t] (x) at ($(i0)$) {$\mathllap{
					\delta \phi^a\hnem(\s_1)
				}$};
			\node[t] (p) at ($(i2)$) {$\mathrlap{
					\delta \tphi_b\hnem(\s_2)
				}$};
			\draw[dprop] (p)--(x);
	\end{tikzpicture}}
	\kern-0.125em\phantom{\small
		\delta p_\n\hnem(\s_2)
	}
	\,\,\,&=\,\,\,
	\delta^a{}_b
	\,
	\Theta_{>}(\s_1,\s_2)
	\,,
\end{align}
reflecting the free theory bracket $\{\phi^a,\tphi_b\}^\circ = \delta^a{}_b$.

In the similar fashion,
the Feynman rules for the gravitational interaction
are found as
\begin{subequations}
\begin{align}
	\label{Grav.sv1(x)}
	\adjustbox{valign=c,raise=-1.2em}{\begin{tikzpicture}[]
			\node (o) at (0,0) {};
			\node (xshift) at (1.3,0) {};
			\node (yshift) at (0,-0.9) {};
			\node (i1) at ($(o)-(xshift)$) {};
			\node (i2) at ($(o)+(xshift)$) {};
			\node (i3) at ($(o)+(yshift)$) {};
			\node[cdot] (v) at ($(o)$) {};
			\node[T] (A) at ($(i3)$) {$3$};
			\node[t] (x) at ($(i2)$) {$\mathrlap{
					\delta x^\m\hnem(\s)
				}$};
			\draw[wiggly] (v)--(A);
			\draw[prop] (x)--(v);
	\end{tikzpicture}}
	\kern-0.125em\phantom{\small
		\delta x^\m\hnem(\s)
	}
	\,\,\,&=\,\,\,
	\bb{
		4i\mem (\varphi_3p)_\m
		\,\frac{
			\eta\hhem\varphi_3p
		}{\a}
	}
	\,
	\mathe^{i K_3\s}
	\,,\\
	\label{Grav.sv1(p)}
	\adjustbox{valign=c,raise=-1.2em}{\begin{tikzpicture}[]
			\node (o) at (0,0) {};
			\node (xshift) at (-1.3,0) {};
			\node (yshift) at (0,-0.9) {};
			\node (i2) at ($(o)+(xshift)$) {};
			\node (i3) at ($(o)+(yshift)$) {};
			\node[cdot] (v) at ($(o)$) {};
			\node[T] (A) at ($(i3)$) {$4$};
			\node[t] (x) at ($(i2)$) {$\mathllap{
					\delta p_\a\hnem(\s)
				}$};
			\draw[wiggly] (v)--(A);
			\draw[prop] (v)--(x);
	\end{tikzpicture}}
	\kern-0.125em\phantom{\small
		\delta p_\a\hnem(\s)
	}
	\,\,\,&=\,\,\,
	\bb{
		8\mem (\eta\hhem\varphi_4)^\a
		\,
		\frac{\eta\hhem\varphi_4p}{\b^2}
	}
	\,
	\mathe^{i K_4\s}
	\,.
\end{align}
\end{subequations}
On the other hand, \eqref{YMcalc.v0}
describes a contact vertex:
\begin{align}
	\label{Grav.sv0}
	\adjustbox{valign=c,raise=-1.5em}{\begin{tikzpicture}[]
			\node (o) at (0,0) {};
			\node (xshift) at (0.5,0) {};
			\node (yshift) at (0,-0.9) {};
			\node (i3) at ($(o)+(yshift)-(xshift)$) {};
			\node (i4) at ($(o)+(yshift)+(xshift)$) {};
			\node[cdot] (v) at ($(o)$) {};
			\node[T] (A) at ($(i3)$) {$3$};
			\node[T] (B) at ($(i4)$) {$4$};
			\draw[wiggly] (v)--(A);
			\draw[wiggly] (v)--(B);
	\end{tikzpicture}}
	\,\,\,&=\,\,\,
	\,\,
	\bb{
		-
		\frac{
			16
			\mem
			(p\hhem\varphi_3\varphi_4p)^2
		}{{K_3}^2t_{34}\vph}
	}
	\bb{
		1 + \frac{\d}{\a\b}
		+ \frac{\d\c_{34}}{\a^2\b}
		- \frac{\d\c_{43}}{\a\b^2}
	}
	\, \frac{i}{K_3 \mplus K_4}
	\,.
\end{align}
Here, we have introduced
\begin{align}
	\label{cparam}
	\c_{34} \,:=\, 
		\frac{\eta\hhem\varphi_3\varphi_4p}{p\hhem\varphi_3\varphi_4p}\,
			K_3
	\,,\quad
	\c_{43} \,:=\, 
		\frac{\eta\hhem\varphi_4\varphi_3p}{p\hhem\varphi_3\varphi_4p}\,
			K_3
	\,.
\end{align}
Note that the parameters
$\a$, $b$, $\c_{34}$, $\c_{43}$, and $\d$
in \eqrefs{cparam}{abdparam}
are dimensionless,
provided that the reference vector $\eta$
carries the dimensions of length.

\subsection{Compton Amplitudes}

Finally,
we can readily obtain the Compton amplitudes.

We start with the comment that
the choice $\eta = p$
for the reference vector
leads to a large simplification.
This choice
kills all vertices
except 
the contact vertices in \eqrefs{YM.sv0}{Grav.sv0},
and the vertex in \eqref{YM.sv1(x)} for Yang-Mills theory.
The parameters $\c_{34}$, $\c_{43}$, and $\d$ also vanish in this limit.
As a result, 
the amplitudes are immediately obtained as
\begin{subequations}
\begin{align}
	\label{Compton.YM}
	\A^\text{YM}_4
	\,&=\,
		- 
		(qc_3)(qc_4)
		\mem
		\bb{
			\frac{
				8\mem p\hhem\varphi_3\varphi_4p
			}{{K_3}^2\vph}
		}
		+
		(qc_{34})\mem
		\bb{
			-\frac{
				8i\mem
				p\hhem\varphi_3\varphi_4p
			}{K_3t_{34}\vph}
		}
	\,,\\
	\label{Compton.Grav}
	\A^\text{Grav}_4
	\,&=\,
		-
		\frac{
			16
			\mem
			(p\hhem\varphi_3\varphi_4p)^2
		}{{K_3}^2t_{34}\vph}
	\,,
\end{align}
\end{subequations}
where the first term in \eqref{Compton.YM}
is
a straightforward shadow
of the abelian computation in
\eqref{QED.A4}.
Note that this simplification could have been 
implemented from the very beginning stage
of evaluating the symplectic perturbations.

To ensure the consistency of our formalism,
however,
let us explicitly check
if the gauge-dependent terms cancel.
For Yang-Mills theory,
it suffices to examine the diagrams with the color channel:
\begin{align}
	\label{YM.colorchannel}
	\adjustbox{valign=c,raise=-1.2em}{\begin{tikzpicture}[]
			\node (o) at (0,0) {};
			\node (xshift) at (1.3,0) {};
			\node (yshift) at (0,-0.9) {};
			\node (i1) at ($(o)$) {};
			\node (i2) at ($(o)+(xshift)$) {};
			\node (i3) at ($(i1)+(yshift)$) {};
			\node (i4) at ($(i2)+(yshift)$) {};
			\node[cdot] (v1) at ($(i1)$) {};
			\node[cdot] (v2) at ($(i2)$) {};
			\node[T] (A3) at ($(i3)$) {$3$};
			\node[T] (A4) at ($(i4)$) {$4$};
			\draw[wiggly] (v1)--(A3);
			\draw[wiggly] (v2)--(A4);
			\draw[dprop] (v2)--(v1);
	\end{tikzpicture}}
	\,\,\,+\,\,\,
	\adjustbox{valign=c,raise=-1.2em}{\begin{tikzpicture}[]
			\node (o) at (0,0) {};
			\node (xshift) at (1.3,0) {};
			\node (yshift) at (0,-0.9) {};
			\node (i1) at ($(o)$) {};
			\node (i2) at ($(o)+(xshift)$) {};
			\node (i3) at ($(i1)+(yshift)$) {};
			\node (i4) at ($(i2)+(yshift)$) {};
			\node[cdot] (v1) at ($(i1)$) {};
			\node[cdot] (v2) at ($(i2)$) {};
			\node[T] (A3) at ($(i3)$) {$4$};
			\node[T] (A4) at ($(i4)$) {$3$};
			\draw[wiggly] (v1)--(A3);
			\draw[wiggly] (v2)--(A4);
			\draw[dprop] (v2)--(v1);
	\end{tikzpicture}}
	\,\,\,\,&=\,\,\,
	-
	\bb{
		-\frac{
			8i\mem
			p\hhem\varphi_3\varphi_4p
		}{K_3t_{34}\vph}
	}
	\bb{
		\frac{\d}{\a\b}
	}\mem
	\BB{
		\tphi\mem [c_3,c_4] \phi
	}
	\,
	\frac{i}{K_3 \mplus K_4}
	\,.
\end{align}
This exactly cancels the gauge-dependent terms in \eqref{YM.sv0}
via the Jacobi identity of the gauge Lie algebra 
$\g = \su(N)$.
In fact, this traces back to the fact that
the Poisson bracket in \eqref{qq}
holds on the support of the Jacobi identity of $\g = \su(N)$:
$\{ qc_3 , qc_4 \}^\circ = qc_{34}$.

For gravity,
the ``kinematic'' channel is given by
\begin{align}
	\label{Grav.channel1}
	\adjustbox{valign=c,raise=-1.2em}{\begin{tikzpicture}[]
			\node (o) at (0,0) {};
			\node (xshift) at (1.3,0) {};
			\node (yshift) at (0,-0.9) {};
			\node (i1) at ($(o)$) {};
			\node (i2) at ($(o)+(xshift)$) {};
			\node (i3) at ($(i1)+(yshift)$) {};
			\node (i4) at ($(i2)+(yshift)$) {};
			\node[cdot] (v1) at ($(i1)$) {};
			\node[cdot] (v2) at ($(i2)$) {};
			\node[T] (A3) at ($(i3)$) {$3$};
			\node[T] (A4) at ($(i4)$) {$4$};
			\draw[wiggly] (v1)--(A3);
			\draw[wiggly] (v2)--(A4);
			\draw[prop] (v2)--(v1);
	\end{tikzpicture}}
	\,\,\,+\,\,\,
	\adjustbox{valign=c,raise=-1.2em}{\begin{tikzpicture}[]
			\node (o) at (0,0) {};
			\node (xshift) at (1.3,0) {};
			\node (yshift) at (0,-0.9) {};
			\node (i1) at ($(o)$) {};
			\node (i2) at ($(o)+(xshift)$) {};
			\node (i3) at ($(i1)+(yshift)$) {};
			\node (i4) at ($(i2)+(yshift)$) {};
			\node[cdot] (v1) at ($(i1)$) {};
			\node[cdot] (v2) at ($(i2)$) {};
			\node[T] (A3) at ($(i3)$) {$4$};
			\node[T] (A4) at ($(i4)$) {$3$};
			\draw[wiggly] (v1)--(A3);
			\draw[wiggly] (v2)--(A4);
			\draw[prop] (v2)--(v1);
	\end{tikzpicture}}
	\,,
\end{align}
which parallels the color channel in \eqref{YM.colorchannel}.
This evaluates to
\begin{align}
	\label{Grav.channel1.calc}
	- \A^\text{Grav}_4
	\,
	\d\,
	\bb{
		- 
		\frac{
			(p\hhem\varphi_3)(\varphi_4\eta)
		}{p\hhem\varphi_3\varphi_4p}
		\frac{K_3}{\a\b^2}
		+
		\frac{
			(p\hhem\varphi_4)(\varphi_3\eta)
		}{p\hhem\varphi_3\varphi_4p}
		\frac{K_3}{\a^2\b}
	}
	\,
	\frac{i}{K_3 \mplus K_4}
	\,,
\end{align}
canceling the $\c_{34}$ and $\c_{43}$ terms
in \eqref{Grav.sv0}.
The other term
is canceled through the second-order channel:
\begin{align}
\begin{split}
	\label{Grav.channel2}
	\adjustbox{valign=c,raise=-1.12em}
	{\begin{tikzpicture}[]
			\node (o) at (0,0) {};
			\node (ynudge) at (0,0.01265) {};
			\node (yshift) at (0,-0.9) {};
			\node (a) at ($(o)+(165:1.4)$) {};
			\node (e) at ($(a)+(yshift)$) {};
			\node (b) at ($(o)+(195:1.0)$) {};
			\node (f) at ($(b)+(yshift)$) {};
			\node[cdot] (v) at ($(o)$) {};
			\node[cdot] (v3) at ($(a)$) {};
			\node[cdot] (v4) at ($(b)$) {};
			\node[T] (A3) at ($(e)$) {$3$};
			\node[T] (A4) at ($(f)$) {$4$};
			\draw[wiggly] (v3)--(A3);
			\draw[wiggly] (v4)--(A4);
			\draw[prop] ($(o)+(ynudge)$)--(v3);
			\draw[prop] ($(o)-(ynudge)$)--(v4);
	\end{tikzpicture}}
	\,\,\,\,\,\,
	\begin{aligned}[t]
	&=\,\,\,\,
		32\mem
			(\varphi_3p) \mdot (\varphi_4p)
		\,
		\frac{\eta\hhem\varphi_3p}{\a}
		\frac{\eta\hhem\varphi_4p}{\b}
		\,
		\frac{1}{{K_3}^2\vph}
		\,
		\frac{i}{K_3 \mplus K_4}
	\,,\\
	&=\,\,\,\,
		- \A^\text{Grav}_4
		\,
		\bb{	
			\frac{\d}{\a\b}
		}
		\,
		\frac{i}{K_3 \mplus K_4}
	\,.
	\end{aligned}
\end{split}
\end{align}

The correspondence between 
Yang-Mills theory and gravity 
in the above calculations
should be clear:
\begin{align}
\begin{split}
	\int d\s\,\,
		q_a\mem A^a{}_\m(\bmx)\mem \dot{\bmx}^\m
	&\qquad\xleftrightarrow{\,}\qquad
	\int d\s\,\,
		p_\a\mem h^\a{}_\m(\bmx)\mem \dot{\bmx}^\m
	\,,\\
	\delta q_a\mem A^a{}_\m(\bmx)\mem \dot{\bmx}^\m
	&\qquad\xleftrightarrow{\,}\qquad
	\delta p_\a\mem h^\a{}_\m(\bmx)\mem \dot{\bmx}^\m
	\,,\\
	q_a\mem (dA^a)_{\m\n}\mem \delta x^\m\mem \dot{\bmx}^\n
	&\qquad\xleftrightarrow{\,}\qquad
	p_\a\mem (de^\a)_{\m\n}\mem \delta x^\m\mem \dot{\bmx}^\n
	\,.
\end{split}
\end{align}
This echoes the isomorphism identified 
at the level of symplectic structures
in \eqref{isom.YM-Grav}.

As the cancellation through the first-order channel in \eqref{YM.colorchannel}
has encoded the color Lie algebra $\g=\su(N)$,
it might be interesting
if a parallel structure can be observed 
for the ``kinematic'' first-order channel in \eqref{Grav.channel1.calc}.
Apparently, the Poisson bracket $\{x,p\}^\circ = 0$ is capable of implementing the diffeomorphism algebra, for instance.
Specifically, since we are identifying the vector polarizations as
$e_I^\m = -\varphi_I^{\m\n} \eta_\n / k_I\mdot \eta$
due to the axial gauge,
this first-order channel 
in  
has described a tensor combination
$
	p_\a\mem\bigbig{
		(\varphi_3)^\a{}_\m\mem (e_4)^\m
		-
		(\varphi_4)^\a{}_\m\mem (e_3)^\m
	}
$.
This suggests a Poincar\'e group action,
although it might be a simple coincidence.

\newpage

\section{Summary and Outlook}

In this paper,
we implemented the worldline formalism
with phase space actions
to derive an efficient framework for computing scattering amplitudes.

In the top-down exposition in \Sec{General},
the phase space worldline formalism
is studied as a symplectic sigma model.
The ideas of 
Hamiltonian and symplectic perturbation theories
are introduced
in terms of worldline field basis.
Several notable
universal features of the Feynman rules 
are pointed out.
In particular, 
the propagator encodes the Poisson bracket
while the symplectic vertex of valence one
exhibits a ``Lorentz force'' formula.

In \Sec{WLF},
it is established that
LSZ reductions can be implemented purely within the first-quantized framework
by employing non-compact worldline topologies.
With the proper identifications of the moduli spaces,
the strict definition of the partition function 
of a relativistic particle
yields
off-shell propagators,
partly on-shell propagators,
and
scattering amplitudes
for
the interval, half-line, and full line topologies,
respectively.
These computations are naturally associated with
various boundary conditions in the phase space.

By building upon these observations,
\Sec{QED}
established 
an efficient formalism
that obtains
scattering amplitudes
by computing
bulk-to-boundary propagators in plane wave backgrounds.
Notably, this framework
directly features a momentum variable in the Feynman rules
and greatly reduces 
post-processing such as
Fourier or Gaussian integrals.
Also, 
the intermediate computations are made manifestly gauge-invariant
by adopting symplectic perturbation theory.
It is concretely and explicitly verified that
this formalism produces
correct results
for clasical Compton amplitudes in electromagnetism
up to six points.

Finally, \Sec{NA}
has demonstrated
how the formalism applies to nonabelian gauge theory and gravity as well.
It is established that
implementing the worldline formalism in 
the on-shell backgrounds of nonlinearly superposed plane waves
computes the Compton scattering amplitudes.
A uniform treatment of gauge theory and gravity
is achieved,
in terms of a parallel at the level of symplectic structures.

The phase space worldline formalism is versatile,
and this paper has explored only a subset of its full potential.
Firstly,
applications to in-in formalism
should be pursued.
In this paper,
the focus was on revisiting the original worldline formalism
as an in-out formalism,
although several aspects of the in-in formalism
are discussed in \Sec{General} and \App{EXPL}.
Given that
an in-in formalism \cite{Mogull:2020sak,Kopp:2022acm,Bohnenblust:2025gir,Comberiati:2022cpm,Wang:2022ntx,Haddad:2024ebn,Jakobsen:2021zvh,Shi:2021qsb}
has found fruitful applications in recent years,
a systematic study from the phase space perspective
can bring some insights,
especially regarding the notion of classical eikonal \cite{Kim:2025hpn,Kim:2024svw,Kim:2024grz}.
Secondly,
obtaining fully quantum amplitudes
is a straightforward future direction.
It will be interesting to see if 
a master formula
is viable 
in terms of bivector polarizations
by employing symplectic perturbation theory.
Thirdly,
it should be interesting to approach double copy structures
in the symplectic perturbation language.
Lastly,
it will be fruitful to extend
the formalism
to more general particles.
For nonabelian and gravitational interactions,
a covariant symplectic perturbation theory
shall be presented in a future work.
For adding spin on the worldline,
one can not only consider
finite-spin models
such as the supersymmetric spinning particles
\cite{Gibbons:1993ap,Haddad:2024ebn,Jakobsen:2021zvh}
but also 
the massive twistor model \cite{ambikerr0,ambikerr1,Kim:2024grz},
which is capable of describing arbitrarily high massive spin.

	\vskip10pt
	\paragraph{Acknowledgements.}
	The author
	is grateful to
		Clifford Cheung
	and
		Zvi Bern
	for motivating comments
	and
		Vinicius Nevoa
	for insightful conversations. 
	The author 
	thanks
		Sangmin Lee
	for sharing his recent related work.
	J.-H.K. is supported by the Department of Energy (Grant No.~DE-SC0011632) and by the Walter Burke Institute for Theoretical Physics.
	 
%	\newpage
	\appendix
	
\section{Exploration of Diagrammatics}
\label{EXPL}

In \Sec{General},
we have derived the Feynman rules
for worldline perturbation theory in phase space
with full generality.
To explicate the universal implications of these Feynman rules
and understand the role of each diagrammatic element,
we run a pedagogical exploration
on a perturbative treatment of Hamiltonian mechanics
in the diagrammatic language.

\subsection{Poisson Bracket}

In this subsection, we investigate the Poisson bracket.
In \Sec{General.feyn.prop}, 
we have observed that the bare propagator encodes the free theory's Poisson bracket,
$\{ \z^i , \z^j \}^\circ = (\omega^\circ{}^{-1})^{ij}$.
A natural follow-up question is then
whether a resummed propagator can achieve the interacting theory's Poisson bracket,
$\{ \z^i , \z^j \} = (\omega{}^{-1}\hnem(\z)\hnem)^{ij}$.

We start with a preliminary remark that
the Poisson bracket
is geometrically a bivector in phase space:
$\omega^{-1} = \frac{1}{2}\mem (\omega^{-1}\hnem(\z)\hnem)^{ij}\, \partial_i \wedge \partial_j$.
As is suggested from the notation,
it describes the inverse of the symplectic form 
in the sense that $(\omega^{-1}\hnem(\z)\hnem)^{ik}\mem \omega_{kj}(\z) = \delta^i{}_j$.
As a result,
the Poisson bivector admits a geometrical series expansion
after the free-interaction split:
\begin{align}
	\label{spt}
	\omega \mem=\mem \omega^\circ + \omega'
	\qiq
	\omega^{-1}
	\mem=\mem
	\omega^\circ{}^{-1}
	- 
	\omega^\circ{}^{-1}
	{\mem \omega' \,}
	\omega^\circ{}^{-1}
	+
	\omega^\circ{}^{-1}
	{\mem \omega' \,}
	\omega^\circ{}^{-1}
	{\mem \omega' \,}
	\omega^\circ{}^{-1}
	- \cdots
	\,.
\end{align}
This means that we have an expansion of
the interacting theory's Poisson bracket
in terms of the free theory's Poisson bracket:
\begin{align}
	\begin{split}
		\label{spt-pb}
		\{ f , g \}
		\mem=\mem
		\{ f , g \}^\circ
		&-
		\{ f , \z^{k_1} \}^\circ
		{\,\hem \omega'_{k_1k_2} \mem}
		\{ \z^{k_2} , g \}^\circ
		\\
		&+
		\{ f , \z^{k_1} \}^\circ
		{\,\hem \omega'_{k_1k_2} \mem}
		\{ \z^{k_2} , \z^{k_3} \}^\circ
		{\,\hem \omega'_{k_3k_4} \mem}
		\{ \z^{k_4} , g \}^\circ
		- \cdots
		\,.
	\end{split}
\end{align}
Clearly,
\eqref{spt-pb} shows
how the symplectic perturbation
modifies the Poisson bracket---%
from canonical to a noncanonical form.
Note that Poisson brackets are defined as
\begin{align}
	\label{pbdef}
	\{ f(\z) , g(\z) \}^\circ 
	\,=\,
	(\omega^\circ{}^{-1})^{ij}
	f_{,i}\hnem(\z)\mem
	g_{,j}\hnem(\z)
	\,,\quad
	\{ f(\z) , g(\z) \}
	\,=\,
	(\omega^{-1}\hnem(\z)\hnem)^{ij}
	f_{,i}\hnem(\z)\mem
	g_{,j}\hnem(\z)
	\,.
\end{align}

The expansion in \eqref{spt}
is a local equation,
equating two tensor expressions at the same point.
Hence, 
in the diagrammatic terms,
it describes no ``long-range'' propagation on the worldline
but a contact interaction.
As a result,
one can identify that
the relevant object is
the pinched symplectic vertex of valence two in \eqref{generic.psv2}.

To see this explicitly,
let us compute the time-ordered correlator between
$\dz^i(\s)$ and $\dz^j(0)$,
which uses the symmetric propagator in \eqref{prop.sym}.
As discussed in \Sec{General.feyn.prop},
examining the simple pole
will reveal the Poisson structure.
To expedite our demonstration,
let us assume a constant perturbation on the symplectic form,
$\omega'_{ij}\hnem(\z) = B_{ij}$,
noting that the expansion in \eqref{spt} 
does not contain any derivatives of $\omega'$.
The Hamiltonian vertices can be effectively ignored,
as they will generate only higher-order poles.

Provided this setup,
the relevant subset of Feynman rules reduces to the following:
\begin{subequations}
	\begin{align}
		\label{Bdemo.feyn1}
		\phantom{\delta \z^i\hnem(\t_1)}
		\adjustbox{valign=c}{\begin{tikzpicture}[]
				\node (o) at (0,0) {};
				\node (xshift) at (0.8,0) {};
				\node (i0) at ($(o)$) {};
				\node (i1) at ($(i0)+1*(xshift)$) {};
				\node (i2) at ($(i0)+2*(xshift)$) {};
				\node (v) at (i1) {};
				\node[t] (L) at ($(i0)$) {$\mathllap{
						\delta \z^i\hnem(\s_1)
					}$};
				\node[t] (R) at ($(i2)$) {$\mathrlap{
						\delta \z^j\hnem(\s_2)
					}$};
				\draw[linear] (R)--(L);
		\end{tikzpicture}}
		\phantom{{\delta \z^j\hnem(\s_2)}}
		\kern-0.125em
		\,\,\,&=\,\,\,
		(\omega^\circ{}^{-1})^{ij}\,
		\Theta(\s_1,\s_2)
		\,,\\[0.3\baselineskip]
		\label{Bdemo.feyn2}
		\phantom{\dz^i\hnem(\s_1)}
		\adjustbox{valign=c}{\begin{tikzpicture}[]
				\node (o) at (0,0) {};
				\node (xshift) at (0.8,0) {};
				\node (i0) at ($(o)$) {};
				\node (i1) at ($(i0)+1*(xshift)$) {};
				\node (i2) at ($(i0)+2*(xshift)$) {};
				\node (v) at (i1) {};
				\node[t] (L) at ($(i0)$) {$\mathllap{
						\delta\z^i\hnem(\s_1)
					}$};
				\node[t] (R) at ($(i2)$) {$\mathrlap{
						\delta\z^j\hnem(\s_2)
					}$};
				\node (p) at ($(v)+(135:0.22)$) {\pin};
				\node (q) at ($(v)+(-45:0.22)$) {\phantom{\pin}};
				\draw[linear] (v)--(L);
				\draw[linear] (v)--(R);
				\node[odot] (V) at ($(v)$) {};
		\end{tikzpicture}}
		\phantom{\dz^j\hnem(\s_2)}
		\kern-0.125em
		+
		\phantom{\dz^i\hnem(\s_1)}
		\adjustbox{valign=c}{\begin{tikzpicture}[]
				\node (o) at (0,0) {};
				\node (xshift) at (0.8,0) {};
				\node (i0) at ($(o)$) {};
				\node (i1) at ($(i0)+1*(xshift)$) {};
				\node (i2) at ($(i0)+2*(xshift)$) {};
				\node (v) at (i1) {};
				\node[t] (L) at ($(i0)$) {$\mathllap{
						\delta\z^i\hnem(\s_1)
					}$};
				\node[t] (R) at ($(i2)$) {$\mathrlap{
						\delta\z^j\hnem(\s_2)
					}$};
				\node (p) at ($(v)+(45:0.22)$) {\pin};
				\node (q) at ($(v)+(-45:0.22)$) {\phantom{\pin}};
				\draw[linear] (v)--(L);
				\draw[linear] (v)--(R);
				\node[odot] (V) at ($(v)$) {};
		\end{tikzpicture}}
		\phantom{\dz^j\hnem(\s_2)}
		\kern-0.125em
		\,\,\,&=\,\,\,
		B_{ij}\,
		\delta'(\s_1{\mem-\,}\s_2)
		\,.
	\end{align}
\end{subequations}
Then, the fat propagator, satisfying
\begin{align}
	\label{prop.red}
	\adjustbox{valign=c}{\begin{tikzpicture}[]
			\node (o) at (0,0) {};
			\node (xshift) at (0.8,0) {};
			\node (i0) at ($(o)$) {};
			\node (i1) at ($(i0)+1*(xshift)$) {};
			\node (i2) at ($(i0)+2*(xshift)$) {};
			\node (v) at (i1) {};
			\node (L) at ($(i0)$) {};
			\node (R) at ($(i2)$) {};
			\draw[linear] (R)--(L);
			\node[arb-blob-red] (V) at ($(v)$) {};
	\end{tikzpicture}}
	\quad=\quad
	\adjustbox{valign=c}{\begin{tikzpicture}[]
			\node (o) at (0,0) {};
			\node (xshift) at (0.65,0) {};
			\node (i0) at ($(o)$) {};
			\node (i1) at ($(i0)+1*(xshift)$) {};
			\node (i2) at ($(i0)+2*(xshift)$) {};
			\node (v) at (i1) {};
			\node (L) at ($(i0)$) {};
			\node (R) at ($(i2)$) {};
			\draw[linear] (R)--(L);
	\end{tikzpicture}}
	\,\,\,+\,\,\,
	\adjustbox{valign=c}{\begin{tikzpicture}[]
			\node (o) at (0,0) {};
			\node (xshift) at (0.8,0) {};
			\node (i0) at ($(o)$) {};
			\node (i1) at ($(i0)+1*(xshift)$) {};
			\node (i2) at ($(i0)+2*(xshift)$) {};
			\node (i3) at ($(i0)+3*(xshift)$) {};
			\node (v) at (i1) {};
			\node (L) at ($(i0)$) {};
			\node (R) at ($(i3)$) {};
			\node (p) at ($(v)+(135:0.22)$) {\pin};
			\node (q) at ($(v)+(-45:0.22)$) {\phantom{\pin}};
			\draw[linear] (R)--(L);
			\node[odot] (V) at ($(v)$) {};
			\node[arb-blob-red] (R) at ($(i2)$) {};
	\end{tikzpicture}}
	\,\,\,+\,\,\,
	\adjustbox{valign=c}{\begin{tikzpicture}[]
			\node (o) at (0,0) {};
			\node (xshift) at (0.8,0) {};
			\node (i0) at ($(o)$) {};
			\node (i1) at ($(i0)+1*(xshift)$) {};
			\node (i2) at ($(i0)+2*(xshift)$) {};
			\node (i3) at ($(i0)+3*(xshift)$) {};
			\node (v) at (i1) {};
			\node (L) at ($(i0)$) {};
			\node (R) at ($(i3)$) {};
			\node (p) at ($(v)+(45:0.22)$) {\pin};
			\node (q) at ($(v)+(-45:0.22)$) {\phantom{\pin}};
			\draw[linear] (R)--(L);
			\node[odot] (V) at ($(v)$) {};
			\node[arb-blob-red] (R) at ($(i2)$) {};
	\end{tikzpicture}}
	\,\,\,
	\,,
\end{align}
is found as
\begin{align}
	\phantom{\delta \z^i\hnem(\t_1)}
	\adjustbox{valign=c}{\begin{tikzpicture}[]
			\node (o) at (0,0) {};
			\node (xshift) at (0.8,0) {};
			\node (i0) at ($(o)$) {};
			\node (i1) at ($(i0)+1*(xshift)$) {};
			\node (i2) at ($(i0)+2*(xshift)$) {};
			\node (v) at (i1) {};
			\node[t] (L) at ($(i0)$) {$\mathllap{
					\delta \z^i\hnem(\s_1)
				}$};
			\node[t] (R) at ($(i2)$) {$\mathrlap{
					\delta \z^j\hnem(\s_2)
				}$};
			\draw[linear] (R)--(L);
			\node[arb-blob-red] (V) at ($(v)$) {};
	\end{tikzpicture}}
	\phantom{{\delta \z^j\hnem(\s_2)}}
	\kern-0.125em
	\mem\,&=\,\,\,
	(	
	\omega^\circ{}^{-1}
	{\mem+\,} \omega^\circ{}^{-1} B\mem \omega^\circ{}^{-1}
	{\mem+\,} \omega^\circ{}^{-1} B\mem \omega^\circ{}^{-1}
	{\mem+\,} \cdots
	)^{ij}\,
	\Theta(\s_1,\s_2)
	\,,
\end{align}
whose tensor factor
readily
reproduces the expansion in \eqref{spt}.

In conclusion,
the insight to be learned
from this analysis
is that the pinched symplectic vertex of valence two
is essential for
constructing
the perturbed Poisson bracket
in the interacting theory.
Specifically,
it implements a pinching mechanism
to adjoin $\omega^\circ{}^{-1}$ and $\omega'$
into a geometric series in \eqref{spt}
without intermediate propagators.

A related exercise is to compute the time-ordered correlator between
two observables in the phase space,
$f(\bmz(\s){\mem+\,}\dz(\s))$
and
$g(\bmz(0){\mem+\,}\dz(0))$,
while retaining quantum contributions.
This computes the quantum product between $f$ and $g$
as operators inserted on the worldline
and derives what is known as the Moyal star product
\cite{Moyal:1949sk,Groenewold:1946kp,Cattaneo:1999fm,kontsevich2003deformation}. 
Using this correspondence, one can obtain
the $\hbar$-deformed version of
the expansion in \eqref{spt-pb},
describing a quantum-corrected Poisson bracket
as the counterpart of the operator commutator.
In this process,
one can in principle keep the derivatives of $\omega'$
and see how they contribute nontrivially
through the pinched vertices.\footnote{
	This seems to produce graphs similar to those studied by
	Kontsevich \cite{kontsevich2003deformation}
	and
	Cattaneo and Felder \cite{Cattaneo:1999fm}.
	However, such an worldline implementation 
	can exhibit higher complexity
	in the sense that
	the symplectic form is taken as fundamental
	instead of the Poisson bivector.
}

\subsection{Hamiltonian Equations of Motion}
\label{sec:General.heom}

In this subsection, we investigate the Hamiltonian equations of motion.
At the end of \Sec{General.feyn.sv},
we have remarked that 
the valence-one vertices in our Feynman rules
encode the leading perturbation on Hamiltonian equations of motion.
To elaborate,
the Hamiltonian equations of motion 
follow from the Poisson bracket in \eqref{pbdef} as
\begin{align}
	\label{heom-dot}
	\dot{\z}^i
	\,=\,
	\{ \z^i , H(\z) \}
	\,=\,
	(\omega^{-1}\hnem(\z)\hhnem)^{ij}\mem H_{,j}\hnem(\z)
	\,,
\end{align}
which are the classical equations of motion 
implied by the phase space Lagrangian $L[\z]$ in \eqref{psL}:
\begin{align}
	\label{vari1}
	\begin{split}
		L[\z {\,+\,} \dz]
		\,=\,
		L^\circ[\z]
		+
		\frac{d}{d\s}\hem \Big(\,\mem{
			\theta_j\hnem(\z)\mem \dz^j
		}\mem\Big)
		&
		+ \dz^i\mem \Big(\,{
			\omega_{ij}(\z)\mem \dot{\z}^j
			- H_{,i}\hnem(\z)
		}\,\Big)
		+ \mathcal{O}(\dz^2)
		\,.
	\end{split}
\end{align}
Clearly, the combination in the second bracket in \eqref{vari1}
splits into
free and interaction parts as
$\omega^\circ_{ij}(\z)\mem \dot{\z}^j - H^\circ_{,i}\hnem(\z)$
and
$\omega'_{ij}(\z)\mem \dot{\z}^j - H'_{,i}\hnem(\z)$,
the latter of which 
is the sum of
the valence-one vertices in \eqrefs{generic.H1}{generic.int1}
when restricted on the background trajectory.
This is due to the very fact that
the first-order variation of a Lagrangian
yields the classical equations of motion,
which is not specific to phase space, in fact
(cf.\,\rrcite{Cheung:2023lnj-EMR,jordan2023,Cheung:2024byb}).

The question we ask in this subsection
is whether this fact
can be derived in a more physical fashion.

To this end, we
investigate the following problem.
Suppose a classical trajectory 
$\s \mapsto \z^i(\s)$
in phase space,
slightly perturbed away from a free theory saddle
$\s \mapsto \bmz^i(\s)$
by interactions.
To expedite our demonstration,
let us suppose 
constant gradient and magnetic forces
in phase space:
\begin{align}
	\label{ebc}
	H'(\z)
	\,=\,
	-\e\, E_i\hem \z^i
	\,,\quad
	\omega'_{ij}\hnem(\z)
	\,=\,	
	\e\, B_{ij}
	\,,
\end{align}
where $\e$ describes a coupling constant.
The goal of this problem is to find the leading deflection $\dz^i(\s)$ 
of the interacting trajectory from the free trajectory:
$\z^i(\s) = \bmz^i(\s) + \e\, \dz^i(\s) + \O(\e^2)$.

Notably, two approaches are viable.
First of all, a straightforward approach
is to perturbatively solve the equations of motion
in the plain and faithful fashion.
Under the assumptions in \eqref{ebc}, 
the Hamiltonian equations of motion read
\begin{align}
	\label{ebc-dotz}
	\dot{\z}^i
	\,=\,
	(\omega^\circ{}^{-1})^{ij}\mem H^\circ_{,j}\hnem(\z)
	-\e\, \Big(\,{
		(\omega^\circ{}^{-1})^{ij}\mem E_j
		+ (\omega^\circ{}^{-1} B\mem \omega^\circ{}^{-1})^{ij}\mem H^\circ_{,j}\hnem(\z)
	}\,\Big)
	+ \O(\e^2)
	\,.
\end{align}
To obtain the equation governing the leading deflection,
we plug in
$\z^i(\s) = \bmz^i(\s) + \e\, \dz^i(\s) + \O(\e^2)$
to \eqref{ebc-dotz}
and extract the $\O(\e^1)$ part:
\begin{align}
	\label{ebc-de}
	-\bigg(\,{
		\omega^\circ_{ij}\,
		\frac{d}{d\s}
		- H_{,ij}^\circ\hnem(\bmz)
	}\mem\bigg)\,
	\dz^j
	\,=\,
	E_i + B_{ij}\mem \dot{\bmz}^j
	\,.
\end{align}
Here, the free theory's equations of motion has been substituted:
$\dot{\bmz}^i = (\omega^\circ{}^{-1})^{ij}\hem H_{,j}\hnem(\bmz)$.
The solution to \eqref{ebc-de} is found by
standard methods of Green's function.

Next, we provide a diagrammatic approach
that utilizes our Feynman rules in \Sec{General.feyn}.
The crucial fact is that
the solution to \eqref{ebc-de}
can also be obtained
by computing the one-point function of the worldline fluctuation \cite{Mogull:2020sak}.
Given a fat ``grey'' propagator,
\begin{align}
	\label{greyprop}
	{\,\,\,
		\adjustbox{valign=c}{\begin{tikzpicture}[]
				\node (L) at (0,0) {};
				\node (R) at (1.39,0) {};
				\node (R2) at (1.41,0) {};
				\draw[fprop] (R)--(L);
				\draw[linear] (R)--(R2);
		\end{tikzpicture}}
		\,\,}
	\,:=\,
	{\,\,
		\adjustbox{valign=c}{\begin{tikzpicture}[]
				\node (o) at (0,0) {};
				\node (xshift) at (-1.4,0) {};
				\node (i0) at ($(o)$) {};
				\node (i1) at ($(i0)+1*(xshift)$) {};
				\draw[qprop] (i0)--(i1);
		\end{tikzpicture}}
		\,\,}
	+
	{\,\,
		\adjustbox{valign=c}{\begin{tikzpicture}[]
				\node (o) at (0,0) {};
				\node (xshift) at (-0.9,0) {};
				\node (i0) at ($(o)$) {};
				\node (i1) at ($(i0)+1*(xshift)$) {};
				\node (i2) at ($(i0)+2*(xshift)$) {};
				\draw[qprop] (i0)--(i1);
				\draw[qprop] (i1)--(i2);
				\node[poly2] (m1) at (i1) {};
		\end{tikzpicture}}
		\,\,}
	+
	{\,\,
		\adjustbox{valign=c}{\begin{tikzpicture}[]
				\node (o) at (0,0) {};
				\node (xshift) at (-0.9,0) {};
				\node (i0) at ($(o)$) {};
				\node (i1) at ($(i0)+1*(xshift)$) {};
				\node (i2) at ($(i0)+2*(xshift)$) {};
				\node (i3) at ($(i0)+3*(xshift)$) {};
				\draw[qprop] (i0)--(i1);
				\draw[qprop] (i1)--(i2);
				\draw[qprop] (i2)--(i3);
				\node[poly2] (m1) at (i1) {};
				\node[poly2] (m2) at (i2) {};
		\end{tikzpicture}}
		\,\,}
	+ \,\, \cdots
	\,,
\end{align}
the one-point function is given by the sum of two diagrams:
\begin{align}
	\label{ebc-sol}
	\dz^i(\s)
	\,\,\,=\,\,\,
	{\phantom{
			\dz^i\hnem(\s)
		}\kern-0.17em
		\adjustbox{valign=c}{\begin{tikzpicture}[]
				\node (o) at (0,0) {};
				\node (xshift) at (0.6,0) {};
				\node (i0) at ($(o)$) {};
				\node (i1) at ($(i0)+1*(xshift)$) {};
				\node (i2) at ($(i0)+2*(xshift)$) {};
				\node (v) at (i2) {};
				\node[t] (L) at ($(i0)$) {$\mathllap{
						\dz^i\hnem(\s)
					}$};
				\draw[fprop] (v)--(L);
				\node[odot] (V) at ($(v)$) {};
		\end{tikzpicture}}\,\mem}
	\,+\,
	{\phantom{
			\dz^i\hnem(\s)
		}\kern-0.17em
		\adjustbox{valign=c}{\begin{tikzpicture}[]
				\node (o) at (0,0) {};
				\node (xshift) at (0.6,0) {};
				\node (i0) at ($(o)$) {};
				\node (i1) at ($(i0)+1*(xshift)$) {};
				\node (i2) at ($(i0)+2*(xshift)$) {};
				\node (v) at (i2) {};
				\node[t] (L) at ($(i0)$) {$\mathllap{
						\dz^i\hnem(\s)
					}$};
				\draw[fprop] (v)--(L);
				\node[poly1] (V) at ($(v)$) {};
		\end{tikzpicture}}\,\mem}
	\,.
\end{align}

Finally,
we stipulate that the two approaches give the same result.
The fat propagator in \eqref{greyprop}
is exactly 
the retarded Green's function for the operator in the left-hand side of \eqref{ebc-de}.
Thus, by demanding that 
the diagram-based approach in \eqref{ebc-sol}
must give the same result as the straightforward calculation in \eqref{ebc-dotz},
one can argue that
the sum of the valence-one vertices in the Feynman rules
is equal to
the right-hand side of \eqref{ebc-de},
which precisely describes the net force in phase space due to the Hamiltonian and symplectic couplings.

The insight can be learned
from this analysis
is that
the elements in the Feynman rules 
find concrete semantics
in the perturbative approach to 
the Hamiltonian equations of motion.
This lesson continues to apply
at higher orders,
as is demonstrated in the next subsection.

\subsection{Impulse from One-Point Function}
\label{sec:General.bgrecursion}

We continue our investigation of the Hamiltonian equations of motion
by proceeding to higher perturbative orders.
In \Sec{sec:General.heom},
we have 
established a physical argument for
the fact that the valence-one vertices
encode leading perturbation in the Hamiltonian equations of motion.
A crucial fact used is that
the one-point function of the worldline fluctuation
computes the deflection from the background trajectory.

In this subsection,
we establish that the above observation 
is simply the lowest order instance
of a more general fact that
the perturbative treatment of classical equations of motion
encodes
tree diagrams
through
Berends-Giele recursion \cite{BerendsGiele}.

To this end, we investigate the following problem.
Suppose a classical trajectory 
$\s \mapsto \z^i(\s)$
in phase space,
slightly perturbed away from a free theory saddle
$\s \mapsto \bmz^i(\s)$
by interactions.
This perturbed trajectory is expanded as
\begin{align}
	\label{deflect}
	\z^i(\s)
	\,=\,
	\bmz^i(\s)
	+ \e^1\, \delta^{(1)}\hnem\z^i
	+ \e^2\, \delta^{(2)}\hnem\z^i
	+ \e^3\, \delta^{(3)}\hnem\z^i
	+ \cdots
	\,.
\end{align}
The goal is to find the deflection $\delta^{(n)}\hnem\z^i$ at each order $n$.
The overall setup is the same as in \Sec{sec:General.heom},
but here,
let us assume $H {\,=\,} H^\circ$ with a completely general $\omega'$.
That is, we suppose a generic symplectic perturbation theory,
given the context that
symplectic perturbation theory
is a less familiar construction
than the Hamiltonian perturbation theory.
Let $\omega'$ be assigned with order $\e^1$
for the order-counting parameter $\e$.

Our strategy is the same:
we start by the straightforward approach
and then implement the diagrammatic approach.

\subsubsection{Perturbative Solution to Equations of Motion}
\label{PSOL.1}

The Hamiltonian equations of motion in \eqref{heom-dot}
can be geometrically interpreted as 
the flow along a vector field $X$ in phase space,
which we refer to as the time evolution vector field:
\begin{align}
	\label{tevX}
	X^i(\z)
	\,=\,
	(\omega^{-1}\hnem(\z)\hnem)^{ij}\mem
	H_{,j}(\z)
	\,.
\end{align}
The vector field in \eqref{tevX}
is expanded
in two aspects.
First, the expansion of the Poisson bracket in \eqref{spt}
unfolds it as
\begin{align}
	\label{eq:spt-tev}
	X^i
	\,=\,
	X^\circ{}^i
	- (\omega^\circ{}^{-1})^{ik_1}\mem
	\omega'_{k_1k_2}\mem
	X^\circ{}^{k_2}
	+ (\omega^\circ{}^{-1})^{ik_1}\mem
	\omega'_{k_1k_2}\mem
	(\omega^\circ{}^{-1})^{k_2k_3}\mem
	\omega'_{k_3k_4}\mem
	X^\circ{}^{k_4}
	- \cdots
	\,.
\end{align}
Second,
plugging in the expansion of the trajectory in \eqref{deflect}
produces tensors of the following form:
\begin{align}
	\label{xtensor}
	{}&{}
	(X^{(n,m,r)}\snem(\zeta)\hnem)^i{}_{j_1\cdots j_m
		|
		(k_{1,1}\cdots k_{1,\r_1})
		\cdots
		(k_{n,1}\cdots k_{n,\r_n})
	}\\
	{}&{}
	:=\mem
	(-1)^n\mem
	\Big(\hem{
		\omega^\circ{}^{-1}\hnem
		\underbrace{
			({
				\partial_{k_{1,1}}
				\nem{\cdots}\mem
				\partial_{k_{1,\r_1}}
			}\hnem\omega'\hnem(\zeta)\hhnem)\mem
			\omega^\circ{}^{-1}
			\cdots\mem
			({
				\partial_{k_{n,r}}
				\nem{\cdots}\mem
				\partial_{k_{n,\r_r}}
			}\hnem\omega'\hnem(\zeta)\hhnem)\mem
			\omega^\circ{}^{-1}
			\hnem}_{\text{$n$ occurrence of $\omega'$}}
		\mem
	}\Big){}^{ij}
	\,H_{,j j_1\cdots j_m}\hhnem(\zeta)
	\,.
\end{align}
Here, $\r_1 + \r_2 + \cdots + \r_n = r$
and $\r_1, \r_2, \cdots \r_n \geq 0$
so that there exist
$\binom{n+r-1}{r}$ such tensors
for fixed $(n,m,r)$.
For simplicity, we abbreviate
\begin{align}
\label{xtensor-abbrv}
	(X^{(n,0,0)}\snem(\zeta)\hnem)^i{}_{|}
	\mem=:
	(X^{(n)}\snem(\zeta)\hnem)^i
	\,,\quad
	(X^{(n,m,0)}\snem(\zeta)\hnem)^i{}_{j_1\cdots j_m|}
	=:
	(X^{(n,m)}\snem(\zeta)\hnem)^i{}_{j_1\cdots j_m}
	\,.
\end{align}

A distinguished role will be played by the matrix,
\begin{align}
	\label{M-matrix}
	M^i{}_j\hnem(\s)
	\,:=\,
		(X^{(0,1)}\snem(\bmz(\s))\hnem)^i{}_j
	\,=\,
		(\omega^\circ{}^{-1})^{ij_1}\mem H_{,jj_1}(\bmz(\s)\hnem)
	\,.
\end{align}
In particular, the fat grey propagator in \eqref{greyprop}
is given by a path-ordered exponential:
\begin{align}
	\phantom{\delta \z^i\hnem(\s)}
	\adjustbox{valign=c}{\begin{tikzpicture}[]
			\node (o) at (0,0) {};
			\node (xshift) at (0.8,0) {};
			\node (i0) at ($(o)$) {};
			\node (i1) at ($(i0)+1*(xshift)$) {};
			\node (i2) at ($(i0)+2*(xshift)$) {};
			\node (v) at (i1) {};
			\node[t] (L) at ($(i0)$) {$\mathllap{
					\delta \z^i\hnem(\s_1)
				}$};
			\node[t] (R) at ($(i2)$) {$\mathrlap{
					\delta \z^j\hnem(\s_2)
				}$};
			\draw[fprop] (R)--(L);
	\end{tikzpicture}}
	\kern-0.125em
	\phantom{{\delta \z^j\hnem(\s')}}
	\,\,\,=\,\,\,
		\bigg[\,{\,
			\mathrm{P}\exp\bb{\,
				\int^{\s_1}_{\s_2} d\s\,\,
					M(\s)
					\nem
			}
			\,
			\omega^\circ{}^{-1}
		}\,\bigg]^{ij}
		\,
		\Theta_>(\s_1,\s_2)
	\,.
\end{align}
It should be also clear 
from Eqs.\,(\ref{ebc-de})-(\ref{ebc-sol})
that the solution to the differential equation
\begin{align}
	\label{zpert}
	\dot{Z}^i(\s)
	-
	M^i{}_j(\s)\mem Z^j(\s)
	\,=\,
		-(\omega^\circ{}^{-1})^{ij}\,
		Y_j(\s)
	\,,
\end{align}
with the retarded boundary condition,
is given by
\begin{align}
	\label{zpert-sol}
	Z^i(\s)
	\,\,\,=\,\,\,
	\int d\s'\,\,
	\Big({
		\phantom{\delta \z^i\hnem(\s)}
		\adjustbox{valign=c}{\begin{tikzpicture}[]
				\node (o) at (0,0) {};
				\node (xshift) at (0.8,0) {};
				\node (i0) at ($(o)$) {};
				\node (i1) at ($(i0)+1*(xshift)$) {};
				\node (i2) at ($(i0)+2*(xshift)$) {};
				\node (v) at (i1) {};
				\node[t] (L) at ($(i0)$) {$\mathllap{
						\delta \z^i\hnem(\s)
					}$};
				\node[t] (R) at ($(i2)$) {$\mathrlap{
						\delta \z^j\hnem(\s')
					}$};
				\draw[fprop] (R)--(L);
		\end{tikzpicture}}
		\kern-0.125em
		\phantom{{\delta \z^j\hnem(\s')}}
	}\Big)
	\,\mem
	Y_j(\s')
	\,.
\end{align}

With these understandings, 
we plug in the expansion in \eqref{deflect}
to time evolution vector field in \eqref{tevX}
and find that the Hamiltonian equations of motion,
$\dot{\z}^i(\s) = X^i(\z(\s)\hnem)$,
boil down to 
a set of
first-order ordinary differential equations
of the form
\begin{align}
	\label{xpert}
	\delta^{(n)}\snem\dot{\zeta}^i\hnem(\s)
	-
	M^i{}_j\hhnem(\s)\,
	\delta^{(n)}\hnem\zeta^j\hnem(\s)
	\,=\,
		- (\omega^\circ{}^{-1})^{ij}\,	(Y^{(n)}\hhnem[\bmz,\delta^{(1)}\snem\zeta,{\cdots}\hem,\delta^{(n-1)}\snem\zeta](\s)\hhnem)_j
	\,.
\end{align}
Explicitly,
the inhomogeneous terms
in the right-hand sides are found as
\begin{subequations}
\label{inhom}
\begin{align}
\label{inhom-1}
&
	(\omega^\circ{}^{-1})^{ij}\,
	(Y^{(1)}\hhnem[\bmz])_j
	\,=\,
		{\textstyle\frac{1}{0!}}\hem
		(X^{(1,0)}\snem(\bmz)\hnem)^i
\,,\\[0.35\baselineskip]
\label{inhom-2}
&
	(\omega^\circ{}^{-1})^{ij}\,
	(Y^{(2)}\hhnem[\bmz,\delta^{(1)}\snem\zeta])_j
	\\
&\nonumber
\begin{aligned}[t]
	\,=\,{}&{}
		{\textstyle\frac{1}{2!}}\hem
		(X^{(0,2)}\snem(\bmz(\t))\hnem)^i{}_{j_1j_2}\,\mem
		\delta^{(1)}\snem\zeta^{j_1}\,
		\delta^{(1)}\snem\zeta^{j_2}
	\\
	{}&{}+
		{\textstyle\frac{1}{1!}}\hem
		(X^{(1,1)}\snem(\bmz(\t))\hnem)^i{}_{j}\,\mem
		\delta^{(1)}\snem\zeta^j
	+
		{\textstyle\frac{1}{0!1!}}\hem
		(X^{(1,0,1)}\snem(\bmz(\t))\hnem)^i{}_{|k}\,\mem
		\delta^{(1)}\snem\zeta^k
	\\
	{}&{}+
		{\textstyle\frac{1}{0!}}\hem
		(X^{(2,0)}\snem(\bmz(\t))\hnem)^i
	\,,
\end{aligned}
\\[0.35\baselineskip]
\label{inhom-3}
&
	(\omega^\circ{}^{-1})^{ij}\,
	(Y^{(3)}\hhnem[\bmz,\delta^{(1)}\snem\zeta,\delta^{(2)}\snem\zeta])_j
\\
&\nonumber
\begin{aligned}[t]
	\,=\,{}&{}
		{\textstyle\frac{2}{2!}}\hem
		(X^{(0,2)}\snem(\bmz(\t))\hnem)^i{}_{j_1j_2}\,\mem
		\delta^{(2)}\snem\zeta^{j_1}\,
		\delta^{(1)}\snem\zeta^{j_2}
	\\
	{}&{}+
		{\textstyle\frac{1}{3!}}\hem
		(X^{(0,3)}\snem(\bmz(\t))\hnem)^i{}_{j_1j_2j_3}\,\mem
		\delta^{(1)}\snem\zeta^{j_1}\,
		\delta^{(1)}\snem\zeta^{j_2}\,
		\delta^{(1)}\snem\zeta^{j_3}
	\\[0.15\baselineskip]
	{}&{}+
		{\textstyle\frac{1}{1!}}\hem
		(X^{(1,1)}\snem(\bmz(\t))\hnem)^i{}_{j}\,\mem
		\delta^{(2)}\snem\zeta^{j}
	+
		{\textstyle\frac{1}{2!}}\hem
		(X^{(1,2)}\snem(\bmz(\t))\hnem)^i{}_{j_1j_2}\,\mem
		\delta^{(1)}\snem\zeta^{j_1}\,
		\delta^{(1)}\snem\zeta^{j_2}
	\\
	{}&{}+
		{\textstyle\frac{1}{1!1!}}\hem
		(X^{(1,1,1)}\snem(\bmz(\t))\hnem)^i{}_{j|k}\,\mem
		\delta^{(1)}\snem\zeta^{j}\,
		\delta^{(1)}\snem\zeta^{k}
	\\
	{}&{}+
		{\textstyle\frac{1}{0!2!}}\hem
		(X^{(1,0,2)}\snem(\bmz(\t))\hnem)^i{}_{|k_1k_2}\,\mem
		\delta^{(1)}\snem\zeta^{k_1}\,
		\delta^{(1)}\snem\zeta^{k_2}
	+
		{\textstyle\frac{1}{0!1!}}\hem
		(X^{(1,0,1)}\snem(\bmz(\t))\hnem)^i{}_{|k}\,\mem
		\delta^{(2)}\snem\zeta^{k}
	\\[0.15\baselineskip]
	{}&{}+
		{\textstyle\frac{1}{1!}}\hem
		(X^{(2,1)}\snem(\bmz(\t))\hnem)^i{}_{j}\,\mem
		\delta^{(1)}\snem\zeta^{j}
	\\
	{}&{}+
		{\textstyle\frac{1}{1!}}\hem
		(X^{(2,0,1)}\snem(\bmz(\t))\hnem)^i{}_{|(k)()}\,\mem
		\delta^{(1)}\snem\zeta^{k}
	+
		{\textstyle\frac{1}{1!}}\hem
		(X^{(2,0,1)}\snem(\bmz(\t))\hnem)^i{}_{|()(k)}\,\mem
		\delta^{(1)}\snem\zeta^{k}
	\\[0.15\baselineskip]
	{}&{}+
		{\textstyle\frac{1}{1!}}\hem
		(X^{(3)}\snem(\bmz(\t))\hnem)^i
	\,.
\end{aligned}
\end{align}
\end{subequations}

The general formula for the ``$Y$-functional''
$Y^{(n)}\hnem[\z,\delta^{(1)}\hnem\z,\cdots,\delta^{(n-1)}\hnem\z]$
is given by
\begin{align}
	\begin{split}
		\label{inhom-general}
		\sum_{n'=0}^n
		\mem
		\sum_{\substack{
				\mathrm{max}(\b) \mem<\mem n
				,
				\\
				|\c_1| + \mem\cdots\mem + |\c_{n'}| + |\b| \,=\, n
		}}
		\frac{\mem
			X^{(
				n'\nem\hhnem,\mem
				l_\b
				,\mem
				l_{\c_1}
				{+{\mem\cdots\mem}+}
				l_{\c_{n'}}\nem
				)}\snem(\bmz)
			\hem
			\big({
				\prod_{b\in\b}
				\delta^{(b)}\snem\z\mem
			}\big)
			\big({
				\prod_{c\in\c}
				\delta^{(c)}\snem\z\mem
			}\big)
			\mem}{\mem
			l_\b!\mem (
			l_{\c_1}
			{+{\mem\cdots\mem}+}
			l_{\c_{n'}}
			\vphantom{0^{0^0}}
			\hhnem)!
			\mem}
		\,,
	\end{split}
\end{align}
where $\b,\c_1,\cdots,\c_{n'}$ are ordered partitions of non-negative integers
$|\b|,|\c_1|,\cdots,|\c_{n'}|$
into non-negative integers,
and
$l_\a$
denotes the length of the partition $\a$.

Clearly, the result in \eqrefs{zpert}{zpert-sol}
applies to \eqref{xpert}:
\begin{align}
	\label{zy}
    \phantom{\small
        \dz^i\hnem(\s)
    }
    {\,\,\,
    \adjustbox{valign=c}{\begin{tikzpicture}[]
        \node[t] (ext) at (0,0) {$\mathllap{
            \dz^i\hnem(\s)
        }$};
        \node[ysqu] (x) at (1.0,0) {$n$};
        \draw[linear] (x)--(ext);
    \end{tikzpicture}}
    \,\,\,}
    \hhem\,=\,\hem
    \int d\s'\,\,
    \Big({
    \phantom{\small
        \dz^i\hnem(\s)
    }
    {\kern-0.125em
    \adjustbox{valign=c}{\begin{tikzpicture}[]
        \node[t] (L) at (0,0) {$\mathllap{
            \dz^i\hnem(\s)
        }$};
        \node (R) at (1.0,0) {};
        \node[t] (R2) at (1.14,0) {$\mathrlap{
            \dz^j\hnem(\s')
        }$};
        \draw[fprop] (R)--(L);
        \draw[linear] (R)--(R2);
    \end{tikzpicture}}
    \kern0.125em}
    \phantom{\small
        \dz^j\hnem(\s')
    }
    }\Big)
    \,
    (Y^{(n)}\snem[\bmz,\delta^{(1)}\snem\z,{\cdots}\hem,\delta^{(n-1)}\snem\z](\s')\hnem)_j
    \,.
\end{align}
Since $Y^{(n)}$
functionally depends on the worldline fluctuations of strictly lower order,
$\delta^{(1)}\hnem\z$, $\cdots$, $\delta^{(n-1)}\hnem\z$,
\eqrefs{inhom-general}{zy} together facilitates a recursion
from which $\delta^{(n)}\hnem\z^i(\s)$
can be obtained
as a functional of the background trajectory $\bmz$.

\subsubsection{Diagrammatic Berends-Giele Recursion}
\label{PSOL.2}

Now let us provide the diagrammatic approach.
The one-point function of the worldline fluctuation
is recursively obtained in the following fashion.
By graphically representing the $n$\textsuperscript{th}-order contribution to the one-point function as
\begin{align}
    \delta^{(n)}\snem\z^i\hnem(\s)
    \,\,=:
    \phantom{\dz^i\hnem(\s)}
    \adjustbox{valign=c}{\begin{tikzpicture}[]
        \node[t] (L) at (0,0) {$\mathllap{
            \dz^i\hnem(\s)
        }$};
        \node (R) at (1.0,0) {};
        \draw[linear] (R)--(L);
        \node[ysqu] (Y) at ($(R)+(0.15,0)$) {$n$};
    \end{tikzpicture}}\,\,
    \,,
\end{align}
the diagrammatics implies the following relations and so on:
\begin{subequations}
    \label{eq:Ms}
\begin{align}
    \label{eq:M1}
    {\,\,\,
    \adjustbox{valign=c}{\begin{tikzpicture}[]
        \node (ext) at (0,0) {};
        \node[ysqu] (x) at (1.0,0) {$1$};
        \draw[linear] (x)--(ext);
    \end{tikzpicture}}
    \,\,\,}
    \mem&=\mem
        {\,\,\,
        \adjustbox{valign=c}{\begin{tikzpicture}[]
            \node (ext) at (0,0) {};
            \node[odot] (v1) at (0.8,0) {};
            \draw[fprop] (v1)--(ext);
        \end{tikzpicture}}
        \,\,\,}
    \,,\\[0.5\baselineskip]
    \label{eq:M2}
    {\,\,\,
    \adjustbox{valign=c}{\begin{tikzpicture}[]
        \node (ext) at (0,0) {};
        \node[ysqu] (x) at (1.0,0) {$2$};
        \draw[linear] (x)--(ext);
    \end{tikzpicture}}
    \,\,\,}
    \mem&=\mem
        {\,\,\,
        \adjustbox{valign=c}{\begin{tikzpicture}[]
            \node (ext) at (0,0) {};
            \node[poly3] (v1) at (0.8,0) {};
            \draw[fprop] (v1)--(ext);
            \node[ysqu] (n11) at ($(v1)+(+60:0.6)$) {$1$};
            \node[ysqu] (n12) at ($(v1)+(-60:0.6)$) {$1$};
            \draw[linear] (n11)--(v1);
            \draw[linear] (n12)--(v1);
        \end{tikzpicture}}
        \,\,\,}
        \cdot\,\, \frac{1}{2!}\,\,
        +
        {\,\,\,
        \adjustbox{valign=c}{\begin{tikzpicture}[]
            \node (ext) at (0,0) {};
            \node[odot] (v1) at (0.8,0) {};
            \draw[fprop] (v1)--(ext);
            \node[ysqu] (n11) at ($(v1)+(0.6,0)$) {$1$};
            \draw[linear] (n11)--(v1);
        \end{tikzpicture}}
        \,\,\,}
        +
        {\,\,\,
        \adjustbox{valign=c}{\begin{tikzpicture}[]
            \node (ext) at (0,0) {};
            \node[odot] (v1) at (0.8,0) {};
            \draw[fprop] (v1)--(ext);
            \node[ysqu] (n11) at ($(v1)+(0.6,0)$) {$1$};
            \draw[linear] (n11)--(v1);
            \node (p) at ($(v1)+(135:0.22)$) {\pin};
        \end{tikzpicture}}
        \,\,\,}
        +
        {\,\,\,
        \adjustbox{valign=c}{\begin{tikzpicture}[]
            \node (ext) at (0,0) {};
            \node[odot] (v1) at (0.8,0) {};
            \draw[fprop] (v1)--(ext);
            \node[ysqu] (n11) at ($(v1)+(0.6,0)$) {$1$};
            \draw[linear] (n11)--(v1);
            \node (p) at ($(v1)+(45:0.22)$) {\pin};
        \end{tikzpicture}}
        \,\,\,}
    \,,\\[0.5\baselineskip]
    \label{eq:M3}
    {\,\,\,
    \adjustbox{valign=c}{\begin{tikzpicture}[]
        \node (ext) at (0,0) {};
        \node[ysqu] (x) at (1.0,0) {$3$};
        \draw[linear] (x)--(ext);
    \end{tikzpicture}}
    \,\,\,}
    \mem&=\mem
    \begin{aligned}[t]
    	&
        {\,\,\,
        \adjustbox{valign=c}{\begin{tikzpicture}[]
            \node (ext) at (0,0) {};
            \node[poly4] (v1) at (0.8,0) {};
            \draw[fprop] (v1)--(ext);
            \node[ysqu] (n11) at ($(v1)+(+90:0.6)$) {$1$};
            \node[ysqu] (n12) at ($(v1)+(-90:0.6)$) {$1$};
            \node[ysqu] (n13) at ($(v1)+(  0:0.6)$) {$1$};
            \draw[linear] (n11)--(v1);
            \draw[linear] (n12)--(v1);
            \draw[linear] (n13)--(v1);
        \end{tikzpicture}}
        \,\,\,}
        \cdot\,\, \frac{1}{3!}\,\,
        +
        {\,\,\,
        \adjustbox{valign=c}{\begin{tikzpicture}[]
            \node (ext) at (0,0) {};
            \node[poly3] (v1) at (0.8,0) {};
            \draw[fprop] (v1)--(ext);
            \node[ysqu] (n11) at ($(v1)+(+60:0.6)$) {$2$};
            \node[ysqu] (n12) at ($(v1)+(-60:0.6)$) {$1$};
            \draw[linear] (n11)--(v1);
            \draw[linear] (n12)--(v1);
        \end{tikzpicture}}
        \,\,\,}
    \\[0.12\baselineskip]
   	&{}+{}
		\Bigg({
	        {\,\,\,
	        \adjustbox{valign=c}{\begin{tikzpicture}[]
	            \node (ext) at (0,0) {};
	            \node[odot] (v1) at (0.8,0) {};
	            \draw[fprop] (v1)--(ext);
	            \node[ysqu] (n11) at ($(v1)+(+60:0.6)$) {$1$};
	            \node[ysqu] (n12) at ($(v1)+(-60:0.6)$) {$1$};
	            \draw[linear] (n11)--(v1);
	            \draw[linear] (n12)--(v1);
	        \end{tikzpicture}}
	        \,\,\,}
	        +
	        {\,\,\,
	        \adjustbox{valign=c}{\begin{tikzpicture}[]
	            \node (ext) at (0,0) {};
	            \node[odot] (v1) at (0.8,0) {};
	            \draw[fprop] (v1)--(ext);
	            \node[ysqu] (n11) at ($(v1)+(+60:0.6)$) {$1$};
	            \node[ysqu] (n12) at ($(v1)+(-60:0.6)$) {$1$};
	            \draw[linear] (n11)--(v1);
	            \draw[linear] (n12)--(v1);
                \node (p) at ($(v1)+(135:0.22)$) {\pin};
	        \end{tikzpicture}}
	        \,\,\,}
 	        +
 	        {\,\,\,
 	        \adjustbox{valign=c}{\begin{tikzpicture}[]
 	            \node (ext) at (0,0) {};
 	            \node[odot] (v1) at (0.8,0) {};
 	            \draw[fprop] (v1)--(ext);
 	            \node[ysqu] (n11) at ($(v1)+(+60:0.6)$) {$1$};
 	            \node[ysqu] (n12) at ($(v1)+(-60:0.6)$) {$1$};
 	            \draw[linear] (n11)--(v1);
 	            \draw[linear] (n12)--(v1);
                 \node (p) at ($(v1)+(-35:0.22)$) {\pin};
 	        \end{tikzpicture}}
 	        \,\,\,}
  	        +
  	        {\,\,\,
  	        \adjustbox{valign=c}{\begin{tikzpicture}[]
  	            \node (ext) at (0,0) {};
  	            \node[odot] (v1) at (0.8,0) {};
  	            \draw[fprop] (v1)--(ext);
  	            \node[ysqu] (n11) at ($(v1)+(+60:0.6)$) {$1$};
  	            \node[ysqu] (n12) at ($(v1)+(-60:0.6)$) {$1$};
  	            \draw[linear] (n11)--(v1);
  	            \draw[linear] (n12)--(v1);
                  \node (p) at ($(v1)+(30:0.22)$) {\pin};
  	        \end{tikzpicture}}
  	        \,\,\,}
        }\Bigg)\,
        \cdot\, \frac{1}{2!}\,\,
        \\[0.15\baselineskip]
        &
        {}+{}
        \bb{
            {\,\,\,
            \adjustbox{valign=c}{\begin{tikzpicture}[]
                \node (ext) at (0,0) {};
                \node[odot] (v1) at (0.8,0) {};
                \draw[fprop] (v1)--(ext);
                \node[ysqu] (n11) at ($(v1)+(0.6,0)$) {$2$};
                \draw[linear] (n11)--(v1);
            \end{tikzpicture}}
            \,\,\,}
            +
            {\,\,\,
            \adjustbox{valign=c}{\begin{tikzpicture}[]
                \node (ext) at (0,0) {};
                \node[odot] (v1) at (0.8,0) {};
                \draw[fprop] (v1)--(ext);
                \node[ysqu] (n11) at ($(v1)+(0.6,0)$) {$2$};
                \draw[linear] (n11)--(v1);
                \node (p) at ($(v1)+(135:0.22)$) {\pin};
            \end{tikzpicture}}
            \,\,\,}
            +
            {\,\,\,
            \adjustbox{valign=c}{\begin{tikzpicture}[]
                \node (ext) at (0,0) {};
                \node[odot] (v1) at (0.8,0) {};
                \draw[fprop] (v1)--(ext);
                \node[ysqu] (n11) at ($(v1)+(0.6,0)$) {$2$};
                \draw[linear] (n11)--(v1);
                \node (p) at ($(v1)+(45:0.22)$) {\pin};
            \end{tikzpicture}}
            \,\,\,}
        }
        \,.
	\end{aligned}
\end{align}
\end{subequations}

It can be seen that
the tree Feynman graphs enumerated in Eqs.\,(\ref{eq:M1})-(\ref{eq:M3})
provides the solution to \eqref{xpert}
for the $Y$-functionals in \eqref{inhom}.
While the pinched symplectic vertices of valence two
implement
the geometric series expansion
of the Hamiltonian vector field
in \eqref{eq:spt-tev}
by concatenating $\omega'$ and $\omega^\circ{}^{-1}$,
the Hamiltonian and symplectic vertices in general
encode the complete expansion in terms of \eqref{xtensor}.
For instance,
the trivalent graph in \eqref{eq:M2} describes 
the first term in \eqref{inhom-2}
via a Hamiltonian vertex,
implementing the $X$-tensor in \eqref{xtensor}
with $m=2$.
The other graphs in \eqref{eq:M2}
together implements the other terms in \eqref{inhom-2},
encoding the $X$-tensors with nontrivial $n$ and $k$ values
via symplectic vertices.

Let us elaborate on
the mechanics of symplectic vertices 
in these recursion relations.
Regarding the second-order worldline fluctuation $\delta^{(2)}\hnem\z^i$,
the $Y$-functional found in \eqref{inhom-2}
seems to have put a source term
$X^{(2,0)}$
that exhibits a higher sequence of $\omega'$.
Apparently, this term is absent in 
the diagrammatic recursion relation in \eqref{eq:M2}.
However, 
a closer look confirms that
it is indeed generated by the pinching of a fat grey propagator.

The diagram in \eqref{eq:M2}
due to the regular symplectic vertex
features a symmetrized term
$\dot{\bmz}^k\mem \omega'_{k(i,j)}$
in the tensor factor.
Meanwhile, the sum of two pinched diagrams
can be rewritten by 
utilizing the following ``commutator'' identity for the pinching:
\begin{align}
	{\,\,
	\adjustbox{valign=c}{\begin{tikzpicture}[]
			\node (o) at (0,0) {};
			\node (xshift) at (0.8,0) {};
			\node (i0) at ($(o)$) {};
			\node (i1) at ($(i0)+1*(xshift)$) {};
			\node (i2) at ($(i0)+2*(xshift)$) {};
			\node (v) at (i1) {};
			\node[t] (L) at ($(i0)$) {};
			\node[t] (R) at ($(i2)$) {};
			\node (p) at ($(v)+(135:0.22)$) {\pin};
			\node (q) at ($(v)+(-45:0.22)$) {\phantom{\pin}};
			\draw[linear] (v)--(L);
			\draw[linear] (v)--(R);
			\node[odot] (V) at ($(v)$) {};
	\end{tikzpicture}}
	\,\,}
	-
	{\,\,
	\adjustbox{valign=c}{\begin{tikzpicture}[]
			\node (o) at (0,0) {};
			\node (xshift) at (0.8,0) {};
			\node (i0) at ($(o)$) {};
			\node (i1) at ($(i0)+1*(xshift)$) {};
			\node (i2) at ($(i0)+2*(xshift)$) {};
			\node (v) at (i1) {};
			\node[t] (L) at ($(i0)$) {};
			\node[t] (R) at ($(i2)$) {};
			\node (p) at ($(v)+(45:0.22)$) {\pin};
			\node (q) at ($(v)+(-45:0.22)$) {\phantom{\pin}};
			\draw[linear] (v)--(L);
			\draw[linear] (v)--(R);
			\node[odot] (V) at ($(v)$) {};
	\end{tikzpicture}}
	\,\,}
	\mem\sim\,\,\,\,
		\frac{1}{2}\,
		\bigg[\,{
			\frac{\partial}{\partial \s_1}
			\,,\,
			\omega'_{ij}\hnem(\bmz(\s_1)\hnem)
		}\,\bigg]
		\,=\,
		\frac{1}{2}\,
			\dot{\bmz}^k\hnem(\s_1)\,
			\omega'_{ij,k}\hnem(\bmz(\s_1)\hnem)
	\,.
\end{align}
Using the Jacobi identity,
this term is replaced with 
the antisymmetrized combination,
$\dot{\bmz}^k\mem \omega'_{k[i,j]}$,
which combines with
the earlier term with the factor $\dot{\bmz}^k\mem \omega'_{k(i,j)}$
to reproduce the $X^{(1,1)}$ term in \eqref{inhom-2}.

As a result, it remains to compute $2$ times the last diagram in \eqref{eq:M2}.
Substituting \eqref{eq:M1},
one pinches a fat grey propagator.
This returns a delta function
plus the $M$-matrix in \eqref{M-matrix}
due to the following identity:
\begin{align}
    \adjustbox{valign=c}{\,\,
    \begin{tikzpicture}[]
        \node (o) at (0,0) {};
        \node (xshift) at (0.65,0) {};
        \node (i0) at ($(o)$) {};
        \node (i1) at ($(i0)+1*(xshift)$) {};
        \node (i2) at ($(i0)+2*(xshift)$) {};
        \node (v) at (i2) {};
        \node[t] (L) at ($(i0)$) {};
        \draw[fprop] (v)--(L);
    \end{tikzpicture}
    \,\,\,}
	=
    \adjustbox{valign=c}{\,\,\,
    \begin{tikzpicture}[]
        \node (o) at (0,0) {};
        \node (xshift) at (1.2,0) {};
        \node (i0) at ($(o)$) {};
        \node (i1) at ($(i0)+1*(xshift)$) {};
        \node (v) at (i1) {};
        \node[t] (L) at ($(i0)$) {};
        \draw[linear] (v)--(L);
    \end{tikzpicture}
    \,\,\,}
    +
    \adjustbox{valign=c}{\,\,\,
    \begin{tikzpicture}[]
        \node (o) at (0,0) {};
        \node (xshift) at (-0.6,0) {};
        \node (i0) at ($(o)$) {};
        \node (i1) at ($(i0)+1.8*(xshift)$) {};
        \node (i2) at ($(i1)+1.55*(xshift)$) {};
        \node (i3) at ($(i1)+1.6*(xshift)$) {};
        \draw[fprop] (i0)--(i1);
        \draw[qprop] (i1)--(i2);
        \node[poly2] (p2) at ($(i1)$) {};
        \draw[linear] (i2)--(i3);
    \end{tikzpicture}
    \,\,\,}
    \,.
\end{align}
The $M$-matrix term then reproduces the $X^{(1,1)}$ term in \eqref{inhom-2},
while the delta function term finally generates the concatenated series
$X^{(2,0)} \sim \omega^\circ{}^{-1} \omega' \omega^\circ{}^{-1} \omega' \omega^\circ{}^{-1} \partial H$.

Using similar rearrangements of terms,
one can check
the equivalence between
the solutions in 
\Secs{PSOL.1}{PSOL.2}
at higher orders.

Note that the discussion here
justifies why 
one would want to use the identity in \eqref{max-convert}
to generate the boundary term in \eqref{Lbd}.
Crucially, the Hamiltonian equations of motion in the bulk 
have no reference to the symplectic potential.
To manifest this property,
it is natural to 
make the zero-valence symplectic vertex
the only vertex that invokes the symplectic potential.

\subsection{Recoil Operator and Compton Amplitude}
\label{RECOIL}

In this final subsection,
we point out that
the recoil operators
in the framework of self-force expansion,
introduced by works \cite{Cheung:2023lnj-EMR,jordan2023,Cheung:2024byb},
admit a unified formula 
in the phase space worldline formalism.
This is essentially a rephrasing of the insight \cite{Cheung:2023lnj-EMR,jordan2023,Cheung:2024byb}
that the vertices of fluctuation valence-one
encode the equations of motion,
but examining it from a top-down, phase space perspective
is helpful for identifying
isomorphic structures between
gauge theory and gravity.

Suppose
a massive scalar particle
coupled to gauge fields or gravity.
Suppose we pursue symplectic perturbation theory
such that the Hamiltonian is simply momentum-squared.
Then the diagrams 
at the second order in the coupling
are
\begin{subequations}
    \label{eq:recoils}
\begin{align}
    \label{eq:recoil1}
    \adjustbox{valign=c}{\begin{tikzpicture}[]
        \node (L) at (0,0) {};
        \node (R) at (1.0,0) {};
        \draw[linear] (R)--(L);
        \node[odot] (l) at (L) {};
        \node[odot] (r) at (R) {};
    \end{tikzpicture}}\,\mem
    \,\,\,&=\,\,\,
        \int d\s_1\mem d\s_2
        \,\,
            \Theta(\s_1,\s_2)\,
            R^{(1)}\hnem(\s_1,\s_2)
    \,,\\
    \label{eq:recoil2}
    \adjustbox{valign=c}{\begin{tikzpicture}[]
        \node (xshift) at (0.8,0) {};
        \node (L) at ($0*(xshift)$) {};
        \node (M) at ($1*(xshift)$) {};
        \node (R) at ($2*(xshift)$) {};
        \draw[linear] (R)--(M)--(L);
        \node[odot] (l) at (L) {};
        \node[poly2] (m) at (M) {};
        \node[odot] (r) at (R) {};
    \end{tikzpicture}}\,\mem
    \,\,\,&=\,\,\,
        \int d\s_1\mem d\s_2
        \,\,
            \Delta(\s_1,\s_2)\,
            R^{(2)}\hnem(\s_1,\s_2)
    \,,
\end{align}
\end{subequations}
where $\Delta(\s_1,\s_2)$ is a second-order propagator.
Physically,
these
encode the leading order recoil
of the particle
due to the interactions.
For instance,
they can describe
the wobble of a 
heavy massive particle 
during
the process of receiving and emitting photons or gravitons,
where worldline fluctuations
propagate between worldline times $\s_1$ and $\s_2$.

The diagrams in
\eqref{eq:recoils}
are the phase space incarnation of the recoil operator.
Notably, the Feynman rules derived in \Sec{General.feyn}
provide master formulae:
\begin{subequations}
    \label{eq:master-recoil}
\begin{align}
    \label{eq:master-recoil-1}
    R^{(1)}\hnem(\s_1,\s_2)
    &=
    	-
	    \dot{\bmz}(\s_1)
	    \,
	    	\omega'(\bmz(\s_1)\hnem)
	    \,
	    \omega^\circ{}^{-1}
	    \,
	    	\omega'(\bmz(\s_2)\hnem)
	    \,
	    \dot{\bmz}\hnem(\s_2)
    \,,\\
    \label{eq:master-recoil-2}
    R^{(2)}\hnem(\s_1,\s_2)
    &=
    	    \dot{\bmz}(\s_1)
	 	    \,
    	    	\omega'(\bmz(\s_1)\hnem)
	 	    \,
	 	    	(\partial\partial H)
	 	    \,
    	    	\omega'(\bmz(\s_2)\hnem)
    	    \,
    	    \dot{\bmz}(\s_2)
    \,,
\end{align}
\end{subequations}
where $(\partial\partial H)$ is a constant tensor
since the Hamiltonian is quadratic.
We have omitted contracted indices to avoid clutter.

The charm of these formulae
is that
they establish a correspondence between
gauge theory and gravity
in terms of 
the elements in the symplectic geometry of particles,
which is useful in spirit of double copy at the worldline level.
In particular, consider the point made in \Sec{NA.ps}.
With such an understanding
of a correspondence in terms of particle symplectic structures,
the recoil operators in
\rrcite{jordan2023,Cheung:2023lnj-EMR}
are systematically reproduced from the master formulae in \eqref{eq:master-recoil},
together with the Yang-Mills generalization.
One simply needs to identify the elements of symplectic geometry such as
$\omega^\circ{}^{-1}$ and $\omega'$.
The gravitational recoil operator in \rcite{Cheung:2023lnj-EMR}
is written in terms of Christoffel symbols,
so to find a match from the tetrad setup,
one needs to not forget adding the first-order contribution from 
$R^{(1)}\hnem(\s_1,\s_2)$.
This first-order channel may seem like an extra complication,
but in terms of observing double copy structures,
it suggestively utilizes a diffeomorphism algebra
in parallel with the color Lie algebra in nonabelian gauge theory.

Note that the term recoil operator
assumes the context that
the vertices can be viewed as functionals of the off-shell fields.
However,
the structures and universality observed in this subsection
are
straightforwardly 
inherited to on-shell computations,
in which case one obtains the Compton amplitudes
in gauge theories and gravity
in a uniform fashion.

\section{Bulk Perturbation Theory}
\label{CK}

Color-kinematics duality
is a remarkable property of
scattering amplitudes 
that draws a precise correspondence between 
perturbative
gauge theory and gravity.
It has origins in open-closed duality in string theory \cite{KLT}
and has been formulated within quantum field theory 
by Bern, Carrasco, and Johansson (BCJ) \cite{BCJ1, BCJ2},
establishing that 
gauge theory amplitudes ``square'' to
gravitational amplitudes.
This squaring relation has been also observed at the level of classical solutions,
including
black holes
\cite{monteiro2014black,Luna:2015paa}
and
gravitational instantons
\cite{Berman:2018hwd,note-sdtn,nja}.
See \rcite{BCJReview} for a comprehensive review.

A working definition of color-kinematics duality
is given in terms of Bi-Adjoint Scalar (BAS) theory.
BAS theory is a theory of a scalar field $\Phi^{a\ta}$,
carrying two Lie algebra indices $a = 1,\cdots,\dim\g$ and $\ta = 1,\cdots,\dim\tilde{\g}$.
The two Lie algebras, $\g$ and $\tilde{\g}$, 
are independent
and must obey the Jacobi identity by definition.
The equations of motion of the BAS field are
\begin{align}
	\label{BAS}
	\Box\mem \Phi^{a\ta}
	\,=\,
		-
		f^a{}_{bc}\mem 
		\tf^\ta{}_{\tb\tc}\,
			\Phi^{b\tb}\mem \Phi^{c\tc}
	\,.
\end{align}
A theory exhibits
color-kinematics duality
at tree level
if it is equivalent to BAS theory
at the level of equations of motion,
for a choice of the Lie algebras $\g$ and $\tilde{\g}$
\cite{cheung2021cck}.

Well-established instances are 
nonlinear sigma model and special Galileon theory
in $d$ dimensions,
which take
$\g = \su(N)$ and $\tg = \sdiff(\R^d)$
and
$\g = \tg = \sdiff(\R^d)$,
respectively \cite{cheung2021cck}.
Here, $\sdiff(\mathbb{R}^d)$ denotes the Lie algebra of volume-preserving diffeomorphisms in $d$ dimensions,
which is infinite-dimensional.
When applied to plane-wave states,
it takes momenta as indices
and thus is referred to as a kinematic algebra.

For gauge theories,
our current understanding on color-kinematics duality
has been far less complete,
and an explicit identification of the kinematic algebra
has been limited to the self-dual sector in four dimensions
\cite{DonalSDYM1,DonalSDYM2,DonalSDYM3,HenrikSDYM,park1990self,husain1994self,plebanski1994self,plebanski1975some}.
Hence, the question of the kinematic algebra in general $d$ dimensions remains unresolved.

A partial progress has been made by Cheung and Mangan \cite{cheung2021cck},
where the prototype theory
is taken as 
BAS theory coupled to a gauge connection $A^a{}_\a$:
\begin{align}
	\label{GBAS}
	D^2 \Phi^{a\ta}
	\,=\,
		- f^a{}_{bc}\mem \tf^\ta{}_{\tb\tc}\,
			\Phi^{b\tb}\mem \Phi^{c\tc}
	\quad\text{where}\quad
	D_\a \Phi^{a\ta}
	\,=\,
		\partial_\a \Phi^{a\ta}
		+ f^a{}_{cb}\mem A^c{}_\a\mem \Phi^{b\ta}
	\,.
\end{align}
In this construction,
the two Lie algebras play asymmetric roles:
the Lie algebra $\g$ is gauged (color index),
whereas
the Lie algebra $\tilde{\g}$ is global (flavor index).
The observation is that Yang-Mills theory and Born-Infeld theory
can be formulated in terms of the gauged BAS equations of motion in \eqref{GBAS},
for $\tg = \so(1,d-1)$ and $\tg = \sdiff(\R^d)$,
respectively.

For gravity, 
we note that
the Penrose wave equation
\cite{Penrose:1960eq,Ryan:1974nt}
describes
an instance of a fully gauged BAS equation
with $\g = \tg = \so(1,d-1)$:
\begin{align}
	\label{GGBAS}
	D^2\mem \Phi^{\ta_1\ta_2}
	+ 2\Phi^{\ta_1\td}\mem \Phi_\td{}^{\ta_2}
	\,=\,
		-\tf^{\ta_1}{}_{\tb_1\tc_1}\mem \tf^{\ta_2}{}_{\tb_2\tc_2}\,
			\Phi^{\tb_1\tb_2}\mem \Phi^{\tc_1\tc_2}
	\,.
\end{align}
Here, all indices are subject to gauge transformations,
being coupled to the spin connection as a Lorentz-valued gauge connection.
Also, the presence of a metric is also required,
defined from a tetrad:
\begin{align}
\begin{split}
	D_\m \Phi^{\ta_1\ta_2}
	\,&=\,
		\partial_\m \Phi^{\ta_1\ta_2}
		+ \tf^{\ta_1}{}_{\tc_1\tb_1}\mem A^{\tc_1}{}_\m\mem \Phi^{\tb_1\ta_2}
		+ \tf^{\ta_2}{}_{\tc_2\tb_2}\mem A^{\tc_2}{}_\m\mem \Phi^{\ta_1\tb_2}
	\,,\\
	D^2 \Phi^{\ta_1\ta_2}
	\,&=\,
		\eta^{\a\b}
		E^\m{}_\a E^\n{}_\b\mem
		D_\m D_\n \Phi^{\ta_1\ta_2}
	\,.
\end{split}
\end{align}

Using these ideas,
the backgrounds
describing the nonlinear superposition of two plane waves
can be readily derived.
In this appendix, let us adopt a convention
such that the bivector polarizations are related to the usual vector polarizations $e_I{}_\a$ as
\begin{align}
	\varphi_I^\ta\, (\Sigma_\ta)_{\a\b}
	\,=\,
	\varphi_I{}_{\a\b}
	\,=\,
		\sqrt{2}\mem (k_I \swedge e_I)_{\a\b}
	\,,
\end{align}
unlike in the main article.
$(\Sigma_\ta)_{\a\b} = -(\Sigma_\ta)_{\b\a}$ describe the generators of the Lorentz algebra $\tg = \so(1,d-1)$
such that
\begin{align}
	(\Sigma_\tc)^\a{}_\b\mem \tf^\tc{}_{\ta\tb}
	\,=\,
		(\Sigma_\ta)^\a{}_\c\mem (\Sigma_\tb)^\c{}_\b
		-
		(\Sigma_\ta)^\a{}_\c\mem (\Sigma_\tb)^\c{}_\b
	\,.
\end{align}

For Yang-Mills theory,
one views the field strength as 
the gauged bi-adjoint field in \eqref{GBAS},
with $\g = \su(N)$ and $\tg = \so(1,d-1)$:
$F^a{}_{\a\b} = F^{a\ta}\mem (\Sigma_\ta)_{\a\b}$.
Employing the axial gauge,
the gauge potential is determined from $F^{a\ta}$ as
\begin{align}
	\label{axialgauge-A}
	A^a{}_\a\mem \eta^\a 
	\,=\,0
	\qiq
	A^a{}_\b
	\,=\,
		\frac{1}{\eta\mdot\partial}\, F^{a\ta}\,
			\eta^\a\mem (\Sigma_\ta)_{\a\b}
	\,.
\end{align}
Then, via Berends-Giele recursion,
one finds that the configuration
\begin{align}
\begin{split}
	\label{F34}
	F^{a\ta}(x)
	\,=\,
	\begin{aligned}[t]
		&
		\frac{ig}{\sqrt{2}}\,
			c_3^a\, \varphi_3^{\ta}\,
				\mathe^{ik_3x}
		+
		\frac{ig}{\sqrt{2}}\,
			c_4^a\, \varphi_4^{\ta}\,
				\mathe^{ik_4x}
		\\
		&{}
		+
		g^2\,
			c_{34}^a\mem
			\bb{
				\varphi_{34}^{\ta}
				- \varphi_3^{\ta}\,
					\frac{
						\eta\hhem\varphi_4\hnem k_3
					}{
						k_4\eta
					}
				+ \varphi_4^{\ta}\,
					\frac{
						\eta\hhem\varphi_3\hnem k_4
					}{
						k_3\eta
					}
			}\mem
				\frac{1}{t_{34}}\,
				\mathe^{i(k_3+k_4)x}
		+ \O(g^3)
	\end{aligned}
\end{split}
\end{align}
solves the Yang-Mills equations up to $\O(g^2)$.
Also,
from \eqrefs{F34}{axialgauge-A}, it follows that
\begin{align}
	\label{A34}
	A^a{}_\b(x)
	\,=\,
		&
		\frac{g}{\sqrt{2}}\,
			c_3^a\, 
			\frac{(\eta\varphi_3)_\b}{k_3\eta}
				\,\mathe^{ik_3x}
		+
		\frac{g}{\sqrt{2}}\,
			c_4^a\, 
			\frac{(\eta\varphi_4)_\b}{k_4\eta}
				\,\mathe^{ik_4x}
		\\
		&{}
		-i
		g^2\,
			c_{34}^a\mem
			\bb{
				\varphi_{34}^{\ta}
				- \varphi_3^{\ta}\,
					\frac{
						\eta\hhem\varphi_4\hnem k_3
					}{
						k_4\eta
					}
				+ \varphi_4^{\ta}\,
					\frac{
						\eta\hhem\varphi_3\hnem k_4
					}{
						k_3\eta
					}
			}\mem
				\frac{(\eta\Sigma_\ta)_\b}{k_3\eta \mplus k_4\eta}
				\mem
				\frac{1}{t_{34}}
				\,\mathe^{i(k_3+k_4)x}
		+ \O(g^3)
		\,,
\nonumber
\end{align}
where 
$(\eta\Sigma_\ta)_\b = \eta^\a (\Sigma_\ta)_{\a\b}$
and
$(\eta\varphi_3)_\b = \eta^\a \varphi_3{}_{\a\b}$,
etc.

For general relativity,
the Riemann curvature in the orthonormal frame
serves as the fully gauged bi-adjoint field in \eqref{GGBAS}
for $\g = \tg = \so(1,d-1)$:
$
	R_{\a\b\c\d}
	=
		R^{\ta\tb}\,
			(\Sigma_\ta)_{\a\b}
			\mem
			(\Sigma_\tb)_{\c\d}
$.
Adopting an axial gauge condition such that
\begin{align}
	A^\ta{}_\m(x)\mem \eta^\m \,=\, 0
	\,,\quad
	e^\a{}_\m(x)\mem \eta^\m
	\,=\,
		\delta^\a{}_\m\mem \eta^\m
	\,,
\end{align}
the spin connection $A^\ta{}_\m$ and the tetrad $E^\m{}_\a = (e^{-1})^\m{}_\a$
are determined as
\begin{align}
\begin{split}
	A^a{}_\n
	\,&=\,
		\frac{1}{\eta\mdot\partial}\,
		\BB{
			R^{\ta\tb}\,
			(\Sigma_\tb)_{\a\b}\mem
				(\delta^\a{}_\m\mem \eta^\m)
				\mem
				e^\b{}_\n
		}
	\,,\\
	E^\m{}_\b
	\,&=\,
		\delta^\m{}_\b
		+ 
		\frac{1}{\eta\mdot\partial}\,
		\BB{
			E^\m{}_\a\mem
			(\Sigma_\ta \eta)^\a\mem
			A^\ta{}_\n\mem
			E^\n{}_\b
		}
	\,,
\end{split}
\end{align}
which follow from the relations
\begin{align}
\begin{split}
	R^{\ta\tb}\mem
	(\Sigma_\tb)_{\c\d}\mem
	e^\c{}_\r\mem e^\d{}_\s
	\,&=\,
		\partial_\r A^\ta{}_\s
		-
		\partial_\s A^\ta{}_\r
		+ \tf^\ta{}_{\tb\tc}\mem A^\tb{}_\r\mem A^\tc{}_\s
	\,,\\
	E^\m{}_\wrap{\b,[\r}\mem e^\b{}_\wrap{\s]}
	\,&=\,
		E^\m{}_\a\,
		(\Sigma_\ta)^\a{}_\b\,
		A^\ta{}_\wrap{[\r}\,
		e^\b{}_\wrap{\s]}
	\,.
\end{split}
\end{align}
As a result,
it can be shown
that the nonlinear superposition of two gravitational waves
is described by
the following configuration:
\begin{align}
\begin{split}
	R^{\ta\tb}
	\,=\,
		&{}
		\frac{g}{2}\, \varphi_3^\ta\mem \varphi_3^\tb
		\,\mathe^{ik_3x}
		+
		\frac{g}{2}\, \varphi_4^\ta\mem \varphi_4^\tb
		\,\mathe^{ik_4x}
	\\
		&{}
		- \frac{g^2}{2}\,
		\left({
		{\renewcommand{\arraystretch}{1.25}
		\begin{array}{l}
			\BB{
				\varphi_{34}
				- \varphi_3\, \frac{\eta\hhem\varphi_4\hnem k_3}{k_4\eta}
			}^\ta
			\BB{
				\varphi_{34}
				- \varphi_3\, \frac{\eta\hhem\varphi_4\hnem k_3}{k_4\eta}
			}^\tb
			\\
			+
			\BB{
				\varphi_{34}
				- \varphi_4\, \frac{\eta\hhem\varphi_3\hnem k_4}{k_3\eta}
			}^\ta
			\BB{
				\varphi_{34}
				- \varphi_4\, \frac{\eta\hhem\varphi_3\hnem k_4}{k_3\eta}
			}^\tb
			\\
			- \varphi_{34}^\ta\mem \varphi_{34}^\tb
			+ \varphi_3\mdot\varphi_4\,
				\BB{
					\varphi_3^\ta\mem \varphi_4^\tb
					+
					\varphi_4^\ta\mem \varphi_3^\tb
				}
		\end{array}}
		}\right)
		\,\frac{1}{t_{34}}\,
			\mathe^{i(k_3+k_4)x}
	\,,
\end{split}
\end{align}
where $\varphi_3\mdot\varphi_4 = \varphi_3{}_\ta\mem \varphi_4^\ta = \frac{1}{2}\, \varphi_3^{\a\b}\mem \varphi_4{}_{\a\b}$.

%	
	% 
%	\newpage
	\let\i\undefined
	\let\c\undefined
	\bibliography{references.bib}
	
\end{document}